\documentclass[twocolumn]{aastex62}

\newcommand{\xh}[1]{[#1/H]}
\newcommand{\xm}[1]{[#1/M]}

\newcommand{\xfe}[1]{[#1/Fe]}
\newcommand{\mh}[0]{[M/H]}
\newcommand{\afe}[0]{[$\alpha$/Fe]}
\newcommand{\am}[0]{[$\alpha$/M]}
\newcommand{\feh}[0]{[Fe/H]}
\newcommand{\teff}[0]{$T_{\rm eff}$}
\newcommand{\logg}[0]{$\log{g}$}
\newcommand{\vmic}[0]{$v_{\rm mic}$}
\newcommand{\vmac}[0]{$v_{\rm mac}$}
\newcommand{\vsini}[0]{$v\sin i$}
\newcommand{\vrad}[0]{$v_{\rm rad}$}

\received{2020}
\revised{2020}
\accepted{2020}
\submitjournal{AJ}
\shorttitle{APOGEE Data and Spectral Analysis}
\shortauthors{J\"onsson et al.}

\begin{document}

\title{APOGEE Data and Spectral Analysis from SDSS Data Release 16:\\Seven Years of Observations Including First Results from APOGEE-South}

\correspondingauthor{Henrik J\"onsson}
\email{henrikj@astro.lu.se}

\author[0000-0002-4912-8609]{Henrik J\"onsson}
\affil{Materials Science and Applied Mathematics, Malm\"o University, SE-205 06 Malm\"o, Sweden}
\affil{Lund Observatory, Department of Astronomy and Theoretical Physics, Lund University, Box 43, SE-22100 Lund, Sweden}

\author[0000-0002-9771-9622]{Jon A. Holtzman}
\affil{New Mexico State University, Las Cruces, NM 88003, USA}

\author{Carlos Allende Prieto}
\affil{Instituto de Astrof\'isica de Canarias (IAC), E-38205 La Laguna, Tenerife, Spain}
\affil{Universidad de La Laguna, Dpto. Astrof\'isica, E-38206 La Laguna, Tenerife, Spain}

\author{Katia Cunha}
\affil{Steward Observatory, The University of Arizona, 933 North Cherry Avenue, Tucson, AZ 85721-0065, USA}
\affil{Observat\'orio Nacional, Rua General Jos\'e Cristino, 77, 20921-400 S\~ao Crist\'ov\~ao, Rio de Janeiro, RJ, Brazil}

\author{D. A. Garc\'ia-Hern\'andez}
\affil{Instituto de Astrof\'isica de Canarias (IAC), E-38205 La Laguna, Tenerife, Spain}
\affil{Universidad de La Laguna, Dpto. Astrof\'isica, E-38206 La Laguna, Tenerife, Spain}

\author{Sten Hasselquist}
\affil{Department of Physics \& Astronomy, University of Utah, Salt Lake City, UT, 84112, USA}
\affil{NSF Astronomy and Astrophysics Postdoctoral Fellow}

\author{Thomas Masseron}
\affil{Instituto de Astrof\'isica de Canarias (IAC), E-38205 La Laguna, Tenerife, Spain}
\affil{Universidad de La Laguna, Dpto. Astrof\'isica, E-38206 La Laguna, Tenerife, Spain}

\author[0000-0001-5832-6933]{Yeisson Osorio}
\affil{Instituto de Astrof\'isica de Canarias (IAC), E-38205 La Laguna, Tenerife, Spain}
\affil{Universidad de La Laguna, Dpto. Astrof\'isica, E-38206 La Laguna, Tenerife, Spain}

\author{Matthew Shetrone}
\affil{University of Texas at Austin, McDonald Observatory, Fort Davis, TX 79734, USA}

\author{Verne Smith}
\affil{National Optical Astronomy Observatory, 950 North Cherry Avenue, Tucson, AZ 85719, USA}

\author[0000-0003-1479-3059]{Guy S. Stringfellow}
\affil{Center for Astrophysics and Space Astronomy, Department of Astrophysical and Planetary Sciences, University of Colorado, Boulder, CO, 80309-0389, USA}

\author{Dmitry Bizyaev}
\affil{Apache Point Observatory and New Mexico State University, PO Box 59, Sunspot, NM 88349-0059, USA}
\affil{Sternberg Astronomical Institute, Moscow State University, Moscow, Russia}

\author{Bengt Edvardsson}
\affil{Theoretical Astrophysics, Division of Astronomy \& Space Physics, Department of Physics and Astronomy, Uppsala University, Box 516, SE-751 20 Uppsala, Sweden}

\author{Steven R. Majewski}
\affil{Department of Astronomy, University of Virginia, P.O. Box 400325, Charlottesville, VA 22904-4325, USA}

\author{Szabolcs M\'esz\'aros}
\affil{ELTE E\"otv\"os Lor\'and University, Gothard Astrophysical Observatory, 9700 Szombathely, Szent Imre h. st. 112, Hungary}
\affil{MTA-ELTE Exoplanet Research Group, 9700 Szombathely, Szent Imre h. st. 112, Hungary}

\author[0000-0002-7883-5425]{Diogo Souto}
\affiliation{Departamento de F\'isica, Universidade Federal de Sergipe, Av. Marechal Rondon, S/N, 49000-000 S\~ao Crist\'ov\~ao, SE, Brazil}
\affiliation{Observat\'orio Nacional/MCTIC, R. Gen. Jos\'e Cristino, 77,  20921-400, Rio de Janeiro, Brazil}

\author{Olga Zamora}
\affil{Instituto de Astrof\'isica de Canarias (IAC), E-38205 La Laguna, Tenerife, Spain}
\affil{Universidad de La Laguna, Dpto. Astrof\'isica, E-38206 La Laguna, Tenerife, Spain}

\author[0000-0002-1691-8217]{Rachael L. Beaton}
\affiliation{Department of Astrophysical Sciences, Princeton University, 4 Ivy Lane, Princeton, NJ~08544}
\affiliation{The Observatories of the Carnegie Institution for Science, 813 Santa Barbara St., Pasadena, CA~91101}

\author{Jo Bovy}
\affil{David A. Dunlap Department of Astronomy and Astrophysics, University of Toronto, 50 St. George Street, Toronto, ON, M5S 3H4, Canada}
\affil{Dunlap Institute for Astronomy and Astrophysics, University of Toronto, 50 St. George Street, Toronto, Ontario, M5S 3H4, Canada}

\author{John Donor}
\affil{Department of Physics \& Astronomy, Texas Christian University, TCU Box 298840, Fort Worth, TX 76129, USA}

\author[0000-0002-7549-7766]{Marc H. Pinsonneault}
\affil{Department of Astronomy, The Ohio State University, 140 West 18th Avenue, Columbus OH 43210, USA}

\author{Vijith Jacob Poovelil}
\affil{Department of Physics \& Astronomy, University of Utah, Salt Lake City, UT, 84112, USA}

\author{Jennifer Sobeck}
\affil{Department of Astronomy, Box 351580, University of Washington, Seattle, WA 98195, USA}

\begin{abstract}
The spectral analysis and data products in Data Release 16 (DR16; December 2019) from the high-resolution near-infrared APOGEE-2/SDSS-IV survey are described. Compared to the previous APOGEE data release (DR14; July 2017), APOGEE DR16 includes about 200\,000 new stellar spectra, of which 100\,000 are from a new southern APOGEE instrument mounted on the 2.5~m du Pont telescope at Las Campanas Observatory in Chile. DR16 includes all data taken up to August 2018, including data released in previous data releases. All of the data have been re-reduced and re-analyzed using the latest pipelines, resulting in a total of 473\,307 spectra of 437\,445 stars. Changes to the analysis methods for this release include, but are not limited to, the use of MARCS model atmospheres for calculation of the entire main grid of synthetic spectra used in the analysis, a new method for filling ``holes'' in the grids due to unconverged model atmospheres, and a new scheme for continuum normalization. Abundances of the neutron capture element Ce are included for the first time. A new scheme for estimating uncertainties of the derived quantities using stars with multiple observations has been applied, and calibrated values of surface gravities for dwarf stars are now supplied. Compared to DR14, the radial velocities derived for this release more closely match those in the {\it Gaia} DR2 data base, and a clear improvement in the spectral analysis of the coolest giants can be seen. The reduced spectra as well as the result of the analysis can be downloaded using links provided in the SDSS DR16 web page.
\end{abstract}


\section{Introduction} \label{sec:intro}
The Apache Point Observatory Galactic Evolution Experiment \citep[APOGEE,][]{2017AJ....154...94M} was originally an infrared stellar spectroscopic survey within SDSS-III \citep[][henceforth APOGEE-1]{2011AJ....142...72E}, and APOGEE-2 is the continuation of the same program within SDSS-IV \citep{2017AJ....154...28B}. For every SDSS data release that has included APOGEE data (beginning with DR10), the survey has re-analyzed the previous (APOGEE-1 and APOGEE-2) spectra using the most up-to-date version of the data reduction and analysis pipelines, and hence SDSS-IV/APOGEE-2 data releases include data taken during the SDSS-III/APOGEE-1 project. In this paper, we present the data and data analysis from the sixteenth SDSS Data Release (DR16). Henceforth, we will use ``APOGEE'' to refer to the full data set that includes data from both SDSS-III and SDSS-IV. The selection of targets for the stars observed within the APOGEE-1 period is described in \citet{2013AJ....146...81Z} and the selection for those in APOGEE-2 are described in \citet{2017AJ....154..198Z}, R. Beaton et al. (in prep), and F. Santana et al. (in prep). 

In previous data releases, all main survey data have been collected using the APOGEE-N (north) instrument \citep{2019PASP..131e5001W} in combination with the 2.5~m Sloan Foundation telescope \citep{2006AJ....131.2332G} at Apache Point Observatory in New Mexico (APO). Henceforth this instrument/telescope combination will be referred to as ``APO~2.5~m''. Using this combination, 300 spectra of different objects within a 3 degree (diameter) field on the sky can be collected. In addition, some spectra have been collected using the NMSU 1.0~m telescope at APO using the same APOGEE instrument with a single object fiber feed (``APO~1.0 m''). With this instrument/telescope combination, only one star can be observed at a time, and it has mainly been used to observe bright targets for validation of the APOGEE spectral analysis. Since February 2017 another, nearly identical APOGEE spectrograph \citep{2019PASP..131e5001W}, APOGEE-S (south), has been operating at the 2.5~m du Pont telescope \citep{1973ApOpt..12.1430B} at Las Campanas Observatory in Chile (``LCO~2.5~m''), enabling observations of the southern sky not accessible from APO. Given the different focal ratio of the du Pont telescope, the field of view is limited to 2 degrees in diameter. DR16 is the first data release of APOGEE that includes data from the southern instrument/telescope. 

Within the APOGEE-1 and APOGEE-2 surveys, subprojects -- and hence their observations -- are classified as {\it core}, {\it goal}, or {\it ancillary} and given different observational priorities. The {\it core} programs focus on the {\it Galactic Evolution Experiment}, while the {\it goal} and {\it ancillary} projects have more specialized science goals. Within the core program are the APOGEE {\it main survey} targets, which are chosen using a well-defined, relatively simple, color and magnitude selection function that is designed to target cooler stars. In addition to the survey targets, the current dataset also contains data from {\it external} contributed programs taken with the southern instrument by the Carnegie Observatories and Chilean community who have access to the duPont telescope; these are ``classical'' observing programs vetted through a Time Allocation Committee outside of SDSS for which the individuals granted time are responsible for preparing the observations, but have agreed to have their data included in the SDSS releases. The final target selection for APOGEE-2N (North, APO) and APOGEE-2S (South, LCO) will be presented in R. Beaton et al. (in prep) and F. Santana et al. (in prep), respectively.

\section{The scope of DR16} \label{sec:scope}
DR16 contains high-resolution (R$\sim$22\,500), multiplexed, near-infrared (15\,140-16\,940~\AA) spectra for about 430\,000 stars covering both the northern and southern sky, from which radial velocities, stellar parameters, and chemical abundances of up to 26 species are determined.

\begin{figure*}
\gridline{\fig{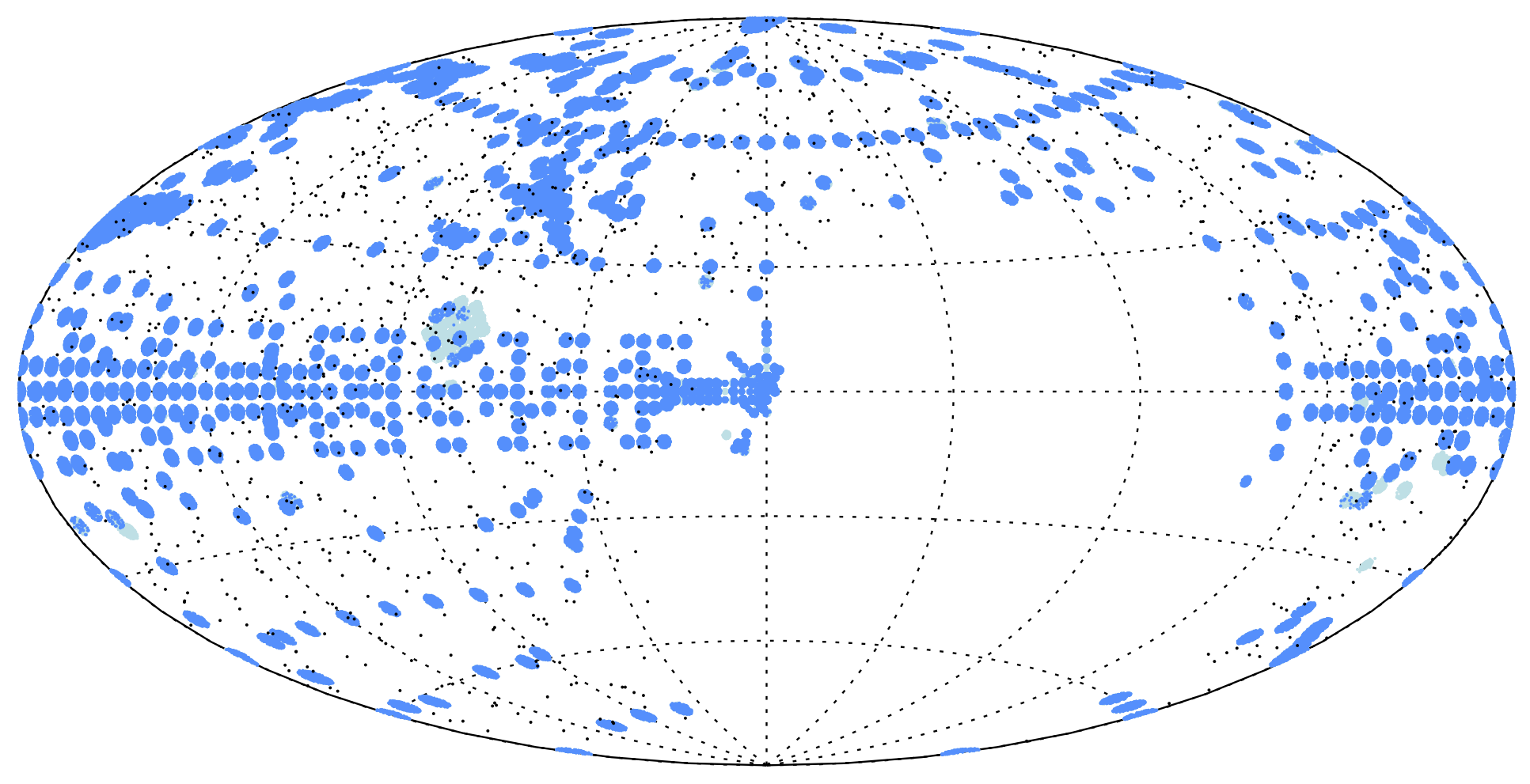}{0.50\textwidth}{}
          \fig{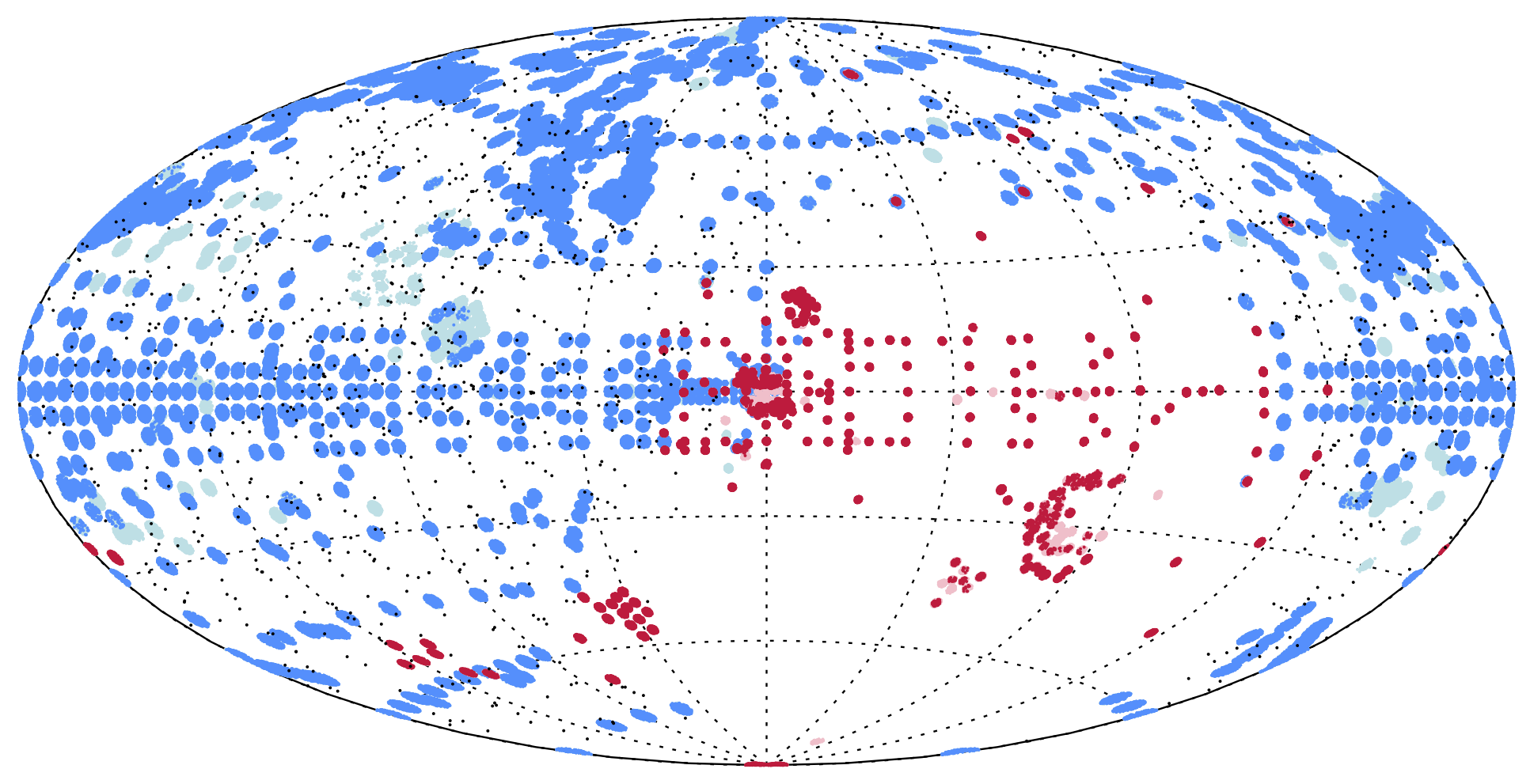}{0.50\textwidth}{}}
\caption{The left figure shows the APOGEE sky coverage of SDSS DR14, while the right figure shows the coverage of DR16. Observations made with APO 2.5 m are plotted in blue, observations made with LCO 2.5 m are plotted in red, and observations made with APO 1.0 m are plotted as small black dots. Observations not meeting the main survey target selection criteria are marked with lighter colors. Note in particular how the new southern instrument delivers a more complete coverage of the bulge region (in the center of the plots), and enables the Magellanic clouds to be observed (the large collection of red points in the lower right-hand corner of the right panel).\label{fig:coverage}}
\end{figure*}

Figure \ref{fig:coverage} shows the DR16 coverage of the sky compared to the sky coverage of the previous APOGEE data release (DR14; APOGEE did not release any new data in the SDSS DR15). The circular footprints of the 300 simultaneous stellar spectral observations that are made with APO~2.5~m and LCO~2.5~m can clearly be seen (henceforth ``fields''), as well as the more scattered, single star APO~1.0~m observations. The targets that meet the main survey target selection criteria (which can be identified in the release by objects that have a EXTRATARG bitmask\footnote{\url{https://www.sdss.org/dr16/algorithms/bitmasks/}} value of 0) have been marked with a darker color. Note that these stars might {\it also} be ``special targets'' from goal, ancillary, or external programs, that happen to meet the survey criteria, see \citet{2013AJ....146...81Z,2017AJ....154..198Z} for details. 

An overview of the different APOGEE data releases is shown in Table \ref{tab:drs}\footnote{The first release, DR10, is excluded in the table since this release only included stellar parameters. DR11 and DR15 did not include any new APOGEE data/analysis.}. DR16 contains spectra and derived data for 437\,445 individual stars. Most stars are observed in multiple observations, ``visits." While individual radial velocities are determined for each visit, the visits are combined for the stellar parameter and abundance analysis. However, some stars are observed as part of multiple fields, i.e., using different instrument/telescope combinations and/or in more than one field center position, and these are analyzed separately; hence some stars have more than one entry in the final (fits) table of analysis results (the allStar-file\footnote{\url{https://www.sdss.org/dr16/irspec/spectro_data/}}). This is the reason that the total number of spectra in Table \ref{tab:drs} has 473\,307 entries for DR16\footnote{For the stars with multiple entries in the allStar-file,  we aid the user by choosing a preferred spectrum of those stars based on the S/N, by setting bit 4 in the EXTRATARG bitmask for the non-preferred spectra.}.

\begin{deluxetable*}{lcccc}
\tablecaption{The APOGEE data releases that include abundance determinations (the first APOGEE release, DR10, included only stellar parameters -- see \citet{2013AJ....146..133M} -- and APOGEE did not release any new data/analysis in the SDSS-III/IV Data Releases 11 and 15). The number of spectra are listed as main survey target \emph{stars}/number of \emph{entries} in the corresponding allStar-file, see text for details.\label{tab:drs}}
\tablehead{      & \colhead{DR12}     & \colhead{DR13}     & \colhead{DR14} & \colhead{DR16}}
\startdata
Release date     & January 2015       & August 2016        & July 2017    & December 2019\\
Data taken up to & July 2014 & July 2014 & July 2016      & August 2018\\
Main survey stars/number of entries & 108\,324/163\,278 & 109\,376/164\,562 & 184\,148/277\,371 & 281\,575/473\,307\\
From APO~2.5~m                      & 108\,324/162\,398 & 109\,376/163\,668 & 184\,148/276\,353 & 225\,095/370\,036\\
From APO~1.0~m                      & 0/880 & 0/894         & 0/1018        & 0/1071\\
From LCO~2.5~m                      & 0/0 & 0/0           & 0/0           & 56\,480/102\,200\\
allStar filename & allStar-v603.fits & allStar-l30e.2.fits & allStar-l31c.2.fits & allStar-r12-l33.fits \\
Reference & \citet{2015AJ....150..148H} & \citet{2018AJ....156..125H} & \citet{2018AJ....156..125H} & This work\\
\enddata    
\end{deluxetable*}

For DR16, we decided to remove the spectra observed during the commissioning of the APOGEE-N instrument in winter-spring 2011, since these are of significantly lower resolution due to initial optical alignment issues with the instrument, and therefore do not meet the survey requirements. Most of these stars have been re-observed after the performance of the instrument was improved in the summer of 2011, but we note that this results in some objects that appeared in previous releases but do not appear in DR16.

For most objects, multiple visits are made to build up signal-to-noise (S/N) and to provide multiple radial velocity (RV) measurements. However, all data taken up to the cutoff date for a given data release are included, even if all of the planned visits for some fields have not been completed. Given this and other issues that might affect S/N, not all spectra in a data release reach the target S/N of 100 per half-resolution element. For DR16, 67\,503 spectra (14\%) have S/N$<70$ and 19\,796 spectra (4\%) have S/N$<30$. These spectra are flagged with the SN\_WARN and SN\_BAD bits, respectively, set in the ASPCAPFLAG bitmask in the allStar-file.

\section{Data reduction} \label{sec:reduction}
 The basics of the reduction pipeline are described in \citet{2015AJ....150..173N}, with subsequent updates for DR13 and DR14 described in \citet{2018AJ....156..125H}. While the data reduction for DR16 (version r12) is similar to that used for the previous data release (DR14, version r8), some updates/changes have been made:
 
\begin{itemize}
    \item Motivated by different cosmetic issues in the detectors for APOGEE-S, some changes were implemented in the construction of pixel masks to improve masking of bad pixels, making the masking more conservative to avoid some poor quality data not being masked, as seen for some spectra in previous data releases.
    \item Several changes were made to provide reduced spectra with approximate relative flux calibration, which was not done for DR14. These include changes in the removal of illumination spectral signatures in the internal and dome flats, and the subsequent use of hot stars on each plate to provide an approximate relative flux calibration. 
    \item Improvements were made to the wavelength calibration routines: rather than using single wavelength calibration frames for the entire survey, a wavelength solution is determined separately for each year of observation from multiple wavelength calibration frames taken throughout the year. The wavelength calibration routines now allow for small relative motions of the three detectors, which appear to occur when the detector assembly is moved to provide for observations at two different detector dither positions as a means to improve sampling of the spectra.
    \item The list of sky lines used to determine the wavelength zero-points of each observation was revised slightly, and the wavelength offsets calculated for each observation (necessary because of the dithering) allow for the small relative motions of the three detectors.  The revised sky line list was also used for the determination of the line spread function (LSF).
    \item A new grid of synthetic spectra used for radial velocity determination was constructed using a subset of the synthetic grid used for stellar parameter and abundance determination (see Section \ref{sec:aspcap}). 
    \item Comparison of each stellar spectrum against the full RV grid was made for RV determination; DR14 had implemented a restriction of the grid based on the observed color of each star, but this was found to lead to some spurious results.
    \item Improvements were made for the removal of telluric lines in APO~1.0~m spectra, which need to be handled differently than the normal multi object observations since there are no concurrent observations of hot stars.
\end{itemize}
For DR16, the organization of the reduced data files has changed from that of previous data releases, to a large extent because of the addition of the LCO data; reductions are now separated into subdirectories based on the telescope and the field names. The data file organization is described in the SDSS data model\footnote{\url{https://www.sdss.org/dr16/irspec/spectro_data/}}. Reduced data frames and spectra are available for download from the SDSS Science Archive Server\footnote{\url{https://data.sdss.org/sas/dr16/}}; the Science Archive Webapp\footnote{\url{https://dr16.sdss.org/}} provides a convenient interface to inspect and download spectra for individual and groups of objects.

\section{Spectral analysis} \label{sec:aspcap}
The heart of the spectral analysis of the APOGEE Stellar Parameter and Chemical Abundance Pipeline \citep[ASPCAP,][]{2016AJ....151..144G} is the program FERRE\footnote{\url{http://github.com/callendeprieto/ferre}} \citep{2006ApJ...636..804A}, which interpolates in a pre-computed grid of synthetic spectra to find the best fitting stellar parameters describing an observed spectrum. Once the stellar parameters have been determined, these (including the ``abundance parameters'', \am, \xm{C}, \xm{N}, see Section \ref{sec:parameters}) are held fixed and the abundances are determined with fits using the same grids, but restricted to windows of the spectra that include lines of the element of interest. A development of FERRE motivated by the large spectral grids of APOGEE is the use of {\it principal component analysis} (PCA) to compress the grids; the actual interpolations are performed in the PCA coefficients to speed up the calculations (see Section \ref{sec:pca}). This type of analysis has been used in all previous APOGEE data releases, and has been previously described in \citet{2016AJ....151..144G}, with updates in \citet{2018AJ....156..125H}. In this section we focus on DR16-specific updates/changes made to previous iterations of ASPCAP described in those papers.

\subsection{Main stellar atmospheric models} \label{sec:models}
\begin{figure*}
\plotone{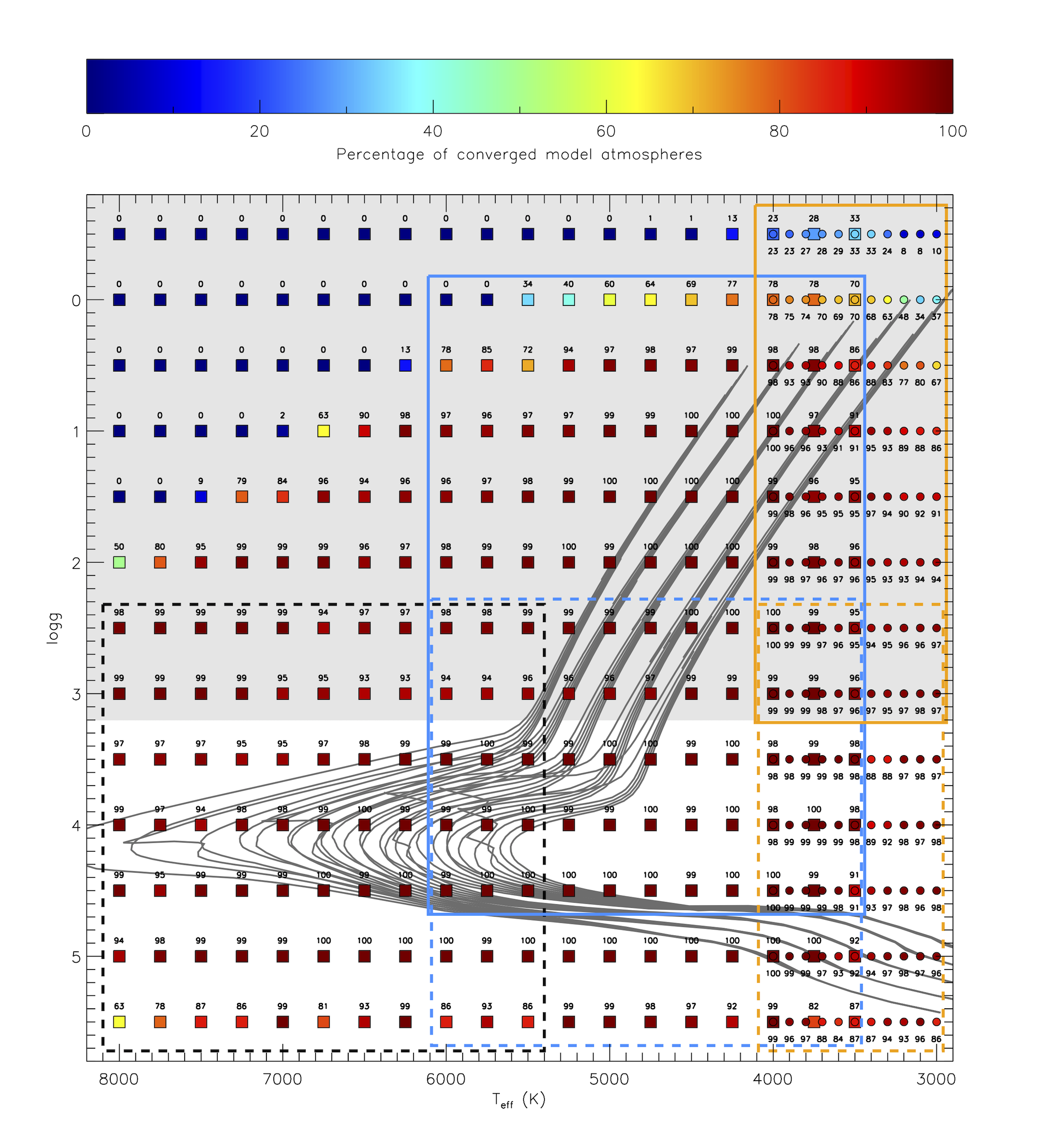}
\caption{The stellar atmosphere grid points used in DR16. Squares mark the warmer, more sparsely spaced model atmospheres, while the circles mark the cooler, more densely spaced model atmospheres in the \teff-\logg~ plane. The small numbers above or below the symbols indicate the percentages of converged models in the \teff-\logg~gridpoint in question. This is also reflected in the color-coding of the points with blue points having many holes, and red no holes. The five subgrids of synthetic spectra are marked with rectangles: the F-, GK-, and M-dwarf subgrids are marked using black, blue, and orange dashed lines respectively, and the GK- and M-giant subgrids are marked using blue and orange solid lines, respectively. The region for which the atmospheric models and synthetic spectra are calculated using spherical geometry are shaded (\logg$\leq 3$).  Isochrones with \mh=-1.5,-1.0,-0.5,0.0,+0.5 and ages 3-8 Gyr are plotted using solid dark gray lines \citep{2012MNRAS.427..127B}. The most metal-rich isochones are the right-most on the giant branch and the upper ones amongst the dwarfs.\label{fig:grids}}
\end{figure*}

In DR14, ATLAS-9 atmospheric models \citep[][and updates]{1979ApJS...40....1K} were used to generate synthetic spectra, but a grid of cooler MARCS-models \citep{2008A&A...486..951G} was used for \teff$<3500$ K, see \citet{2012AJ....144..120M,2015AJ....149..181Z,2018AJ....156..125H} for details. For DR16, we (B. Edvardsson) computed a new all-MARCS grid of atmospheric models and these were used exclusively (apart from stars with \teff$> 8000$~K, see Section \ref{sec:bagrid}). The main motivation for this change is to avoid the discontinuity between the two subgrids seen in DR14-data (compare Figure \ref{fig:coolgiants}); MARCS models are required to handle the lowest effective temperatures of our targets since the ATLAS grid has a lower limit of 3500 K. Additionally, a transition to MARCS models has made it possible to use spherical models for \logg$\leq 3$. Figure \ref{fig:grids} shows the location of models in the \teff-\logg-plane; the grid has finer spacing for 3000 K $\le$\teff$\le 4000$ K where the model structure changes more with specified \teff.

For every grid point shown in the \teff-\logg~plane \mh~is varied from -2.50 to +1.00 in steps of 0.25 dex (15 steps), \afe~ (which includes changes in O, Ne, Mg, Si, S, Ar, Ca, and Ti) is varied between -1.0 to +1.0 in steps of 0.25 dex  (9 steps) and \xfe{C} is varied between -1.00 to +1.00 in steps of 0.25 dex (9 steps), meaning that every grid point shown in Figure \ref{fig:grids} in fact represents 1215 model atmospheres. This adds up to 300\,105 attempted calculated atmospheric models for the warmer grid, and 173\,745 models in the cooler, finer spaced grid, and 442\,260 models in total (there are some overlapping grid points in the two grids, see Figure \ref{fig:grids}). However, only 358\,123 of these models converged, leading to 84\,137 holes in our grid. The fraction of holes in the \teff-\logg-plane is shown in the small numbers as well as in the color-coding in Figure \ref{fig:grids}. In general, most of the holes are in regions of the grid where we do not expect many stars, for example with high \teff~and low \logg, and/or with elemental abundances near the grid edges in the abundance dimension in question. However, of particular interest is that of the models with \logg$=-0.5$, only 1\% of all models converged. 

The holes in the model atmospheric grid obviously translate to holes in the grid of synthetic spectra, but as described in Section \ref{sec:rbf}, these holes are filled before the analysis of data using Radial Basis Functions (RBF) interpolation (and extrapolation) in flux-space.

\subsection{Line list} \label{sec:line list}
In DR14 we used the atomic and molecular line lists described in \citet{2015ApJS..221...24S} (this set of lists is internally labeled as 20150714 based on the date of adoption, in the format YYYYMMDD). In short, these line lists were based on a thorough, up-to-date literature review and evaluation by comparing to observed high-resolution spectra of standard stars \citep{2013ApJ...765...16S}. For the atomic lines, the transition probabilities were adjusted within the quoted uncertainties to match the spectra of the Sun and Arcturus \citep{1991aass.book.....L,1995iaas.book.....H}. For DR16, we decided to launch another literature review to find possibly newer, more accurate line data. This has led to the addition of lines and/or updates of atomic data for almost all atomic species compared to the DR14 line list, and also several updates regarding molecular transitions. Most notably, our line lists now include transitions from Ce~II \citep{2017ApJ...844..145C}, more transitions from Nd~II \citep{,2016ApJ...833...81H}, and the FeH molecule \citep{2010AJ....140..919H}. The line list and its creation is thoroughly described in Smith et al.~(in prep).

For very limited parts of the spectrum we were not able to fit the Sun and/or Arcturus well in this process. Reasons for this could be missing transitions in our line list, and/or too small uncertainties cited in the atomic data reference, which limited our code from adjusting the transition probability. These regions have been masked out in subsequent analysis, and therefore have not affected our results.

The resulting DR16 line list is internally labeled 20180901.

\subsection{Main synthetic spectra} \label{sec:maingrids}
As in DR14, the synthetic spectra for the main spectral grids were made using Turbospectrum \citep{1998A&A...330.1109A,2012ascl.soft05004P}. 
Plane parallel and spherical radiative transfer was used, consistent with the model atmosphere in question.

To ensure regular dimensions in the grid of synthetic spectra (same range in \logg~for all values of \teff) and to enable the entire grid to be loaded in memory during the running of FERRE, the grid of synthetic spectra has, as in previous data releases, been divided into subgrids in ASPCAP \citep{2015AJ....149..181Z}. The division is somewhat different in DR16 compared to the previous data release and is shown in Figure \ref{fig:grids}: the solid green and red lines mark what we label the GK and M giant grids, respectively, while the dashed blue, green and red lines indicate the F, GK, and M dwarf grids, respectively.   

In the calculation of synthetic spectra we change some of the dimensionality compared to the dimensions of the grid of the atmospheric models, and also in several instances compared to the grids used for DR14:
\begin{itemize}
\item We do not use the models with \afe$=-1.00$, limiting the grid of synthetic spectra to 8 steps in \afe~between -0.75 to +1.00. 
\item We add a microturbulent velocity dimension having values of 0.3, 0.6, 1.2, 2.4, and 4.8 km/s (5 steps). In the calculated model atmospheres, a value of 1.0 km/s is used for models with \logg$>3$ and a value of 2.0 km/s for models with \logg$\leq 3$. 
\item In the giant subgrids, we add gridpoints with \xfe{C}$=-1.25$ and \xfe{C}$=-1.50$ using the otherwise appropriate atmospheric model with \xfe{C}$=-1.00$, for a total of 11 steps in \xfe{C} between -1.50 and +1.00.
\item In the dwarf subgrids we do not use all the available models in the \xfe{C} dimension, restricting \xfe{C} from -0.50 to +0.50 in steps of 0.25 (5 steps).
\item In the giant subgrids we add a \xfe{N}-dimension from -0.50 to +2.00 in steps of 0.50 (6 steps), while we go from -0.50 to +1.50 in steps of 0.50 (5 steps) in the dwarf subgrids, using the otherwise appropriate atmospheric model. The nitrogen abundance is not expected to affect the model atmosphere structure, so the N abundance was varied in the synthesis only.
\item In the dwarf subgrids, we add a projected rotational velocity (\vsini) dimension with values of 1.5, 3.0, 6.0, 12.0, 24.0, 48.0, 96.0 km/s (7 steps), using the rotational line broadening from \citet{gray2005observation} using a linear limb-darkening coefficient appropriate for the near-IR, $\epsilon=0.25$.
\item In the giant subgrids, where there is no rotational broadening, we adopt a macroturbulent velocity broadening with the same prescription as that used for DR14: \vmac$=10^{(0.471-0.254{\rm [M/H]})}$
\end{itemize}

The final dimensionality of the  different subgrids are listed in Table \ref{tab:grids}. For the dwarf subgrids a solar value of $^{12}$C/$^{13}$C$=89.9$ \citep{2003ApJ...591.1220L} is used when calculating the synthetic spectra, while for the giant grids a carbon isotopic ratio such that $^{12}$C/$^{13}$C$=15$ has been adopted.
The single value of $^{12}$C/$^{13}$C=15 represents a typical isotopic ratio in red giants within a mass range of M$\sim$1-2M$_{\odot}$ spanning a moderate range of metallicities, from [Fe/H]$\sim$-1.0 to +0.3.  \citet{2019A&A...621A..24L} present a set of stellar models to probe red giant mixing and compare theoretical values of $^{12}$C/$^{13}$C with observations from a number of studies of open and globular clusters; the \citet{2019A&A...621A..24L}-models include additional mixing mechanisms from both stellar rotation and thermohaline mixing.  The observed values of $^{12}$C/$^{13}$C from the various globular and open clusters, which have red giant masses ranging from M$\sim$0.9 to 2.5M$_{\odot}$, have values between $\sim$5 to 25, with 15 being a representative value \citep[see][ Figure 12 for a summary of the range of $^{12}$C/$^{13}$C as a function of red giant mass for both the observations of cluster and field red giants, along with predictions from their stellar models]{2019A&A...621A..24L}.

\begin{deluxetable*}{l|c|c|c|c|c|}
\tablecaption{The dimensionality and parameter ranges of the final subgrids of synthetic spectra. The step size and number of steps are shown in the parentheses. \label{tab:grids}}
\tablehead{
\colhead{} & \colhead{GK giant} & \colhead{M giant} & \colhead{F dwarf} & \colhead{GK dwarf} & \colhead{M dwarf}
}
\startdata
\teff & 3500 ... 6000    (250, 11)  & 3000 ... 4000    (100, 11)  & 5500 ... 8000    (250, 11)  & 3500 ... 6000    (250, 11)  & 3000 ... 4000    (100, 11)  \\ 
\logg & +0.0 ... +4.5    (0.5, 10)  & -0.5 ... +3.0    (0.5, 8)   & +2.5 ... +5.5    (0.5, 7)   & +2.5 ... +5.5    (0.5, 7)   & +2.5 ... +5.5    (0.5, 7)   \\
\mh & -2.50 ... +1.00  (0.25, 15) & -2.50 ... +1.00  (0.25, 15) & -2.50 ... +1.00  (0.25, 15) & -2.50 ... +1.00  (0.25, 15) & -2.50 ... +1.00  (0.25, 15) \\
\afe & -0.75 ... 1.00   (0.25, 8)  & -0.75 ... +1.00  (0.25, 8)  & -0.75 ... 1.00   (0.25, 8)  & -0.75 ... 1.00   (0.25, 8)  & -0.75 ... 1.00   (0.25, 8)  \\
\xfe{C} & -1.50 ... +1.00  (0.25, 11) & -1.50 ... +1.00  (0.25, 11) & -0.50 ... +0.50  (0.25, 5)  & -0.50 ... +0.50  (0.25, 5)  & -0.50 ... +0.50  (0.25, 5)  \\
\xfe{N} & -0.50 ... +2.00  (0.50, 6)  & -0.50 ... +2.00  (0.50, 6)  & -0.50 ... +1.50  (0.50, 5)  & -0.50 ... +1.50  (0.50, 5)  & -0.50 ... +1.50  (0.50, 5)  \\
\vmic & 0.3, 0.6, 1.2, 2.4, 4.8  (5) & 0.3, 0.6, 1.2, 2.4, 4.8  (5) & 0.3, 0.6, 1.2, 2.4, 4.8  (5) & 0.3, 0.6, 1.2, 2.4, 4.8  (5) & 0.3, 0.6, 1.2, 2.4, 4.8  (5)\\
\vsini & \multicolumn{2}{|c|}{1.5 (1)} & \multicolumn{3}{|c|}{1.5, 3.0, 6.0, 12.0, 24.0, 48.0, 96.0 (7)}\\
N & 4\,356\,000 & 3\,484\,800 & 8\,085\,000 & 8\,085\,000 & 8\,085\,000 \\
\enddata    
\end{deluxetable*}

For DR14, four differently smoothed grids were created to roughly match the different Line Spread Functions (LSFs) of the different fibers in the APO instrument. For the DR16 grids, we have made a corresponding characterization of the LSFs for the LCO instrument, so each subgrid has eight different versions; the appropriate one is used when analyzing a particular spectrum taken with a given instrument and mean fiber. This issue and procedure is described in more depth in \citet{2018AJ....156..125H}. While the use of four different LSFs for each instrument significantly reduces the dependence of parameters on fiber number, some low level dependence may still remain, see, e.g. \citet{2018ApJ...853..198N}.

\subsubsection{Filling of ``holes''} \label{sec:rbf}

One of the difficulties of computing model atmospheres is the possible lack of convergence of their iteration algorithm. This issue affects both ATLAS and MARCS atmospheres \citep{2012AJ....144..120M}, and is usually solved by interpolating in the atmospheric structure space. However, it may be more accurate to interpolate in the flux space of the synthetic spectra \citep{2013MNRAS.430.3285M}.

In DR14, the holes in the grid of synthetic spectra were filled by spectral syntheses using the ``closest'' neighboring model atmosphere according to a metric specified in \citet{2018AJ....156..125H}. This can be extremely inaccurate if the number of holes is significant. For DR16, we instead implemented radial basis function (RBF) interpolation to fill the missing synthetic spectra in the grids.

The RBF is a real-valued function whose value depends only on the distance from the known points, and works in any number of D dimensions ($D\geq1$) \citep{buhmann}. The interpolated value is represented as a sum of N radial basis functions (where N is the number of known points). These functions are strictly positive definite functions, and the most widely used definitions are Gaussian, multiquadric, polyharmonic spline, or thin plate spline. We chose the multiquadric form defined below, as it is the most versatile when used with sparse datasets like ours while still achieving the necessary accuracy.  Each of the RBF functions are associated with a different known point $x_i$, weighted by an appropriate coefficient $w_i$, and scaled by the parameter $r_0$:

$$y(x) = \sum_{i=1}^{N} w_{i} \cdot (||x - x_{i}||^2 + {r_0}^2)^{0.5}$$

The known points in our case are the synthetic spectra calculated with effective temperature, metallicity, surface gravity, etc., of the converged model atmospheres. Determining the $w_i$ weights can be accomplished by solving a system of N linear equations, but round-off errors grow large and the required computation time becomes unfeasible long for high values of N, since the computation complexity scales as O(N$^3$).

Therefore, many iterative methods have been developed to reduce the required computation time. One such method is a Krylov subspace algorithm developed by \citet{faul2005} for multiquadric interpolation in multiple dimensions, which scales as O(N$^2$), a significant improvement compared to direct methods. We implemented this algorithm based on a previous implementation in Matlab available from \citet{gumerov2007} who also further optimized Faul et al.'s algorithm by reducing its complexity to O(N*logN). Faul et al.'s algorithm includes two main steps:

1. a precondition phase that depends only on the distances between the known points and a parameter, $q$, which is the power of the Lagrange functions of the interpolation, and

2. an iteration phase that provides the desired weights for the interpolation. 

In the preconditioning phase \citet{faul2005} carefully select a set of $q$ points for each known point to construct the preconditioner. This preconditioner is used to build a set of directions in the Krylov space for the iteration phase. Larger $q$ values will result in fewer iterations (of order $\sim10$ depending on the particular problem), but calculating the preconditioner takes significantly longer. In general cases, when $q<<$N,  a good compromise is to have $q$ around 30-50 to limit the computation time of the preconditioner.

In APOGEE's case we need a different approach. While the spectra depend on the atmospheric parameters, in a single spectrum the flux only depends on the wavelength, so we do not need to compute the preconditioning phase for every single wavelength. This allows us to save significant computation time by using the same preconditioning for every frequency by selecting $q=N$. While this increases the complexity of the preconditioning phase, the overall time to determine the weights for the entire spectra is reduced significantly, because the choice of $q=N$ makes the algorithm converge in only 2 or 3 iterations. 

A given APOGEE spectral subgrid contains of order 1\,000\,000 spectra (see Table \ref{tab:grids}). While the \citet{gumerov2007} algorithm can handle such large number of points in reasonable time, our internal testing showed that the accuracy of how well we can recover missing models degrades significantly when N$>2000$. It is our goal to be able to recover spectra with 0.01-0.02 or better in normalized flux, an accuracy that is possible to achieve only if we can select 4 or 5 known points in each dimension. For this reason we chose to implement Faul et al.'s method for simplicity and for the fact that it is faster than the \citet{gumerov2007} approach when N$<2000-3000$. 

\begin{figure}
\epsscale{1.18}
\plotone{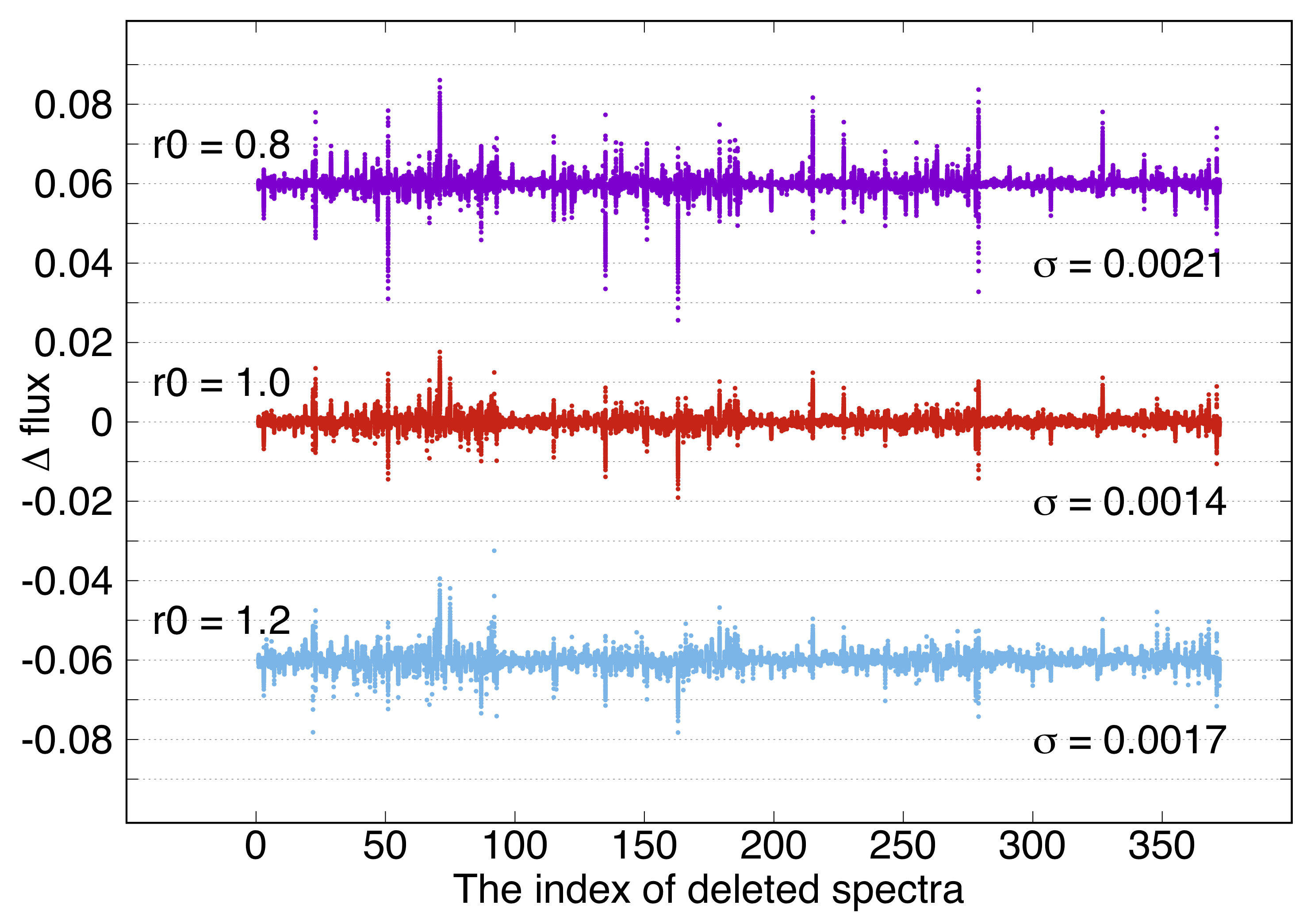}
\caption{Examples of the interpolation error. Of 4096 known spectra in a small subgrid of the larger GK giant grid, 372 were deleted and then re-created using interpolation based on the remaining spectra, using different $r_0$ values, with the aim of  evaluating the overall accuracy. The $r_0$=0.8 and 1.2 cases are shifted up and down in the plot, to aid visibility. On the x-axis are the 372 deleted spectra, and on the y-axis differences between the interpolated and original spectrum for \emph{all} wavelengths are plotted, i.e. there are thousands of points for every spectrum (x-axis value) and every choice or $r_0$ (0.8, 1.0, 1.2). The stated $\sigma$ is the standard deviation around the mean value. We chose $r_0=1$ because higher $r_0$ values do not improve the accuracy, but add computation time.\label{fig:rbf}}
\end{figure}

To fill each hole, we use a small grid of models around the hole, where the size of this grid depends on the location in parameter space, but generally has 3-5 points in each dimension. We determine the RBF coefficients for this grid from the filled points and use them to fill the missing point. The shape of the RBF is controlled through the $r_0$ scale factor, which is recommended to be greater than the minimal distance between points, and significantly less than the maximal distance. It is important to note that no established method exists for determining what is the best scale-factor in terms of accuracy. The best way to evaluate the uncertainties is to temporarily delete known spectra from the grid, re-create them with interpolation, and compare the interpolated spectra with the original ones. After extensive testing of this type, we found that $r_0=1$ provided the best accuracy for all grids, except the F-dwarf-subgrid where we chose $r_0=0.5$. An example of interpolation errors in one of these tests for three different values of $r_0$ is shown in Figure \ref{fig:rbf}.

The full grids of synthetic spectra are internally labeled ``l33'' (DR14 used ``l31c''), and are available for download from the Science Archive Server\footnote{\url{https://data.sdss.org/sas/dr16/apogee/spectro/speclib/synth}}. These are available in a series of fits-files, as well as in the FERRE-format described in the code's manual \citep[see also ][]{2018A&A...618A..25A}.

\subsection{Addition of a subgrid for hot stars} \label{sec:bagrid}
For DR16 we added a subgrid suitable for hot stars (\teff$> 8000$~K), thereby analyzing the more featureless spectra of stars that mainly were targeted for removal of telluric lines in the spectra of main survey target stars. The model atmospheres used for this grid are ATLAS9, the line list is the atomic DR13/14 line list (20150714), and the spectral synthesis code used is Synspec \citep{1988CoPhC..52..103H,2017arXiv170601859H}. The final subgrid only has four grid dimensions; 7000~K$\leq$\teff$\leq$20\,000~K in steps of 500~K (27 steps), $3.0\leq$\logg$\leq5$ in steps of 0.5 dex (5 steps), $-2.5\leq$\mh$\leq1.0$ in steps of 0.5 dex (8 steps), and a projected rotational velocity (\vsini) dimension with values of 1.5, 3.0, 6.0, 12.0, 24.0, 48.0, 96.0 km/s (7 steps).

The analysis of these spectra is extremely challenging; after all, these stars were targeted to show as few spectral features as possible and often hydrogen lines are the only strong features. Still, at least providing an estimate of the basic stellar parameters for these stars might be useful for some science applications. However, it should be noted that these values are not fully evaluated and should be used with caution, preferably by users familiar with hot stars and their spectra.

\subsection{Dimensionality reduction using Principal Component Analysis (PCA)} \label{sec:pca}
Even after dividing the total number of synthetic spectra into subgrids, these are still too large to hold in memory. Hence we have, as in previous APOGEE data releases, used PCA to reduce the dimensionality of the subgrids. Previously, this was done by splitting the APOGEE spectra in 30 pieces and using 30 PCA components for every piece, giving 900 PCA 
components, which provides almost a factor of 10 reduction in grid size. Tests on synthetic data, comparing the reconstructed spectra to originally calculated spectra, have shown that better accuracy is achieved with the same total number of PCA parameters, but by dividing the spectra in 12 pieces and using 75 PCA components for each piece, so this was implemented for the DR16 grids. Interpolation is done in the PCA coefficients, and the resulting values are multiplied by the PCA basis functions to create an interpolated spectrum.

\subsection{Coarse characterization} \label{sec:course}
In DR14, we did an initial coarse characterization of all stellar spectra to decide which synthetic spectra subgrid(s) to use when performing the stellar parameter determination. This coarse characterization was made by passing all stars through reduced-size F-dwarf, GK-giant, and M-giant grids with \xm{C}=\xm{N}=0. Based on the outcome of these runs, the spectrum was finally analyzed using the subgrid that yielded the best fit, or two in the case of cases where the best fit was near a grid edge. After the proper subgrid(s) to be used was determined, FERRE was run with 12 different starting positions (to avoid being trapped within local minima) distributed evenly in \teff, \logg, and \mh~
in the chosen subgrid(s), and the final stellar parameters were set to the best fitting of these 12 (or 2x12 in the case of border line cases) runs. A more thorough description of this process can be found in \citet{2018AJ....156..125H}.

In DR16, we instead created one, large ``coarse'' grid with dimensions 3000~K$\leq$\teff$\leq$8000~K (11 steps of 500~K), $0\leq$\logg$\leq5$ (6 steps of 1 dex), $-2.5\leq$\mh$\leq1.0$ (8 steps of 0.5 dex), $-0.5\leq$\am$\leq1.0$ (4 steps of 0.5 dex), $-0.5\leq$\xm{C}$\leq0.5$ (5 steps of 0.25 dex), $-0.5\leq$\xm{N}$\leq1.0$ (4 steps of 0.5 dex), 5 steps of \vmic; 0.3, 0.6, 1.2, 2.4, 4.8 km/s, and 7 steps of \vsini; 1.5, 3.0, 6.0, 12.0, 24.0, 48.0, 96.0 km/s, that was used to decide what ``fine'' subgrid(s) to use when analyzing the spectrum. Furthermore, the derived values of the stellar parameters from the ``coarse'' run were adopted as starting values when doing the second ``fine'' run with FERRE. This means that in the new scheme, we run FERRE significantly fewer times for every star (1 coarse and 1 or 2 fine), as compared to DR14 (3 coarse and 12 or 24 fine). This led to a reduction in analysis-time, something that is sorely needed as the data set increases for every release (see Table \ref{tab:drs}). However, in addition, we changed the choice of minimizing algorithm in FERRE from the default Nelder-Mead algorithm  \citep{nelder1965}, identified in the code with the option ALGOR=1, to the Unconstrained Optimization BY Quadratic Approximation or UOBYQA  \citep{Powell2002}, ALGOR=3 in FERRE, and this led to a compensating \emph{increase} in analysis time.  Both algorithms perform numerical optimization without the need for the explicit evaluation of derivatives,  but while Nelder-Mead indicates a prescription for the motion of the vertices of a simplex in the search space that on convergence contains the minimum of the objective function (the $\chi^2$ in our case), UOBYQA builds quadratic models for minimizing the objective function within trust regions. These changes were motivated by tests analyzing synthetic spectra -- that then of course have ``known'' stellar parameters -- which showed that the new scheme produces more accurate results.

\subsection{Continuum normalization} \label{sec:continuum}
For DR16, a revised scheme was used to normalize the spectra. First, the reduction process was improved to provide spectra with smoother variations and less ``wiggles'' (see Section \ref{sec:reduction}), helping the normalization of the observed spectra when comparing to the synthetic spectra. In addition, the observed spectra have been slightly continuum-adjusted for the final analysis, based on the fit from the ``coarse'' fit of stellar parameters. The ratio of the observed spectra to the best-fit ``coarse'' model spectrum was smoothed with a broad median filter (with a width of 750 pixels) and the observed spectrum was divided by the smoothed residual before being passed to the ``fine'' run. Manual inspection of spectra and their final, ``fine'' stellar parameter fits have shown this scheme to greatly improve the continuum fits, and perhaps more importantly, to homogenize the APOGEE-N and APOGEE-S data and decrease the spread in derived stellar parameters/abundances for stars observed with both APOGEE instruments. Finally, both these corrected observed spectra and the synthetic spectra are normalized with a fourth order polynomial in the wavelength region covered by each of the three APOGEE detectors. 

For DR16, this final continuum normalization is now made inside FERRE, allowing for rejection of the same pixels (e.g., those contaminated by night sky emission) in the observed and synthetic spectra, based on pixels flagged in the observed spectrum. In previous data releases, the continuum fit of the observed spectrum was made ignoring flagged pixels, while the continuum normalization of the synthetic spectra used all pixels, leading to possible inconsistencies for some spectra.

\subsection{Element ``windows''} \label{sec:windows}
After the stellar parameters (and ``abundance parameters'') have been determined, these are held fixed for additional runs of FERRE to determine the elemental abundances. For these, only windows in the spectra that are sensitive to the element in question are used, and only the most relevant abundance dimension of the grid is varied; \mh~(for Na, Al, P, K, V, Cr, Mn, Fe, Co, Ni, Cu, Ge, Rb, Ce, Nd, and Yb), \am~(for O, Mg, Si, S, Ca, Ti, and Ti II), \xm{C} (for C and C I), or \xm{N} (for N). The windows are chosen based on where our synthetic spectra are sensitive to a given element, and at the same time \emph{not} sensitive to another element in the same abundance group. Based on this, different weights are assigned to pixels in different abundance windows, just as in DR14.

In DR16, however, we performed some test analyses using one window at a time for a subset of spectra for the elements with less than 10 windows, with the aim of weeding out windows that produced deviant results for one reason or another, possibly caused by bad/missing atomic data in the window, unrecognized blends, or 3D/NLTE-effects. These analyses were run on a validation sample, which consists of spectra with high S/N, and including stars from across the HR-diagram, stars in the Kepler field, stars with independently determined stellar parameters and abundances, etc.

Based on manual inspection of the derived ``window-abundances'' as compared to each other, and to expected astrophysical trends in the solar neighborhood, and as a function of \teff~in open clusters, some of the windows used in DR14 were removed for Al, P, S, Ti, V, Cr, Mn, Co, and Yb. The windows and their weights used for DR16 are provided in Table \ref{tab:windows}.

\begin{deluxetable*}{cccccccccccccccc}
\tablecaption{Windows and weights used in the determination of stellar abundances. This is only an excerpt of the table to show its form and content. The complete table is available in electronic form. \label{tab:windows}}
\tablehead{
\colhead{Wavelength} & \colhead{C} & \colhead{C I} & \colhead{N} & \colhead{O} & \colhead{Na} & \colhead{Mg} & \colhead{Al} & \colhead{Si} & \colhead{P} & \colhead{S} & \colhead{K} & \colhead{Ca} & \colhead{Ti} & \colhead{Ti II} & \colhead{\nodata}\\
\colhead{(\AA, vacuum)}
}
\startdata
15152.211 & 0.000 & 0.000 & 0.300 & 0.000 & 0.000 & 0.000 & 0.000 & 0.000 & 0.000 & 0.000 & 0.000 & 0.000 & 0.000 & 0.000 & \nodata \\
15152.420 & 0.000 & 0.000 & 0.354 & 0.000 & 0.000 & 0.000 & 0.000 & 0.000 & 0.000 & 0.000 & 0.000 & 0.000 & 0.000 & 0.000 & \nodata \\
15152.629 & 0.000 & 0.000 & 0.263 & 0.000 & 0.000 & 0.000 & 0.000 & 0.000 & 0.000 & 0.000 & 0.000 & 0.000 & 0.000 & 0.000 & \nodata \\
15152.839 & 0.000 & 0.000 & 0.129 & 0.000 & 0.000 & 0.000 & 0.000 & 0.000 & 0.000 & 0.000 & 0.000 & 0.000 & 0.000 & 0.000 & \nodata \\
15153.048 & 0.000 & 0.000 & 0.045 & 0.000 & 0.000 & 0.000 & 0.000 & 0.000 & 0.000 & 0.000 & 0.000 & 0.000 & 0.000 & 0.000 & \nodata \\
15153.257 & 0.000 & 0.000 & 0.015 & 0.000 & 0.000 & 0.000 & 0.000 & 0.000 & 0.000 & 0.000 & 0.000 & 0.000 & 0.000 & 0.000 & \nodata \\
15153.467 & 0.000 & 0.000 & 0.013 & 0.000 & 0.000 & 0.000 & 0.000 & 0.000 & 0.000 & 0.000 & 0.000 & 0.000 & 0.000 & 0.000 & \nodata \\
15153.676 & 0.000 & 0.000 & 0.025 & 0.000 & 0.000 & 0.000 & 0.000 & 0.000 & 0.000 & 0.000 & 0.000 & 0.000 & 0.000 & 0.000 & \nodata \\
15153.885 & 0.000 & 0.000 & 0.042 & 0.000 & 0.000 & 0.000 & 0.000 & 0.000 & 0.000 & 0.000 & 0.000 & 0.000 & 0.000 & 0.000 & \nodata \\
15154.095 & 0.000 & 0.000 & 0.061 & 0.000 & 0.000 & 0.000 & 0.000 & 0.000 & 0.000 & 0.000 & 0.000 & 0.000 & 0.000 & 0.000 & \nodata \\
\nodata & \nodata & \nodata & \nodata & \nodata & \nodata & \nodata & \nodata & \nodata & \nodata & \nodata & \nodata & \nodata & \nodata & \nodata & \nodata \\
\enddata
\end{deluxetable*}

For the elemental abundance determination for DR16, we have used the new TIE option in FERRE for elements that were fit using the \mh~dimension of the grid. Using this dimension, abundances of \textit{all} elements are varied together during the fit. The TIE option allows the \am, \xm{C}, and \xm{N}~dimensions to be varied oppositely in lockstep, such that the abundances of C, N, and the $\alpha$ elements are \textit{not} varied as the best-fitting abundance from the \mh~variation is determined.

\subsection{Other updates} \label{sec:updates}
We updated FERRE from version 4.7.1 to the latest version at the time of production, 4.8.5. The updates to the code between these releases are rather minor, but include the important TIE option. 

The data were all processed on the SDSS cluster at the University of Utah, which is comprised of 27 nodes with 16 cores each. For processing with FERRE, two jobs are run on each node at once to accommodate the significant memory usage required to load a single subgrid, but the multiprocessing option in FERRE is used to run 16 threads simultaneously for each job. The total processing time is approximately 8-10 hours per field for fields with a single cohort of $\sim$ 160 stars.

\section{Results} \label{sec:results}
In this section, we describe how the APOGEE DR16 results are presented, and the calibrations that were applied. A subsequent section (Section \ref{sec:discussion}) describes some of the validation and attempts to assess accuracy and precision of derived quantities.

The radial velocities, stellar parameters, and abundances for all stars are supplied in a FITS file referred to as the allStar file. For DR16, this file is called allStar-r12-l33.fits\footnote{ \url{https://data.sdss.org/sas/dr16/apogee/spectro/aspcap/r12/l33/allStar-r12-l33.fits}} (reduction version r12 analyzed with the spectral libraries l33).

\subsection{Radial velocities} \label{sec:velocities}
The radial velocities are provided in the VHELIO\_AVG entry in the allStar file. As in DR14, these velocities are given in the solar system barycentric frame, not the heliocentric frame as the name suggests; the naming convention has been maintained from earlier releases for historical reasons. For stars that have been observed with multiple visits, the scatter of the individually derived radial velocities is provided in VSCATTER. This can be used, for example, to filter out possible binary systems.

\subsection{Stellar parameters} \label{sec:parameters}
As in previous data releases, and as described in previous sections in this paper, the ASPCAP stellar parameters include the ``classic'' spectroscopic stellar parameters \teff, \logg, \mh, \vmic, and \vsini ~(for dwarfs; a prescribed \vmac ~in the case of giants) as well as some initial estimate of abundances; \am, \xm{C}, and \xm{N}. The ``abundance parameters'' are needed for several reasons; for many of our cool, metal-rich targets, CNO-bearing molecular lines cover more or less the entire APOGEE spectral region and a correct modelling of these is required to fit the classical stellar parameters. Furthermore, since the $\alpha$-elements are important electron donors, modelling these correctly as the stellar parameters are determined is necessary, and, additionally, some of our targets have carbon abundances far enough from solar that the atmospheric structure is altered. These ``abundance parameters'' are determined from a global fit of the entire spectrum simultaneously with the other stellar parameters. 

In the second stellar abundance measurement stage, these abundances are redetermined using windows in the spectra covering only spectral lines sensitive to the abundance in question (see Section \ref{sec:abundances}). For that reason we recommend the use of these ``windowed'' abundances in most cases, but even so, the ``abundance parameters'' are stored and can be found in the FPARAM-array as well as in the ALPHA\_M tag.

As in previous data releases, some of the spectroscopically determined stellar parameters have been calibrated to match other, independent measurement of the parameters. These calibrations have varied over the data releases, and we include below a description of what has been done in DR16.

The spectroscopic and calibrated abundance parameters are provided in the FPARAM  and PARAM arrays; there are 9 entries in these arrays for each star, corresponding to \teff, \logg, $\log(v_{\rm mic})$, \xm{C}, \xm{N}, \am, log(\vsini), and $O$ (currently unused).
Many of these are also split out into appropriately named tags in the allStar file, as described below.

\subsubsection{Effective temperature, \teff}
\begin{figure}
\epsscale{1.30}
\plotone{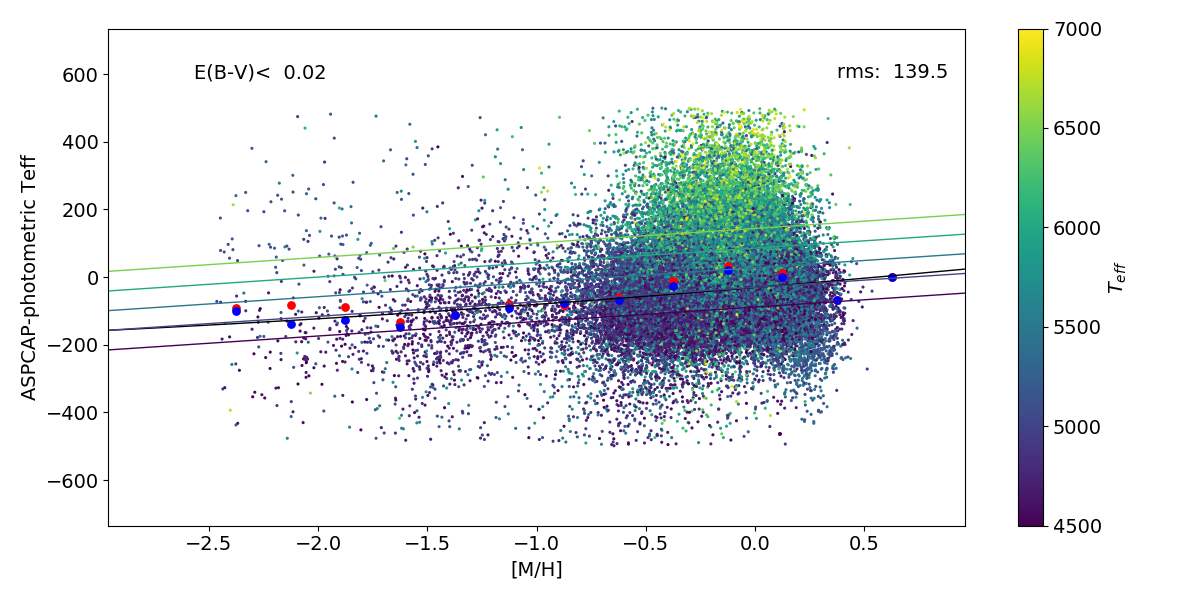}
\caption{Difference between spectroscopic DR16 \teff~and photometric \teff~from \citet{2009A&A...497..497G} as a function of metallicity. Large red and blue points show mean and median differences in bins of metallicity. The adopted \teff ~calibration is a function of \mh ~and \teff, and is indicated by the colored lines. }
\label{fig:teff_cal}
\end{figure}

The spectroscopic \teff~for all stars have been calibrated to the photometric scale of \citet{2009A&A...497..497G} (GHB) using linear relations as a function of metallicity and effective temperature: 
\begin{eqnarray}
    T_{\rm eff,cal} = T_{\rm eff} + 610.81 -4.275\cdot[{\rm M/H}]'- \nonumber \\
    0.116\cdot T_{\rm eff}'
\end{eqnarray}
where \mh ~and $T_{\rm eff}$ are the uncalibrated values of \mh ~and \teff, and the ``primed'' values are clipped to lie in the range $-2.5 < [M/H]' <0.75$ and $4500<T_{\rm eff}'< 7000$. The clipping is applied since the bulk of the stars in GHB fall within these limits, so we prefer not to extrapolate; outside of these ranges, the offsets from the end of the valid range were applied.
Figure \ref{fig:teff_cal} shows the data from which this relation was derived. 

The spectroscopically determined \teff~is given in a new TEFF\_SPEC tag while the calibrated \teff, as in previous data releases, can be found in the TEFF tag.

\subsubsection{Surface gravity, \logg}

As in DR14, the spectroscopic \logg~for giant stars have been calibrated using relations determined from stars in the Kepler field for which asteroseismic surface gravities are available \citep{Pinsonneault:2018vh}. As with previous data releases, we find that the relationship between the spectroscopic and asteroseismic values is complex; in particular, we find different offsets for red clump and red giant stars that occur in similar locations in a \teff-\logg~ diagram.

New for DR16 is that we also provide calibrated surface gravities for dwarfs, for which we use a combination of techniques: for warmer dwarfs we have asteroseismic values that we use, while for cooler dwarfs we derive an approximate calibration using isochrones. 

\begin{figure*}
\plotone{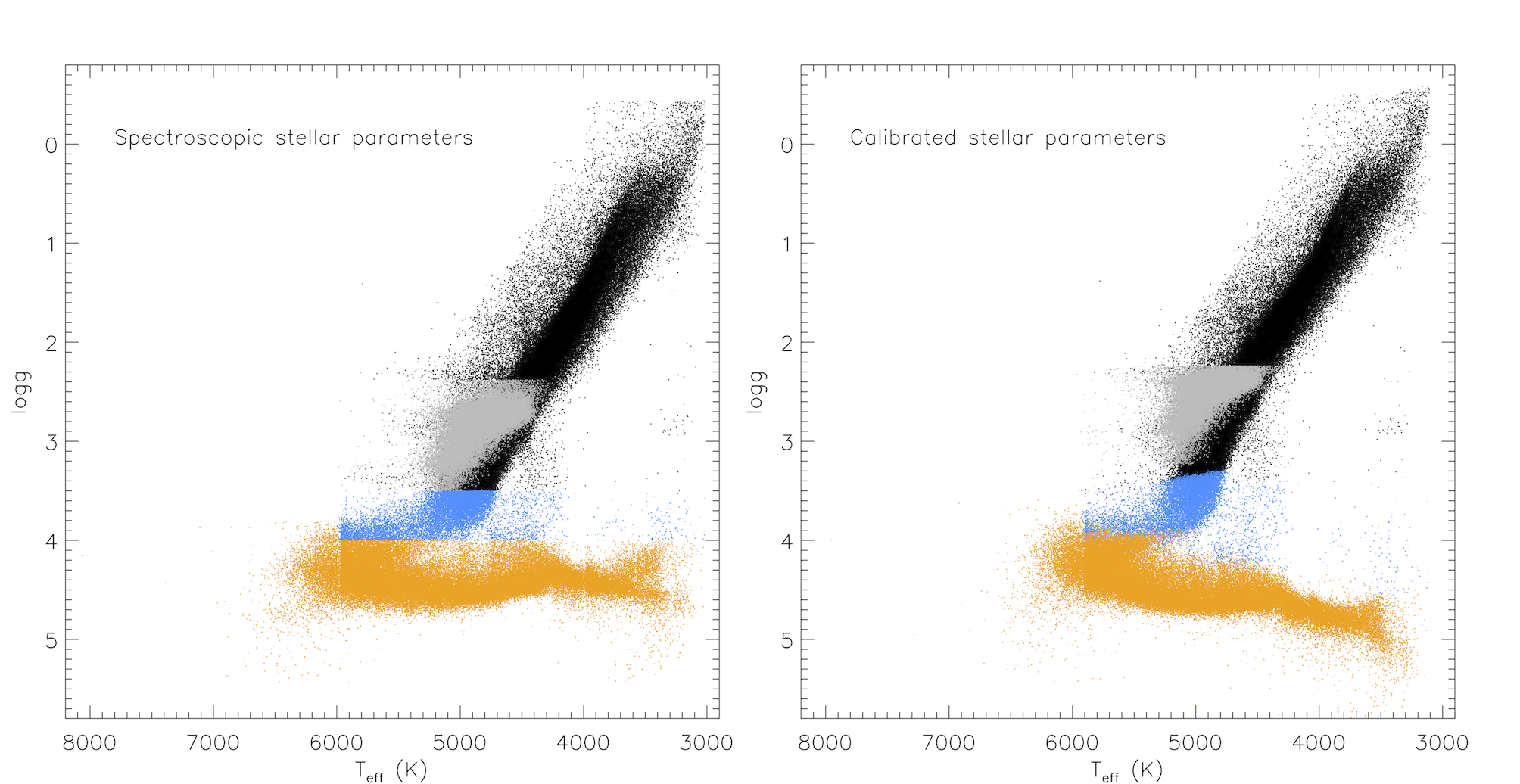}
\caption{Classification of stars into the \logg~ ``calibration-categories''; RGB (black), RC (gray), dwarf (orange), and RGB/dwarf (blue). The left figure shows how the categories were chosen from the spectroscopic stellar parameters, and the right figure shows where the categories end up after calibration. Note in particular the rather sharp RC-RGB ``grid edge'' at \logg$\sim 2.24$~ in the calibrated parameters. This figure does not demonstrate the dependence of the RC/RGB-classification  on \mh ~and [C/N] at a given \teff ~and  \logg.\label{fig:rcrgb}}
\end{figure*}

The classification of stars into these different ``calibration-categories'' was done according to the following criteria:
\begin{itemize}
  \item All stars with uncalibrated \logg$>4$ or \teff$>6000$~K are considered dwarf stars. 
  \item Stars with uncalibrated $2.38<$ \logg $<3.5$ and $$[C/N] > 0.04 - 0.46\cdot [M/H] -0.0028\cdot dT$$ are considered red clump stars. Here $dT$ is defined as 
  \begin{eqnarray}
  dT = T_{\rm eff,spec} - \nonumber \\
  (4400 + 552.6 \cdot (\log{g}_{\rm spec}-2.5)  - 324.6 \cdot [M/H] \nonumber
  \end{eqnarray}
  \item RGB-stars are defined as the stars with  uncalibrated $\log{g} < 3.5$ and $T_{\rm eff}<6000$~K that do not fall in the red clump category, as defined above.
  \item for stars with uncalibrated $3.5<$\logg$<4.0$ and \teff$<6000$~K, a correction is determined using both the RGB and dwarf calibrations, and a weighted correction is adopted based on \logg.
\end{itemize}
These classifications are shown in Figure \ref{fig:rcrgb} in the \teff - \logg ~plane, although this does not show the dependence of the RC/RGB-classification on \mh~ and [C/N].

The calibration relations for dwarf, RC, and RGB stars are respectively:

Dwarf stars:
\begin{eqnarray}
\log{g}_{\rm cal} = \log{g} - \nonumber \\
(-0.947 + 1.886\cdot10^{-4}\cdot T_{\rm eff,spec} +   0.410\cdot [M/H])
\end{eqnarray}

Red clump stars:
\begin{eqnarray}
\log{g}_{\rm cal} = \log{g} - \nonumber \\
( -4.532 + 3.222\cdot \log{g} -0.528\cdot (\log{g})^2)
\end{eqnarray}

Red giant stars:
\begin{eqnarray}
\log{g}_{\rm cal} = \log{g} - \nonumber \\
( -0.441 +  0.7588\cdot \log{g}' - 0.2667\cdot (\log{g'})^2 \nonumber \\
+ 0.02819\cdot (\log{g}')^3 + 0.1346\cdot [M/H]')
\end{eqnarray}
where 
\begin{eqnarray}
\log{g}'          = \log{g} {\rm \ for } \log{g} \ge 1.2795 \nonumber \\
\log{g}'          = 1.2795  {\rm \ for } \log{g} < 1.2795 \nonumber \\
\left[M/H\right]' = [M/H]   {\rm \ for \  [M/H] } \le 0.5 \nonumber \\
\left[M/H\right]' = 0.5     {\rm \ for \  [M/H] } > 0.5 \nonumber
\end{eqnarray}
where the fixed value of $\log{g}'$ at low surface gravity and $[M/H]'$ at high metallicity avoids extrapolation into a region where there are few calibrators.

The functional forms for these calibrations were determined from inspection of the relations between spectroscopic, asteroseismic, and isochrone surface gravities. While these capture a significant portion of the relationships, small trends with other parameters may certainly exist, and the calibrated surface gravities cannot be assumed to be more accurate than $\sim 0.05$ dex.

We note that no smooth transition is implemented between the RGB and RC calibrations resulting in a small discontinuity in \logg at the transition value. Based on the asteroseimsic results, we find that  93\% of the RGB stars and 96\% of the RC stars are classified correctly by our procedure. For the incorrectly classified stars, the calibrated surface gravities will be systematically off. However, since we do the abundance analysis using the uncalibrated parameters, the abundances are unaffected.

The spectroscopic \logg~is given in the LOGG\_SPEC tag in the allStar file, while the calibrated \logg, as in previous data releases, can be found in the LOGG tag. 

\subsubsection{The abundance parameters; \mh, \am, \xm{C}, and \xm{N}}
In DR16, the abundance parameters \xm{C}, and \xm{N}~are not calibrated. The \am~ parameter is calibrated by the application of a zero-point shift of 0.033 dex for giants and 0.01 dex for dwarfs so that the mean of solar metallicity stars in the solar neighborhood has \am=0.0 (see Section \ref{sec:abundances} and Table \ref{tab:abundcal}). 
The  \mh~parameter is also provided in the M\_H tag, and the calibrated \am~parameter is provided in the ALPHA\_M tag. We note that, due to an inadvertent error, the values in the M\_H tag (and the corresponding entry in the PARAM array) differ from the values in the FPARAM array by 0.003 and 0.0004 dex for giants and dwarfs, respectively.

\subsection{Stellar abundances} \label{sec:abundances}
In DR16, the abundance determination of 26 species is attempted; C, C I, N, O, Na, Mg, Al, Si, P, S, K, Ca, Ti, Ti II, V, Cr, Mn, Fe, Co, Ni, Cu, Ge, Rb, Ce, Nd, and Yb. Note that, as in previous data releases, the uncalibrated spectroscopic stellar parameters were used when determining the stellar abundances. The reason for this is that the spectroscopic parameters give the best general fit to the stellar spectrum, and thereby give the best description of possible blends when determining the abundances from the abundance windows. 

All of the ``raw'' abundance measurements for all stars are presented in the FELEM array, in which the order of the array elements for each star is by atomic number, with entries as listed above. Note that, in this array, the abundances for different elements are given with respect to either the total metals or to hydrogen, depending on which grid dimension was used during the fit.

In previous data releases, a \teff-dependent calibration was applied to each individual elemental abundance to remove apparent trends in the uncalibrated abundances, based on observations of star clusters. For DR16 no such calibration is applied because, with the modification to the abundance pipeline, the trends with effective temperature for most elements have reduced amplitude in the cluster sample as compared with previous data processing. That being said, inspection of the full data set suggests that some trends of abundances with stellar parameters can exist for some elements, such that users need to exercise caution when comparing abundances across different regions of stellar parameters space (see Section \ref{sec:loggtrends}).

The only calibration applied to the DR16 abundances is a zero-point shift to force stars with solar \mh~in the solar neighborhood to have a mean \xm{X}=0. This is done separately for giants and dwarfs, where ``giants'' in this case are defined as stars with $\log g < 2+ (T_{eff}-3500)/650$ and \logg$<4$ and \teff$<7000$~K, and all others are defined as dwarfs. More specifically, we average the ``raw'' abundances of all stars within 0.5 kpc of the Sun, based on {\it Gaia} DR2 parallaxes (\citealt{2016A&A...595A...1G,2018A&A...616A...1G,2018A&A...616A...2L}), and with $-0.05<$\mh$<0.05$, and subtract this value from the ``raw'' \xm{X} of all stars. The applied shifts are tabulated in Table \ref{tab:abundcal} (compare Table 5 in \citet{2018AJ....156..125H} for the shifts applied in DR13 and DR14); they are generally small (of the order of hundredths of a dex), but are substantial for a handful of elements such as Al, P, V, and Mn. Note that this calibration is a zero-point offset only. Formally, using bracket notation (\xfe{X}) suggests that the abundances are relative to those of the Sun; we did not choose this procedure because many of the lines/elements that we measure in cooler stars are very weak in the solar spectrum, so an APOGEE-based solar abundance measurement has significant uncertainties. Instead, we build upon many results reported in the literature that suggest that the mean \xfe{X} in solar neighborhood stars is close to solar at solar abundance (\citealt[][among others]{2003MNRAS.340..304R,2012A&A...545A..32A,2014A&A...562A..71B}). Small intrinsic spread in \xfe{X} at solar abundance as found by \citet{2018ApJ...865...68B} will still be reflected in the calibrated abundances, as we only apply a single mean offset to all stars. 

The calibrated abundances are provided in the X\_H and X\_M arrays in the allStar file, where the difference between these is just the value of M\_H. For further discussion about the APOGEE abundance scale, see Section \ref{sec:abundancescale}.

\subsubsection{''Named'' abundance tags, X\_FE}
\label{sec:namedtags}

In addition to the abundances in the X\_H and X\_M arrays, we provide
abundances in ``named'' X\_FE abundance tags, e.g., C\_FE, N\_FE, O\_FE, etc., where we provide abundances relative to iron. These are simply calculated by subtracting the \xh{Fe} abundance from the \xh{X} abundance for each element. 

However, we populate the X\_FE tags only for stars that we believe the abundances are the most reliable, and do not populate them for abundances that are expected to have large uncertainties or the possibility of significant systematic error. There are a number of reasons why a X\_FE tag could be unpopulated (i.e., has a value of -9999.99) :

\begin{itemize}

    \item We do not populate the X\_FE tags if any bit in the corresponding ELEMFLAG is set. This means that if the estimated uncertainty (see Section \ref{sec:uncertainties}) is larger than 0.2 dex, or if the \teff\  is outside the range in which we think the abundances are reliable (see Section \ref{sec:loggtrends}), then then corresponding X\_FE tag is not populated.
    
    \item For carbon, nitrogen, and iron, the corresponding named tags (C\_FE, N\_FE, FE\_H) are not populated if the elemental window abundance deviates significantly (more than 0.25 dex for C and N, more than 0.1 dex for Fe) from the corresponding ``abundance parameter'' (\xm{C}, \xm{N}, and \mh). This behavior is not expected, so these objects are flagged with a PARAM\_MISMATCH bit in the corresponding ELEMFLAG.
    Since this can affect FE\_H, the implication is that \emph{none} of the named tags (C\_FE, N\_FE, O\_FE, etc.) will be populated for such a star, since the named tags give abundance relative to iron. The bulk of the stars that show this behavior are cool, metal-rich giants, so users are warned that using the named tags will lead to a bias against these stars in a sample. For use cases where such biases may be relevant, users may wish to calculate abundances relative to iron from the X\_H or X\_M arrays, recognizing the possibility of some systematic uncertainties for the subset of stars with a PARAM\_MISMATCH bit set.

    \item We do not populate the X\_FE tags for stars with $H>14.6$, since for these the RV determination of the individual visits might fail, leading to bad combination of the spectra, compare Section \ref{sec:rv_val}.
    
    \item We do not populate the CE\_FE tag for stars with \vrad$>120$, because for these stars, the window for the single Ce line that is used shifts into wavelengths that fall in one of the gaps between the APOGEE detectors.
    
    \item We do not populate the named tags for several unreliable elements, including \emph{all} abundances of Ge, Rb, and Yb because the few lines available are so weak/blended that we cannot determine these abundances reliably\footnote{In addition to the Rb line being very weak, an incorrect wavelength of 15289.966~\AA~(air) from the Kurucz line list was used when constructing the spectral grids, instead of the correct 15289.480~\AA (air), rendering the Rb abundances in DR16 useless.}. The Nd abundances are also completely removed in the ND\_FE tag, but in this case the reason is mainly limitations in the current methodology; the available Nd lines are all blended with lines that also vary in the \mh~dimension, which means that we cannot distinguish the Nd-contribution to the absorption line from the contribution from the blending element. The abundances for these four elements were also removed in the named tags in DR14.
\end{itemize}

As a result of these criteria, users should be aware that using abundances from the named tags will yield a sample with additional biases over those present from selection effects, in exchange for getting a sample with abundances that are expected to be more reliable. The abundances in the X\_M and X\_H arrays are not subject to these additional biases, but may be less reliable for some stars. 

\begin{deluxetable}{lrr}
\tablecaption{The determined abundances are zero-point shifted to make stars with solar M\_H in the solar neighborhood have \xm{X}=0. Below is the list of the applied shifts for giant and dwarf stars, respectively. For Na, P, Ti II, and Ce no calibrated abundances are given for dwarfs because of large uncertainties, see Section \ref{sec:elements}.\label{tab:abundcal}}
\tablehead{\colhead{Element} & \colhead{Giants} & \colhead{Dwarfs}}
\startdata
\xm{C}     &   0.000 & +0.003   \\  
\xm{C I}   &   0.000 & -0.003   \\  
\xm{N}     &   0.000 & +0.002   \\  
\xm{O}     &  -0.022 & -0.001   \\  
\xm{Na}    &  -0.022 &  \nodata \\  
\xm{Mg}    &  -0.009 & +0.041   \\  
\xm{Al}    &  -0.148 & -0.043   \\  
\xm{Si}    &  -0.038 & +0.026   \\  
\xm{P}     &  +0.183 &  \nodata \\  
\xm{S}     &  -0.040 & -0.054   \\  
\xm{K}     &  +0.090 & +0.108   \\  
\xm{Ca}    &  -0.002 & -0.035   \\  
\xm{Ti}    &  -0.009 & +0.027   \\  
\xm{Ti II} &  -0.249 &  \nodata \\  
\xm{V}     &  +0.192 & -0.026   \\  
\xm{Cr}    &  +0.020 & -0.065   \\  
\xm{Mn}    &  +0.121 & +0.145   \\  
\xm{Fe}    &   0.000 &  0.000   \\  
\xm{Co}    &  -0.027 & +0.079   \\  
\xm{Ni}    &  -0.016 & -0.043   \\  
\xm{Cu}    &  +0.018 & +0.103   \\  
\xm{Ce}    &  -0.070 &  \nodata \\  
\mh        &   0.000 & +0.003   \\  
\am        &  -0.033 & -0.011   \\
\enddata
\end{deluxetable}

\subsubsection{The abundance scale}\label{sec:abundancescale}
The solar abundance scale of DR16 is complex, but, in general, we are likely to be close to the scale of \citet{2007SSRv..130..105G} for many elements. The relevant steps in making this a hard question are re-iterated below: 

\begin{itemize}

    \item When constructing the line list for the analysis, we adjust the atomic data to fit a spectrum of the Sun with the \citet{2007SSRv..130..105G} abundances (and the parameters \teff=5777 K, \logg=4.44, [Fe/H]=0.00, \vmic=1.10 km/s), but only within the quoted uncertainties of the source of the data. Moreover, we simultaneously adjust the atomic data to also fit a spectrum of Arcturus (with the parameters \teff=4286 K, \logg=1.66, [Fe/H]=-0.52, \vmic=1.74 km/s), and abundances from the literature (see Smith et al. in prep for details). Molecular data are not adjusted.
   
    \item The chemical abundances in the stellar atmosphere models and the spectral synthesis calculations are specified relative to the solar abundance scale of \citet{2007SSRv..130..105G}.
   
    \item The calibrated abundances have been zero-point corrected so that solar-metallicity stars in the solar neighborhood have \xm{X}=0; see Table \ref{tab:abundcal}. We do not calibrate directly to the Sun because it is not typical of the stars in the APOGEE sample, and because abundances of many elements are not well determined in stars with effective temperature as high as that of the Sun. Note, however, that the calibration offsets are small for many elements, as shown in Table \ref{tab:abundcal}. C and N abundances have not been calibrated for giants since those abundances are expected to be affected by the star's evolution and not follow Galactic chemical evolution.

\end{itemize}

We stress that the uncalibrated abundances derived for giants from molecular lines -- C, N, O -- are not adjusted in any way and, provided the molecular data do not have systematic uncertainties, those abundances should be at least close to the \citet{2007SSRv..130..105G} scale.
Regarding the uncalibrated abundances derived from atomic lines, the abundance scale varies from element to element. For elements that have strong features in the Sun, the adjustments to the atomic data do not depend much on the fitting of the Arcturus spectrum/abundances, and if these same features happen to have high weight in the ASPCAP analysis, the abundance scale should be close to that of \citet{2007SSRv..130..105G}. For elements whose abundance determination rely more on lines whose $\log gf$-values were more adjusted using the Arcturus spectrum, the absolute abundance scale is less well known. The fact that the adjustments to the atomic data depend on Arcturus as well as the Sun is a significant motivation for calibrating the derived spectroscopic stellar abundances based on the solar neighborhood solar metallicity stars.  C and N in giants do not have any calibration applied and should -- if we assume that the molecular data used does not have any systematics -- be at least close to the \citet{2007SSRv..130..105G} scale. For all other calibrated abundances our philosophy is that they are provided on a ``true bracket'' (i.e., relative) scale in the spectroscopic sense, where abundances are simply presented in a ratio to our own, undetermined, unspecified solar abundance.

A check on our solar reference scale is provided by our analysis of the solar spectrum reflected off the asteroid Vesta (see Table \ref{tab:sun}). However, we stress again that the Sun is not a typical star within the APOGEE sample, and that these values cannot be taken as deviations from the \citet{2007SSRv..130..105G} scale for the main sample of APOGEE.

\begin{deluxetable}{lrr}
\tablecaption{The determined stellar abundances for our solar spectrum reflected off the asteroid Vesta. \label{tab:sun}}
\tablehead{
\colhead{Element} & \colhead{Spectroscopic} & \colhead{Calibrated} 
}
\startdata
$[{\rm C/M}]$   &     0.02 &     0.02 \\
$[{\rm C~I/M}]$ &     0.03 &     0.03 \\
$[{\rm N/M}]$   &     0.18 &     0.18 \\
$[{\rm O/M}]$   &     0.05 &     0.05 \\
$[{\rm Na/M}]$  &    -0.02 &    -0.02 \\
$[{\rm Mg/M}]$  &    -0.05 &    -0.01 \\
$[{\rm Al/M}]$  &     0.05 &     0.01 \\
$[{\rm Si/M}]$  &    -0.03 &    -0.01 \\
$[{\rm P/M}]$   &    -0.21 &    -0.21 \\
$[{\rm S/M}]$   &     0.03 &    -0.03 \\
$[{\rm K/M}]$   &    -0.18 &    -0.07 \\
$[{\rm Ca/M}]$  &     0.02 &    -0.02 \\
$[{\rm Ti/M}]$  &    -0.06 &    -0.03 \\
$[{\rm Ti~II/M}]$ &   0.11 &     0.11 \\
$[{\rm V/M}]$   &     0.00 &    -0.02 \\
$[{\rm Cr/M}]$  &     0.10 &     0.04 \\
$[{\rm Mn/M}]$  &    -0.10 &     0.04 \\
$[{\rm Fe/M}]$  &    -0.01 &     0.00 \\
$[{\rm Co/M}]$  &     0.21 &     0.29 \\
$[{\rm Ni/M}]$  &     0.05 &     0.01 \\
$[{\rm Cu/M}]$  &    -0.16 &    -0.05 \\
$[{\rm Ce/M}]$  &    -0.12 &    -0.12 \\
\enddata
\end{deluxetable}

\subsection{Uncertainties} \label{sec:uncertainties}
As in DR14, we find that the uncertainties for parameters and abundances returned by the fitting routine in FERRE are unrealistically low in most cases. As a result, we take an alternate approach to derive empirical uncertainties, and adopt for the final uncertainties the larger of the FERRE and empirical uncertainty estimates.

For \teff~and \logg, we estimate uncertainties from the scatter around the calibration relations, parameterized as a function of \teff, \mh, and S/N, which captures the main dependencies of the scatter. The form of the adopted uncertainty parameterization is:
\begin{equation}
    \ln \sigma = A + B \cdot T_{\rm eff}'' + C \cdot S/N'' + D \cdot [M/H]  \label{eq:param_unc}
\end{equation}
where $T_{\rm eff}'' =$ TEFF\_SPEC-4500 and $S/N'' = $ SNREV-100 for SNREV$\leq 200$, otherwise it is capped at a value of 100; the coefficients are presented in Table \ref{tab:param_uncertainties}. The final uncertainties are presented in the TEFF\_ERR and LOGG\_ERR tags in the allStar file.

\begin{deluxetable*}{lrrrrr}
\tablecaption{The coefficients describing the supplied uncertainties in stellar parameters, compare Equation \ref{eq:param_unc}. Due to the parameterization, $e^{\rm A}$ can be taken as a measure of a typical uncertainty for a star with \teff=4500~K, \mh=0.0, and S/N=100. \label{tab:param_uncertainties}}
\tablehead{
\colhead{Parameter}  & \colhead{A}  & \colhead{B}  & \colhead{C}  & \colhead{D} & \colhead{$e^{\rm A}$}
}
\startdata
\teff          &$  4.583$ &$  2.965\cdot 10^{-4}$ &$ -2.177\cdot 10^{-3}$ & $-0.117$ & 98\\
\logg ~(dwarfs)&$ -2.327$ &$ -1.349\cdot 10^{-4}$ &$  2.269\cdot 10^{-4}$ & $-0.306$ & 0.10\\
\logg ~(RC)    &$ -3.444$ &$  9.584\cdot 10^{-4}$ &$ -5.617\cdot 10^{-4}$ & $-0.181$ & 0.03\\
\logg ~(RGB)   &$ -2.923$ &$  2.296\cdot 10^{-4}$ &$  6.900\cdot 10^{-4}$ & $-0.277$ & 0.05\\
\enddata    
\end{deluxetable*}

For the uncertainties for the derived stellar abundances, we adopted a new scheme in DR16 using repeat observations of the same star. As mentioned in Section \ref{sec:scope}, there are a moderate number of stars that were observed in multiple overlapping fields with different field centers, and since the reduction and analysis pipeline is built on processing field-by-field, these stars are completely independently analyzed more than once by ASPCAP. The differences between the derived abundances from the different visits provides some information about the uncertainties. These repeat observations include stars covering a large region in stellar parameter space. To supplement the coverage in S/N, several individual visit spectra of cluster stars were processed using ASPCAP. The differences are larger when the S/N is lower and also in regions of parameter space where lines are generally weaker (lower \mh~ and higher \teff). These ``repeat abundances'' and their deviations as a function of \teff, \mh, and S/N were used to estimate uncertainties for the entire sample of stars. Specifically, the differences between pairs of measurements were tabulated for all of the repeats, along with the mean \teff, \mh, and S/N (only pairs with S/N the same within 20\% were considered). A fit was then performed to these differences (multiplied by $\sqrt{\pi}/2$ to provide an unbiased estimator of the standard deviation, $\sigma$) using the form:
\begin{equation}
    \ln \sigma = A + B \cdot T_{\rm eff}' + C \cdot S/N' + D \cdot [M/H] + E \cdot (T_{\rm eff}')^2 \label{eq:unc}
\end{equation}
where $T_{\rm eff}' =$ TEFF\_SPEC-4500 and $S/N' = $ SNREV-100 for SNREV$\leq 200$, otherwise it is capped at a value of 100. The functional form was chosen to provide a reasonable match to the observed distribution of measured differences, but it is still just an approximation. We adopt the same uncertainties for both X\_H and X\_M in the X\_H\_ERR and X\_M\_ERR arrays.

\begin{figure*}
\epsscale{1.18}
\plotone{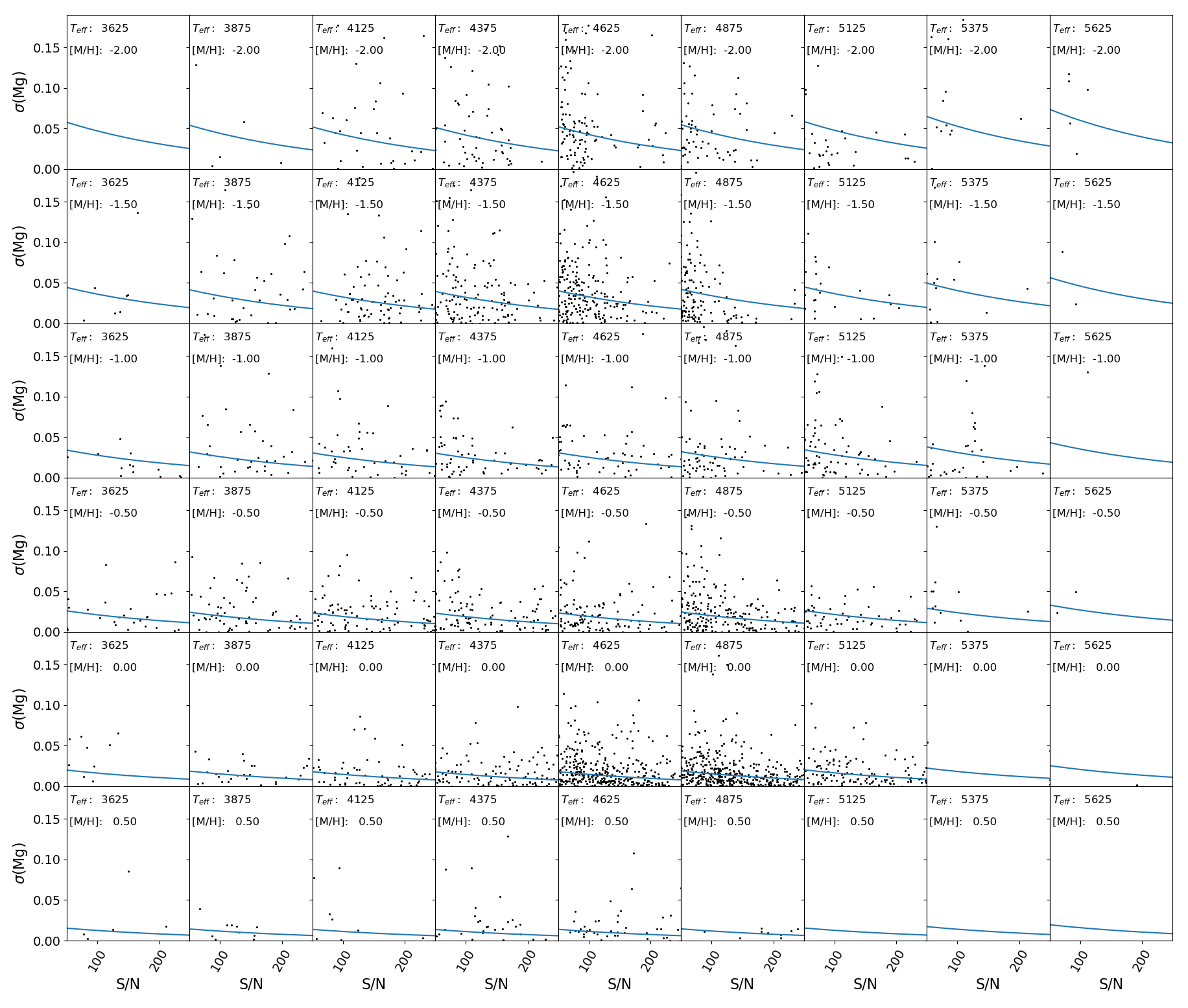}
\caption{Uncertainties for Mg as derived from repeat observations. Different subpanels show observations in different bins of \teff~ (250 K wide)  and \mh~(0.5 dex wide). In each subpanel, points are plotted as a function of S/N, and individual points are the (noisy) uncertainty derived from a single pair. The lines represent the global fit to the full data set using Equation \ref{eq:unc}; since it is a global fit, the fit can be shown in subpanels even if there are no data points.}
\label{fig:Mg_unc}
\end{figure*}

Figure \ref{fig:Mg_unc} shows an example of the methodology for Mg. Each panel in the plot shows data for different bins in \teff~ and \mh, where individual points are the differences for different pairs of observation/measurements; note that these individual differences are noisy by definition. The curves show the function fits to the data as described by equation \ref{eq:unc}.

\begin{figure*}
\plotone{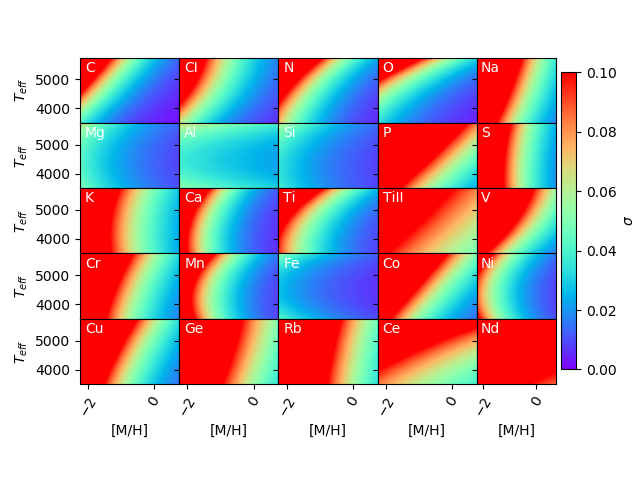}
\caption{Fits for uncertainties (in dex) in all elemental abundances as a function of \teff~ and \mh, at a S/N of 125.}
\label{fig:all_unc}
\end{figure*}

Figure \ref{fig:all_unc} shows a graphical summary of the derived uncertainties for all elements as a function of \teff~ and \mh, for a S/N of 125. The coefficients for all elements are presented for dwarfs in Table \ref{tab:unc-dwarfs} and giants in Table \ref{tab:unc-giants}.

It is important to note that the uncertainties estimated in this way depend only on \teff, \mh, and S/N. Individual stars may have larger uncertainties, if pixels with important information have larger uncertainties, e.g., if they happen to fall near a sky line in all of the visit spectra for the star, but this is not reflected in the tabulated uncertainties unless the uncertainty returned by the fitting routine is larger than that estimated by Equation \ref{eq:unc}.

\begin{deluxetable*}{lrrrrrr}
\tablecaption{The coefficients describing the supplied uncertainties in stellar abundances for dwarfs stars, compare Equation \ref{eq:unc}. Due to the parameterization, $e^{\rm A}$ can be taken as a measure of a typical uncertainty for a star with \teff=4500~K, \mh=0.0, and S/N=100.\label{tab:unc-dwarfs}}
\tablehead{
\colhead{Element}  & \colhead{A}  & \colhead{B}  & \colhead{C}  & \colhead{D}  & \colhead{E} & \colhead{$e^{\rm A}$}
}
\startdata
C     & $ -4.414$      & $  7.793\cdot 10^{-4}$  & $ -3.952\cdot 10^{-3}$  & $ -0.9646$  & $ -9.109\cdot 10^{-8}$  & 0.01\\
C I   & $ -3.610$      & $  7.099\cdot 10^{-5}$  & $ -3.449\cdot 10^{-3}$  & $ -0.6467$  & $  7.871\cdot 10^{-9}$  & 0.03\\
N     & $ -3.000$      & $ -1.600\cdot 10^{-4}$  & $ -2.796\cdot 10^{-3}$  & $ -0.4285$  & $  2.185\cdot 10^{-7}$  & 0.05\\
O     & $ -4.120$      & $  8.330\cdot 10^{-4}$  & $ -4.158\cdot 10^{-3}$  & $ -0.8709$  & $ -8.326\cdot 10^{-8}$  & 0.02\\
Na    & $ -2.504$      & $  5.198\cdot 10^{-4}$  & $ -5.516\cdot 10^{-3}$  & $ -1.004 $  & $ -5.023\cdot 10^{-8}$  & 0.08\\
Mg    & $ -4.304$      & $ -7.000\cdot 10^{-5}$  & $ -3.714\cdot 10^{-3}$  & $ -0.5233$  & $  1.943\cdot 10^{-7}$  & 0.01\\
Al    & $ -3.535$      & $ -5.748\cdot 10^{-5}$  & $ -6.185\cdot 10^{-3}$  & $ -0.2101$  & $  1.925\cdot 10^{-7}$  & 0.03\\
Si    & $ -4.416$      & $  9.626\cdot 10^{-6}$  & $ -3.966\cdot 10^{-3}$  & $ -0.5371$  & $  1.680\cdot 10^{-7}$  & 0.01\\
P     & $ -2.255$      & $  4.740\cdot 10^{-4}$  & $ -5.199\cdot 10^{-3}$  & $ -0.4891$  & $ -2.459\cdot 10^{-8}$  & 0.10\\
S     & $ -2.704$      & $ -1.969\cdot 10^{-4}$  & $ -3.551\cdot 10^{-3}$  & $ -0.5832$  & $  9.759\cdot 10^{-8}$  & 0.07\\
K     & $ -3.651$      & $  4.178\cdot 10^{-4}$  & $ -4.610\cdot 10^{-3}$  & $ -0.5761$  & $  9.917\cdot 10^{-8}$  & 0.03\\
Ca    & $ -4.152$      & $  1.103\cdot 10^{-4}$  & $ -4.836\cdot 10^{-3}$  & $ -0.5673$  & $  2.111\cdot 10^{-7}$  & 0.02\\
Ti    & $ -3.499$      & $  7.504\cdot 10^{-4}$  & $ -4.072\cdot 10^{-3}$  & $ -0.6025$  & $  5.027\cdot 10^{-9}$  & 0.03\\
Ti II & $ -2.456$      & $  4.921\cdot 10^{-4}$  & $ -4.640\cdot 10^{-3}$  & $ -0.3222$  & $ -5.522\cdot 10^{-8}$  & 0.09\\
V     & $ -2.820$      & $  5.491\cdot 10^{-4}$  & $ -4.447\cdot 10^{-3}$  & $ -0.7493$  & $ -5.033\cdot 10^{-8}$  & 0.06\\
Cr    & $ -3.032$      & $  4.725\cdot 10^{-4}$  & $ -5.024\cdot 10^{-3}$  & $ -0.8113$  & $  2.298\cdot 10^{-8}$  & 0.05\\
Mn    & $ -3.715$      & $ -1.693\cdot 10^{-4}$  & $ -3.860\cdot 10^{-3}$  & $ -0.9168$  & $  2.271\cdot 10^{-7}$  & 0.02\\
Fe    & $ -4.495$      & $ -1.658\cdot 10^{-4}$  & $ -3.634\cdot 10^{-3}$  & $ -0.6034$  & $  2.586\cdot 10^{-7}$  & 0.01\\
Co    & $ -1.953$      & $  3.190\cdot 10^{-4}$  & $ -3.384\cdot 10^{-3}$  & $ -0.5297$  & $ -3.746\cdot 10^{-8}$  & 0.14\\
Ni    & $ -3.819$      & $ -2.766\cdot 10^{-4}$  & $ -4.217\cdot 10^{-3}$  & $ -0.7018$  & $  2.978\cdot 10^{-7}$  & 0.02\\
Cu    & $ -2.672$      & $  3.301\cdot 10^{-4}$  & $ -4.262\cdot 10^{-3}$  & $ -0.5880$  & $  3.239\cdot 10^{-8}$  & 0.07\\
Ge    & $ -2.276$      & $ -1.093\cdot 10^{-5}$  & $ -5.608\cdot 10^{-3}$  & $ -0.6357$  & $  1.058\cdot 10^{-7}$  & 0.10\\
Rb    & $ -2.401$      & $  1.689\cdot 10^{-4}$  & $ -4.290\cdot 10^{-3}$  & $ -0.7583$  & $  5.474\cdot 10^{-8}$  & 0.09\\
Ce    & $ -2.341$      & $  8.454\cdot 10^{-4}$  & $ -3.843\cdot 10^{-3}$  & $ -0.9028$  & $ -2.464\cdot 10^{-7}$  & 0.10\\
Nd    & $ -2.159$      & $  3.614\cdot 10^{-4}$  & $ -4.754\cdot 10^{-3}$  & $ -0.2314$  & $ -7.274\cdot 10^{-8}$  & 0.12\\
Yb    & $ -2.225$      & $  8.632\cdot 10^{-5}$  & $ -2.746\cdot 10^{-3}$  & $ -0.4159$  & $  5.910\cdot 10^{-8}$  & 0.11\\
M     & $ -4.571$      & $ -1.225\cdot 10^{-4}$  & $ -3.564\cdot 10^{-3}$  & $ -0.7405$  & $  2.338\cdot 10^{-7}$  & 0.01\\
$\alpha$  & $ -5.320$  & $  3.462\cdot 10^{-4}$  & $ -2.692\cdot 10^{-3}$  & $ -1.210 $  & $  1.181\cdot 10^{-7}$  & 0.01\\
\enddata    
\end{deluxetable*}

\begin{deluxetable*}{lrrrrrr}
\tablecaption{The coefficients describing the supplied uncertainties in stellar abundances for giant stars, compare Equation \ref{eq:unc}. Due to the parameterization, $e^{\rm A}$ can be taken as a measure of a typical uncertainty for a star with \teff=4500~K, \mh=0.0, and S/N=100.\label{tab:unc-giants}}
\tablehead{
\colhead{Element}  & \colhead{A}  & \colhead{B}  & \colhead{C}  & \colhead{D}  & \colhead{E} & \colhead{$e^{\rm A}$}
}
\startdata
C     & $ -4.363$      & $  8.162\cdot 10^{-4}$  & $ -4.265\cdot 10^{-3}$  & $ -0.9649$  & $ -6.485\cdot 10^{-8}$ & 0.01\\
C I   & $ -3.764$      & $  5.137\cdot 10^{-4}$  & $ -4.379\cdot 10^{-3}$  & $ -0.8004$  & $ -1.450\cdot 10^{-7}$ & 0.02\\
N     & $ -3.942$      & $  6.218\cdot 10^{-4}$  & $ -3.558\cdot 10^{-3}$  & $ -0.8631$  & $  1.307\cdot 10^{-7}$ & 0.02\\
O     & $ -4.115$      & $  8.491\cdot 10^{-4}$  & $ -4.604\cdot 10^{-3}$  & $ -0.6873$  & $  1.902\cdot 10^{-7}$ & 0.02\\
Na    & $ -2.981$      & $  3.366\cdot 10^{-4}$  & $ -4.071\cdot 10^{-3}$  & $ -0.7673$  & $ -3.861\cdot 10^{-8}$ & 0.05\\
Mg    & $ -4.231$      & $  6.616\cdot 10^{-5}$  & $ -4.164\cdot 10^{-3}$  & $ -0.5313$  & $  2.218\cdot 10^{-7}$ & 0.01\\
Al    & $ -3.544$      & $  3.438\cdot 10^{-5}$  & $ -5.541\cdot 10^{-3}$  & $ -0.1858$  & $  3.756\cdot 10^{-7}$ & 0.03\\
Si    & $ -4.288$      & $  1.662\cdot 10^{-4}$  & $ -4.279\cdot 10^{-3}$  & $ -0.4886$  & $  2.282\cdot 10^{-7}$ & 0.01\\
P     & $ -2.564$      & $  5.701\cdot 10^{-4}$  & $ -4.494\cdot 10^{-3}$  & $ -0.6192$  & $ -2.407\cdot 10^{-8}$ & 0.08\\
S     & $ -3.166$      & $  8.896\cdot 10^{-5}$  & $ -4.740\cdot 10^{-3}$  & $ -0.8095$  & $  6.315\cdot 10^{-8}$ & 0.04\\
K     & $ -3.083$      & $  6.806\cdot 10^{-5}$  & $ -5.288\cdot 10^{-3}$  & $ -0.6897$  & $  8.034\cdot 10^{-8}$ & 0.05\\
Ca    & $ -4.138$      & $  2.729\cdot 10^{-4}$  & $ -5.101\cdot 10^{-3}$  & $ -1.025 $  & $  2.007\cdot 10^{-7}$ & 0.02\\
Ti    & $ -3.800$      & $  4.627\cdot 10^{-4}$  & $ -4.950\cdot 10^{-3}$  & $ -0.8037$  & $  2.508\cdot 10^{-7}$ & 0.02\\
Ti II & $ -2.441$      & $  2.467\cdot 10^{-4}$  & $ -4.997\cdot 10^{-3}$  & $ -0.2710$  & $ -1.524\cdot 10^{-8}$ & 0.09\\
V     & $ -2.729$      & $  5.757\cdot 10^{-4}$  & $ -5.256\cdot 10^{-3}$  & $ -0.6792$  & $ -1.444\cdot 10^{-7}$ & 0.07\\
Cr    & $ -3.103$      & $  2.797\cdot 10^{-4}$  & $ -5.079\cdot 10^{-3}$  & $ -0.7941$  & $  4.822\cdot 10^{-8}$ & 0.04\\
Mn    & $ -3.924$      & $  1.974\cdot 10^{-4}$  & $ -4.998\cdot 10^{-3}$  & $ -0.9505$  & $  2.664\cdot 10^{-7}$ & 0.02\\
Fe    & $ -4.590$      & $ -6.248\cdot 10^{-7}$  & $ -3.842\cdot 10^{-3}$  & $ -0.4917$  & $  4.590\cdot 10^{-7}$ & 0.01\\
Co    & $ -3.005$      & $  5.860\cdot 10^{-4}$  & $ -5.176\cdot 10^{-3}$  & $ -0.7629$  & $ -5.215\cdot 10^{-9}$ & 0.05\\
Ni    & $ -4.058$      & $  1.110\cdot 10^{-4}$  & $ -4.878\cdot 10^{-3}$  & $ -0.7812$  & $  3.155\cdot 10^{-7}$ & 0.02\\
Cu    & $ -3.112$      & $  3.691\cdot 10^{-4}$  & $ -4.793\cdot 10^{-3}$  & $ -0.8306$  & $  3.571\cdot 10^{-8}$ & 0.04\\
Ge    & $ -2.561$      & $  1.750\cdot 10^{-4}$  & $ -4.898\cdot 10^{-3}$  & $ -0.6220$  & $ -2.326\cdot 10^{-8}$ & 0.08\\
Rb    & $ -2.595$      & $  1.294\cdot 10^{-4}$  & $ -4.953\cdot 10^{-3}$  & $ -0.6973$  & $  1.799\cdot 10^{-8}$ & 0.07\\
Ce    & $ -2.540$      & $  4.835\cdot 10^{-4}$  & $ -3.824\cdot 10^{-3}$  & $ -0.2557$  & $  7.455\cdot 10^{-8}$ & 0.08\\
Nd    & $ -2.040$      & $  9.005\cdot 10^{-5}$  & $ -2.517\cdot 10^{-3}$  & $ -0.1347$  & $ -1.526\cdot 10^{-7}$ & 0.13\\
Yb    & $ -2.092$      & $  3.253\cdot 10^{-4}$  & $ -1.741\cdot 10^{-3}$  & $ -0.3196$  & $ -1.519\cdot 10^{-7}$ & 0.12\\
M     & $ -4.693$      & $  6.115\cdot 10^{-5}$  & $ -4.073\cdot 10^{-3}$  & $ -0.5070$  & $  4.116\cdot 10^{-7}$ & 0.01\\
$\alpha$ & $ -4.763$   & $  1.564\cdot 10^{-4}$  & $ -3.170\cdot 10^{-3}$  & $ -0.6181$  & $  2.295\cdot 10^{-7}$ & 0.01\\
\enddata    
\end{deluxetable*}

The uncertainties are included in the summary files in the X\_H\_ERR and X\_M\_ERR arrays, and in the individual X\_FE\_ERR tags.

\subsection{Data model revisions} \label{sec:datamodel}
Several changes have been made to the summary allStar file since DR14:
\begin{itemize}
    \item The uncalibrated, spectroscopically determined \teff~and \logg~ are now presented in ``named tags'' called TEFF\_SPEC and LOGG\_SPEC.
    \item Data from {\it Gaia} DR2 \citep{2016A&A...595A...1G,2018A&A...616A...1G} have been added as a service for the user, and are provided in the tags\\ 
    GAIA\_SOURCE\_ID,  GAIA\_PARALLAX,\\
    GAIA\_PARALLAX\_ERROR,  GAIA\_PMRA,\\
    GAIA\_PMRA\_ERROR,  GAIA\_PMDEC,\\
    GAIA\_PMDEC\_ERROR,\\
    GAIA\_PHOT\_G\_MEAN\_MAG,\\
    GAIA\_PHOT\_BP\_MEAN\_MAG,\\
    GAIA\_PHOT\_RP\_MEAN\_MAG,\\
    GAIA\_RADIAL\_VELOCITY, and\\
    GAIA\_RADIAL\_VELOCITY\_ERROR\\
    To perform the cross-match between APOGEE and {\it Gaia}, the 2MASS cross-match provided by the {\it Gaia} collaboration was used when available, otherwise a positional match, taking the brightest star within 3", was made.
    \item Distances from \citet{2018AJ....156...58B} have been added as a service for the user, and are provided in the tags GAIA\_R\_EST, GAIA\_R\_LO, and GAIA\_R\_HI
    \item New bits have been added in the PARAMFLAG for OTHER\_WARN and ERR\_WARN.
    \item New bits have been added to PARAMFLAG[1] that provide information about which \logg~ calibration was applied: LOGG\_CAL\_RC (red clump, RC), LOGG\_CAL\_RGB (red giant branch, RGB), LOGG\_CAL\_MS (main sequence, MS), or \\ LOGG\_CAL\_RGB\_MS (RGB-MS transition).
    \item New bits have been added in APOGEE2\_TARGET2 and APOGEE2\_TARGET3 targeting flags.
\end{itemize}

In addition, a new allStarLite file is provided that eliminates some of the information in the allStar file, but contains much of the information likely to be of interest for most users. This file is about a third of the size of the full allStar file. It eliminates information about the individual visits that went into the combined frame, as well as several of the arrays that contain uncalibrated stellar parameters and abundances.

The full datamodel for the two allStar-file versions is described in the online documentation\footnote{\url{https://www.sdss.org/dr16/irspec/spectro_data/}}.

\subsection{Caveats}

\subsubsection{Incorrect IDs in some Value Added Catalogs}

SDSS data releases are first distributed internally to the collaboration. This was done for APOGEE DR16 on 2019 June 28. However, it was subsequently discovered that, due to a technical problem in the data processing, 128 faint stars were tagged with the wrong APOGEE\_ID in the internally released allStar file. This was corrected for the public data release. However, some of the {\it value added catalogs} released with DR16 are based on the old, incorrect file. The affected stars are listed in the online documentation.\footnote{\url{https://www.sdss.org/wp-content/uploads/2019/12/incorrect_ids_dr16.txt}}

\section{Discussion} \label{sec:discussion}
This section enumerates and discusses some of the important features of the APOGEE DR16 data about which users should be aware, including the APOGEE bitmasks, validation of derived stellar properties (e.g., radial velocities, atmospheric parameters, elemental abundances), and a summary of these properties.

\subsection{Bitmasks}
Users of the data should pay attention to the bitmask flagging to sort out spectra/stars that are likely inaccurately analyzed by the pipeline. This flagging-system is the same as in DR14, and is described in \citet[][Tables 10-12]{2018AJ....156..125H} and in the online documentation.\footnote{\url{https://www.sdss.org/dr16/irspec/}}

While there are multiple flags for different quantities, two particularly useful ones are the EXTRATARG bitmask, which allows the user to select only the main survey sample (those objects with EXTRATARG==0) and the ASPCAPFLAG STAR\_BAD bit, which is a ``collection''-bit that flags stars for which the derived quantities are unreliable, both from the data point of view, as well as from the determined stellar parameters point of view.
For users of the spectra, the STARFLAG bit may be of utility to identify potentially problematic object spectra.

We note that there is a small bug in the DR16 summary allStar file ASPCAPFLAG bit related to stars falling near a grid edge. For such stars, the appropriate bit (GRIDEDGE\_BAD or GRIDEDGE\_WARN) is set in the relevant PARAMFLAG, \emph{but} this is not propagated, as it should be, into triggering parameter bits in ASPCAPFLAG (e.g., TEFF\_BAD or TEFF\_WARN if a GRIDEDGE bit is set in the \teff\  PARAMFLAG). However, if GRIDEDGE\_BAD is set in any parameter, the STAR\_BAD bit is correctly triggered in ASPCAPFLAG.

\subsection{Radial velocities}\label{sec:rv_val}

\begin{figure*}
\plotone{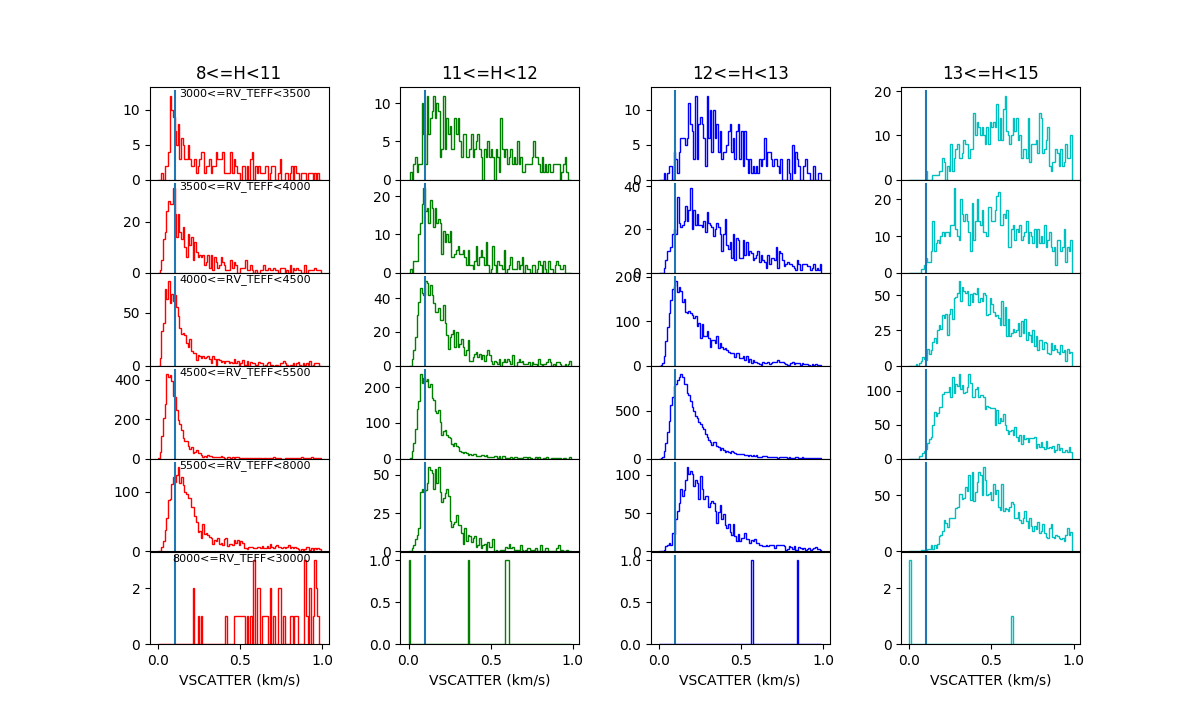}
\caption{Histograms of VSCATTER for stars with NVISITS$>$5 for stars in different bins of apparent magnitude (as a proxy for S/N, since not all visit spectra for an individual star have the same S/N)} and \teff. The vertical lines in each panel are at 100 m/s.
\label{fig:dr16_rv_scatter}
\end{figure*}

The precision of the radial velocities is a function of S/N, \teff, and \mh, with higher precision for brighter, cooler, and more metal-rich stars. This is demonstrated in Figure \ref{fig:dr16_rv_scatter}, which shows histograms of VSCATTER for stars with more than 5 visits. Judging from the peak of the histograms, the RV precision is of order 100 m/s for the best-measured stars and is better than 500 m/s for almost all stars with \teff$<$ 8000 K.

To assess the accuracy of the radial velocities, we compare the APOGEE radial velocities to a set of high-quality literature values accurate to $\sim$ 30 m/s \citep{2002ApJS..141..503N,2012arXiv1207.6212C} as well as to radial velocities from {\it Gaia} \citep{2016A&A...595A...1G,2018A&A...616A...1G}. We compare stars with spectra meeting the criteria APOGEE S/N$>30$, APOGEE VSCATTER$<1$~km/s, APOGEE VERR$<1$~km/s, and {\it Gaia} RV error $<1$~km/s. From Table \ref{tab:rvval}, we find that the APOGEE DR16 radial velocities are offset from the values of \citet{2002ApJS..141..503N} and \citet{2012arXiv1207.6212C} by $\sim0.4$ km/s, as was the case for the DR12 radial velocities \citep{2015AJ....150..173N}. The DR14 radial velocities showed a smaller offset of $\sim$ 0.2 km/s to these two comparison studies \citep{2018AJ....156..125H}.

\begin{figure*}
\plotone{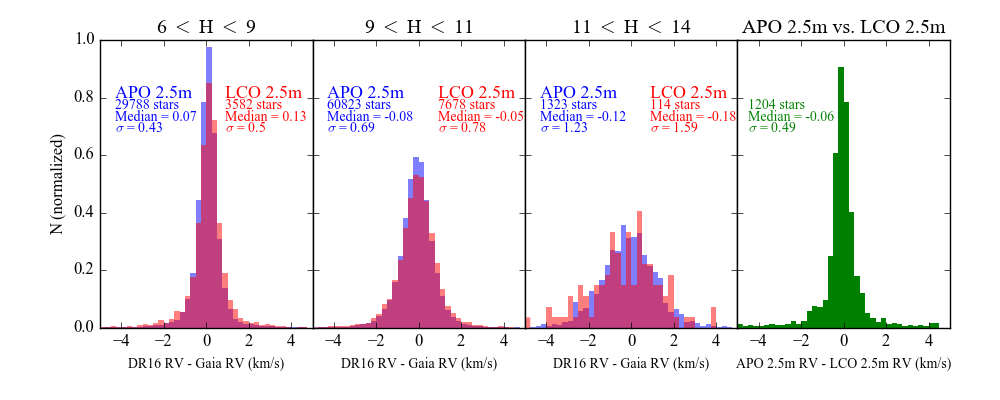}
\caption{Left three panels: RV differences between APOGEE and {\it Gaia} for stars with both APOGEE and {\it Gaia} RV uncertainties $<$ 1 km/s. The blue histogram shows the differences for the northern instrument and the red histogram shows the differences for the southern instrument.
Right panel: RV differences between APO 2.5m and LCO 2.5m for stars observed in both.}
\label{fig:dr16_rv_gaia}
\end{figure*}

However, the {\it Gaia} DR2 radial velocities \citep{2018A&A...616A...5C,2018A&A...616A...6S,2019A&A...622A.205K} are in better agreement with the DR16 APOGEE values than the DR14 APOGEE values. We show the distributions of RV differences between DR16 and {\it Gaia} in the left three panels of Figure \ref{fig:dr16_rv_gaia}. The bright stars ($H<11$) agree to 0.13 km/s or better for both the northern and southern instruments. This offset is larger for the fainter stars ($11<H<14$), with the northern instrument RVs offset by $-0.12$ km/s and the southern instrument RVs offset by $-0.18$ km/s. The standard deviations of the differences plotted in Figure \ref{fig:dr16_rv_gaia} are driven by the individual {\it Gaia} RV uncertainties, which are typically 0.35 km/s, 0.62 km/s, and 0.84 km/s from the bright bin to the faint bin, respectively.

The differences DR14-DR16 are likely because of the updated wavelength calibration (see Section \ref{sec:reduction}). Despite the better agreement with {\it Gaia}, we note that the RVs from RAVE DR5 were found to be offset from the {\it Gaia} RVs by a similar magnitude and direction as the APOGEE DR14 RVs (\citealt{2018AJ....156..170D,2018RNAAS...2..194S}), and this offset persists in RAVE DR6 \citep{2020arXiv200204377S}. Note that  RV offsets at this level, if real, are unlikely to have a significant effect on stellar parameters and abundances because they are a small fraction of the instrumental resolution of APOGEE, which is $\Delta v=c/R \sim 13$ km/s.

\begin{deluxetable*}{ccccc}
\tablecaption{Differences in determined radial velocities between DR16 and literature values for different telescope/instrument combinations. The listed numbers denote the mean and the robust standard deviation (the median absolute deviation divided by 0.67449), and the number in parenthesis is the number of overlapping stars. \label{tab:rvval}}
\tablehead{
\colhead{ } & \colhead{Magnitudes} & \colhead{APO~1~m} & \colhead{APO~2.5~m} & \colhead{LCO~2.5~m}
}
\startdata
$\Delta$\vrad (Nidever+02)       & $2.3< H <5.4$   & $0.44 \pm 0.18$ km/s (6)  & \nodata & \nodata \\
$\Delta$\vrad (Chubak+12)        & $2.9< H <10.2$  & $0.45 \pm 0.69$ km/s (78) & $0.48 \pm 0.66$ km/s (33)      & $0.35  \pm 0.26$ km/s (3) \\
$\Delta$\vrad ({\it Gaia} DR2)   & $6< H <9$       & $0.28 \pm 0.29$ km/s (70) & $0.07 \pm 0.43$ km/s (29\,788) & $0.13  \pm 0.50$ km/s (3582) \\
$\Delta$\vrad ({\it Gaia} DR2)   & $9< H <11$      & \nodata              & $-0.08 \pm 0.69$ km/s (60\,823)     & $-0.05 \pm 0.78$ km/s (7678) \\
$\Delta$\vrad ({\it Gaia} DR2)   & $11< H <14$     & \nodata              & $-0.12 \pm 1.23$ km/s (1323)        & $-0.18 \pm 1.59$ km/s (114) \\
\enddata
\end{deluxetable*}

As another measure of precision and accuracy, we also compare the radial velocities derived for stars observed from both the northern and southern instruments. The results of this comparison are shown in the right panel of Figure \ref{fig:dr16_rv_gaia} and in Table \ref{tab:rvval}. We find a sample of 1204 stars observed from both hemispheres that meet the criteria in both observations of S/N $>30$, VSCATTER$<1$~km/s, and VERR$<1$~km/s. From this sample, we find a radial velocity offset of 0.06 km/s, such that the stars observed from the LCO~2.5~m have slightly lower radial velocities than the same stars observed using the APO~2.5~m. 

For the faintest stars in DR16, the individual visit spectra can have low S/N, and, as a result, the radial velocity determination can fail. In many, but not all, cases, such objects are flagged as having a bad or suspect RV combination. Users who are working with data for stars with $H>14.5$ need to be very careful, as incorrect RVs leads to incorrect spectral combination, which invalidates any subsequent analysis. To minimize inadvertent usage of invalid data, the named abundance tags have not been populated for stars with $H>14.6$.

\subsection{Stellar parameters and the Kiel diagram}
\begin{figure*}
\plotone{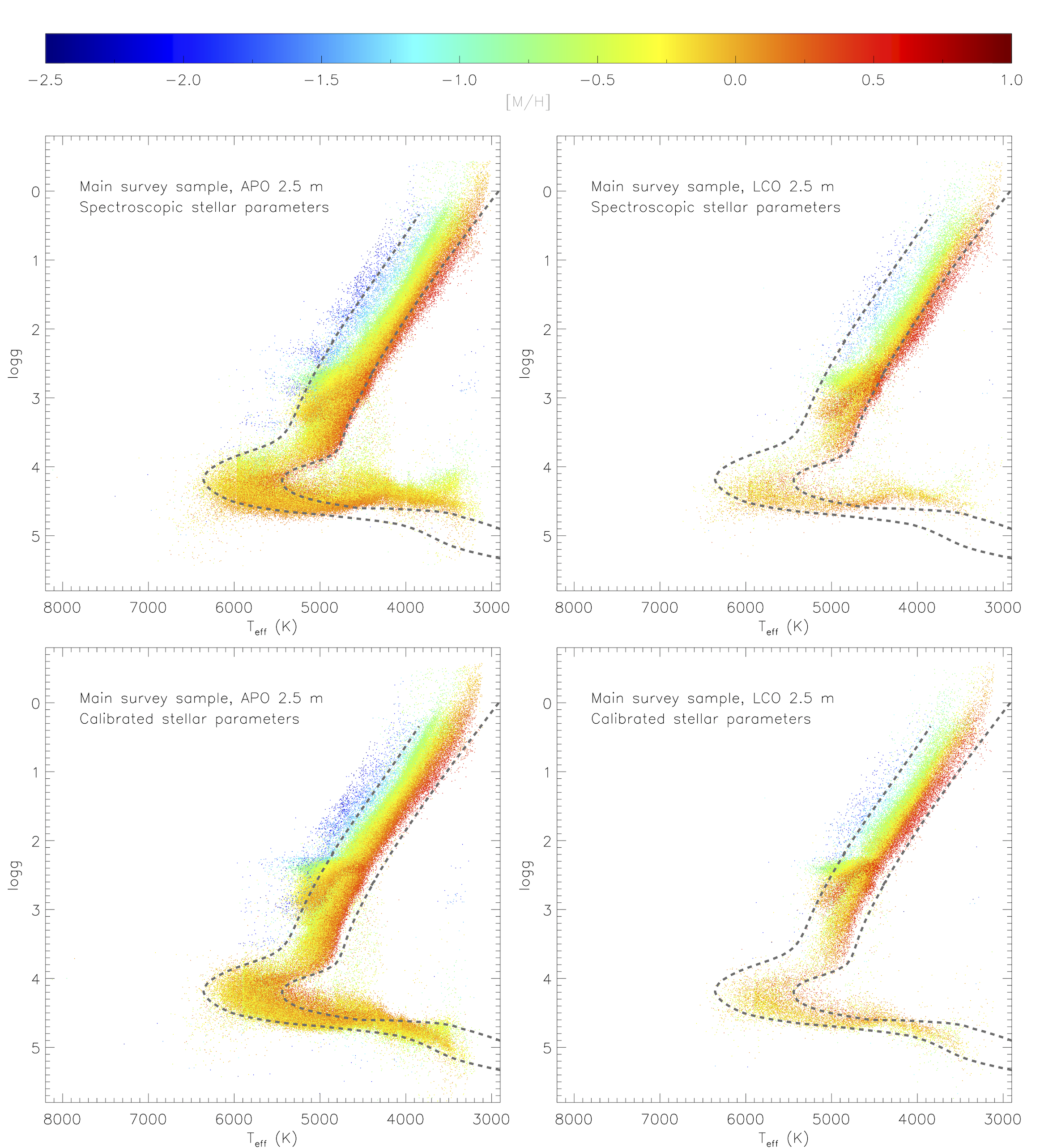}
\caption{Kiel diagrams of the main survey sample collected using APO 2.5 m (left) and LCO 2.5 m (right), plotted using the spectroscopic (top) and the calibrated (bottom) stellar parameters. Stars with the STAR\_BAD bit set in the ASPCAPFLAG have been filtered out. The dashed isochrones are the same as in Figure \ref{fig:grids}. \label{fig:hrd}}
\end{figure*}

Figure \ref{fig:hrd} shows the Kiel diagrams (\teff  ~vs. \logg) of the main survey sample collected from the APO~2.5~m and LCO~2.5~m systems, plotted using the spectroscopic (top panels) and calibrated stellar parameters (bottom panels). Overall, as with previous APOGEE data releases, the general trends are consistent with expectations from isochrones.

\begin{figure*}
\plotone{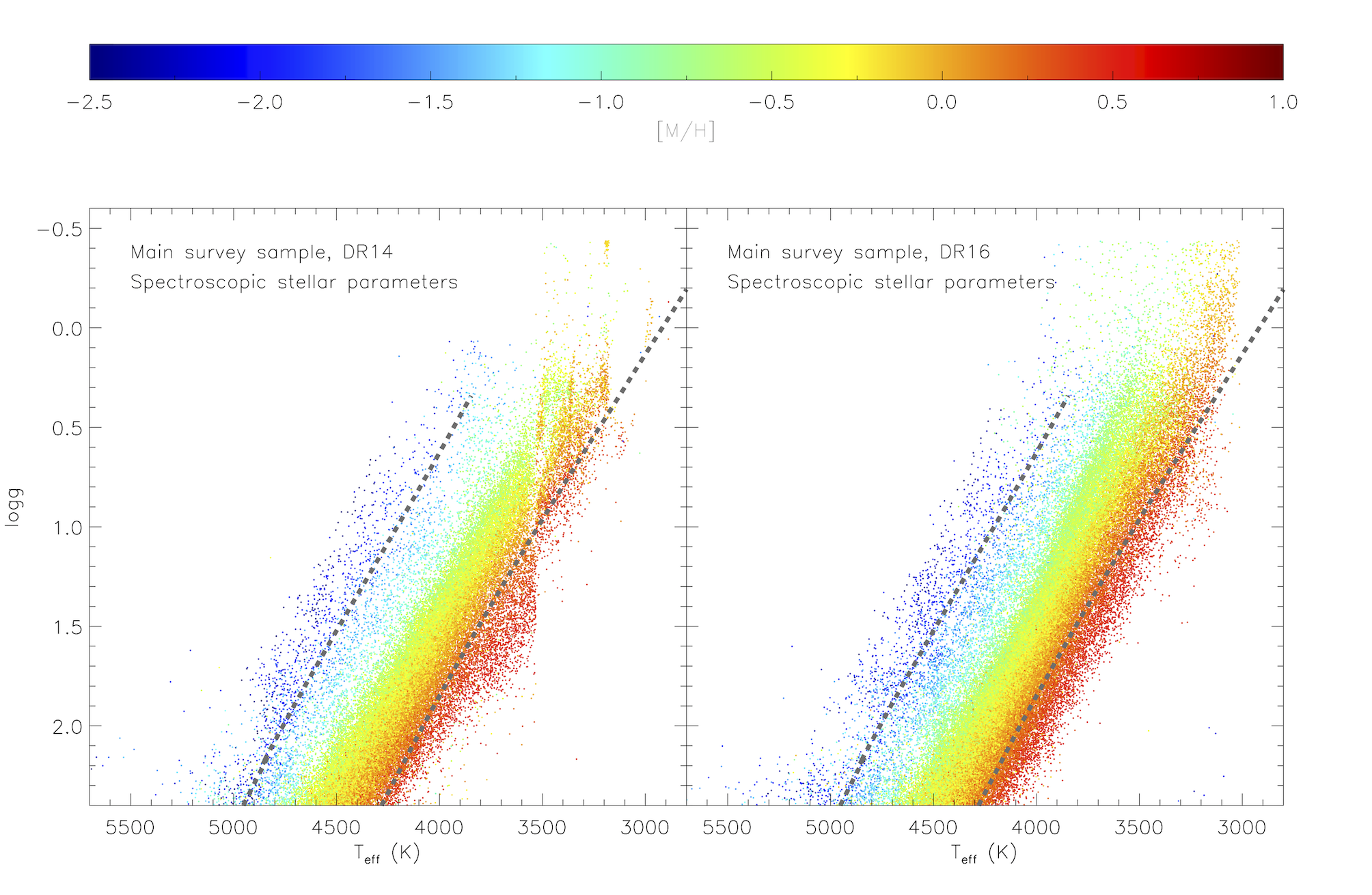}
\caption{Kiel diagrams of the coolest giants for DR14 (left) and DR16 (right). Several improvements can be seen (see text for details). The dashed isochrones are the same as in Figure \ref{fig:grids}.\label{fig:coolgiants}}
\end{figure*}

One of the most significant improvements of DR16 compared to DR14 is that of the accuracy and consistency of determined stellar parameters for the coolest giants (and in turn, abundances derived for these stars). Figure \ref{fig:coolgiants} shows the Kiel diagrams using the spectroscopic stellar parameters for these stars. The DR16 Kiel diagram looks much better than that from DR14 in several aspects. First, the sub-grid edge at \teff$=3500$~K seen in DR14 has disappeared. This change is due to DR16 using only MARCS models, while DR14 used ATLAS-9 models above \teff$=3500$~K, and MARCS below. Second, the unexpected spread of \logg ~values for the most metal-rich giants with \teff ~just above 3500~K in DR14 has disappeared in DR16. It is possible that this improvement can be attributed to the fact that the MARCS models are spherical, while the ATLAS-9 models are plane-parallel. And finally -- and perhaps most strikingly -- the ``clumpiness'' of the Kiel diagram for giants with \teff$<3500$~K in DR14 is gone. This is likely due to the new treatment of holes in the grid, as discussed in Section \ref{sec:rbf}.

All in all, these improvements of the stellar parameters make us more confident in the analysis of the coolest giants in the DR16 data, something that should be welcomed by many users, especially because the coolest giants are in many cases the most distant stars. However, note that we do not have any ``external'' abundances of such cool stars to use for the estimation of accuracy, so users interested in the very coolest giants should still be cautious.

However, comparison of the uncalibrated and calibrated diagrams, along with consideration of the calibration relation, demonstrates several systematic issues with the spectroscopic measurements, which we attempt to compensate for with calibration:

\begin{itemize}
    \item The spectroscopic surface gravities for cooler dwarfs are systematically low with respect to isochrones.
    \item The spectroscopic surface gravities for red giant stars are systematically high, and this difference is larger for red clump stars than for stars on the red giant branch.
\end{itemize}

While we have provided calibrated surface gravities that correct for these effects, we caution that abundances that are sensitive to surface gravity, in particular, could have systematic offsets, especially since we use the spectroscopic surface gravities in the abundance analysis, as described in Section \ref{sec:abundances}. Such elements include, but might not be limited to, Ti II and Ce II.

As described in Section \ref{sec:course}, we have taken great care in choosing which final subgrid to use for analysis of a particular star, and flagged stars that in the final analysis end up close to a grid edge. Even so, there are some overdensities of stars corresponding to the locations that are just outside the flagged regions which are plausibly identified with interpolation issues or, possibly, with unusual spectra. Users should be cautious with stars with parameters close to grid edges (compare Figure \ref{fig:grids}).

\subsection{Chemical abundance trends} \label{sec:trends}
\begin{figure*}
\plotone{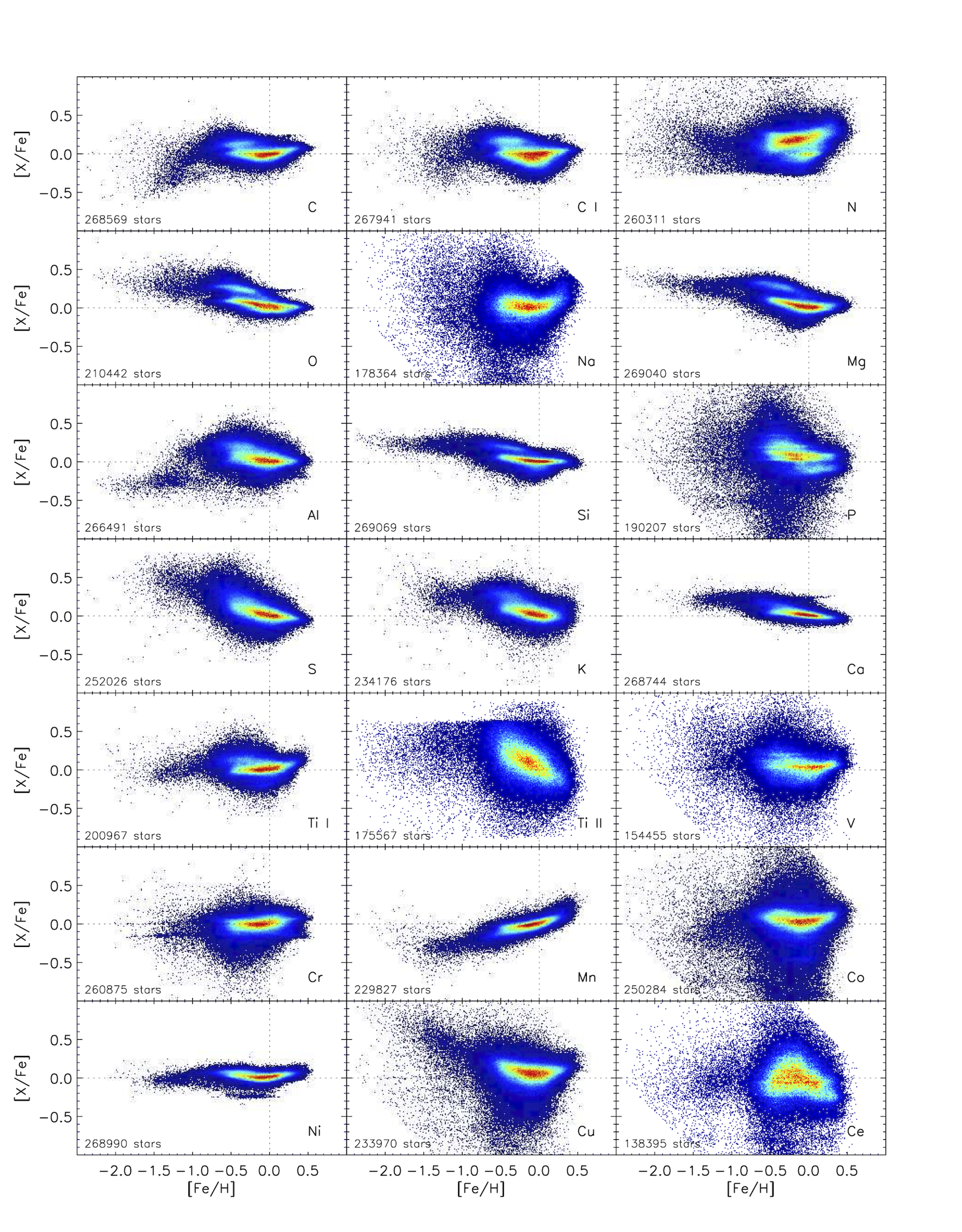}
\caption{Element trends from the ``named'' element tags in DR16 (X\_FE and FE\_H) for all main survey stars (EXTRATARG==0) without the STAR\_BAD bit of the ASPCAPFLAG set. This amounts to 272\,120 stars of very different types from all over the Galaxy. In each panel only stars with the relevant element flag and iron flag equal to zero (X\_FE\_FLAG==0 and FE\_H\_FLAG==0) are included, which is the reason for the different number of stars in the different panels.\label{fig:trends}}
\end{figure*}

Figure \ref{fig:trends} presents the relations between \xfe{X} and \xh{Fe} for main survey stars that do not have the STAR\_BAD flag set, which yields 272\,120 stars. Furthermore, in each panel only stars with the relevant element flag and iron flag equal to zero (X\_FE\_FLAG==0 and FE\_H\_FLAG==0) are plotted. Since different stars have flags set for different elements, the individual panels in the plot have different number of stars. Note that no general cuts on stellar parameters nor position on the sky have been made for this plot, which therefore includes giant and dwarf stars of all effective temperatures and in all parts of the Galaxy. 

Numerous studies have presented diagrams like this, largely for stars in the solar neighborhood, so there are some expectations for what we might expect to see in the APOGEE data.  Based on those expectations, we highlight several areas of difference/concern:

\begin{itemize}
    \item Most of the $\alpha$ elements show a plateau at lower metallicities, as expected, but the value of the plateau --- which, e.g., in \xfe{O} and \xfe{Mg}, is at about 0.3 dex --- is lower than that found in, for example, the optical studies in Section \ref{sec:optical}.
    \item The \xfe{O} vs. \feh~ panel shows a thin ``finger'' of stars sticking out of the high \xfe{O} vs. \feh-trend with \xfe{O}$\sim 0.25$ and \feh$>0$. The same peculiarity is seen in Ca and also in the \am~vs. \mh~ parameter trend, but not as clearly in the \xfe{Si}~vs. \feh~trend, and not at all in the \xfe{Mg}~vs. \feh~trend. The same stars showing this behavior in some of the $\alpha$-elements have a similar feature in C, C I, and \xm{C}, but oddly enough, no discernible matching peculiarity in N. This feature was present in DR14 \citep{2019ApJ...870..138Z}, and seems to affect a small fraction of cool giants. Our estimation is that about 4\% of the ``main'' stellar sample of giants with supersolar metallicity and \teff$<4000$~K -- or of the order of 400 stars (i.e., on the order of 0.1\% of the DR16 stars) -- show these unexpectedly high oxygen abundances. The tightness of the feature in abundance space makes it likely that this is caused by an error in the abundance determination, and that this does not represent a peculiar stellar population. However, despite thorough investigation of the abundance analysis, we could not find the origin of this artifact; thus it is still an open question whether the feature arises from something real in the observed spectra that is not discernible by eye (perhaps leading to inaccurately derived parameters/abundances, but still with some underlying physical origin), or if it simply is the result of some intricate interplay between the different ingredients of the analysis that is leading to incorrect results. Therefore, we recommend caution when using the data for this set of stars.
    \item There is a group of stars with lower-than-expected \xfe{Ni}~values creating a thin horizontal sequence in the \xfe{Ni}~vs. \feh~plot at \xfe{Ni}$\sim-0.25$, which leads to an apparent bimodality. The same type of feature can be seen in Cr, to some extent in Al, and possibly in Co. These kinds of features were not present in DR14, and (post-release of DR16) have been traced to be related to the implementation of the TIE-option in FERRE. While there is little doubt that using the TIE-option is more correct, it seems -- together with the PCA, interpolation in the grid, the minimization, etc. -- to make the $\chi^2$-surface of some solutions shallow enough that there may be multiple minima, and measurements seem to cluster around these values. This problem -- similarly to the [O/Fe]-[Fe/H] ``finger'' discussed above -- seems to  affect giants with \teff$<4000$~K exclusively. However, the two issues do not affect {\it the same stars}, which indicates that the two peculiarities have different causes.
\end{itemize}

\subsection{Abundance trends with stellar parameters} \label{sec:loggtrends}
The primary targets of the APOGEE main survey are red giant stars. For a single population of stars of a given age and metallicity, one expects that stars should have the same \xfe{X} along the giant branch, with the exception of elements that are modified by mixing with material from the stellar interior (see discussion of N below). For a population of mixed age, there might be some astrophysical spread in \xfe{X} at any location on the giant branch, but still no mean trends of abundance along the red giant branch.

\begin{figure*}
\plotone{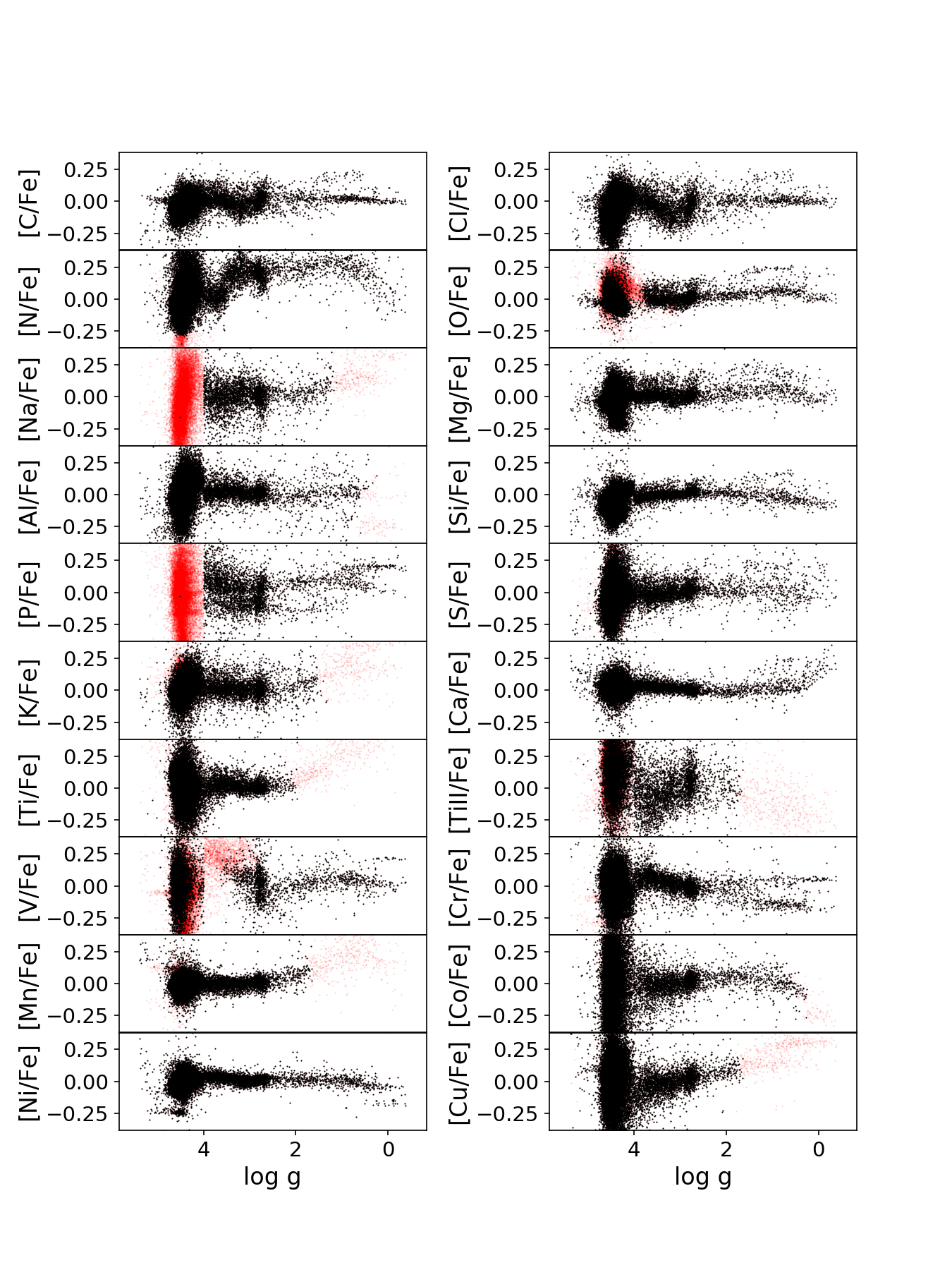}
\caption{\xfe{X} versus surface gravity for stars of near-solar metallicity in the solar neighborhood.  Black points represent calibrated data from the names X\_FE tags, while red points show calibrated data that are in the X\_H and X\_M arrays but not populated in the named X\_FE tags; the latter are converted to be relative to Fe.}\label{fig:logg_abun}
\end{figure*}

Figure \ref{fig:logg_abun} shows the abundances \xfe{X} as a function of surface gravity for stars with near-solar ($-0.1<$\feh$<0.1$) metallicity within 0.5 kpc of the Sun (based on {\it Gaia} parallaxes). These show that, while there is general consistency along the red giant branch, there are some variations. For several elements, the variations are sufficiently large at some surface gravities that they are suspect, and we have chosen in these cases not to populate the named X\_FE tags, although the calibrated abundances are available in the X\_M and X\_H arrays; the unpopulated values are shown in Figure \ref{fig:logg_abun} as red points, while the black points show the populated values. Note that the choices of what regions were chosen to be populated was made by visual inspection, and there is some qualitative judgement involved in the choices that were made.

While the unpopulated X\_FE values remove many of the regions of concern for the abundances, several residual effects remain:
\begin{itemize}
    \item As noted above, a few elements (C, O, Si, Al, P, Cr, Ni) show bimodalities at low \logg/low \teff.
    \item While most elements show relatively little change with \logg, there is still some low-level variation. As a result, if looking for subtle changes in elemental abundances between different data sets, it may be important to ensure that stars of similar \logg ~are being compared.
\end{itemize}

The source of variation of abundance along the giant branch is not fully understood. Possible causes include the use of uncalibrated surface gravities in the abundance determination or changing NLTE or 3D effects along the giant branch.

\subsection{``Non-standard'' abundance ratios}
Since ASPCAP determines the stellar parameters by fitting the entire spectrum, stars with abundance patterns deviating from those used in the calculation of the synthetic spectra might be inaccurately analyzed by the pipeline. For example, second generation globular cluster stars are believed to be oxygen-poor, and since all the $\alpha$-elements are varied together, ASPCAP will struggle to fit a spectrum of a relatively oxygen-poor/calcium-rich star, leading to inaccuracies in the determined stellar parameters and, in turn, in all the subsequently determined abundances. This is a problem intrinsic to our analysis method that has been there from the initial APOGEE DR10 release, and is described in more detail in \citet{2018AJ....156..126J}.

\subsection{Parameter and abundance precision: Stars observed in more than one field}

As already mentioned, many stars have been observed multiple times using both the APO~2.5~m as well as LCO~2.5~m telescope/instrument combinations, and, as described in Section \ref{sec:uncertainties}, the independent ASPCAP results from these different observations of the same stars have been used to estimate the uncertainty in the derived stellar abundances. However, these independent measurements can also be used as a way of assessing and validating the precision of the analysis pipeline, as done in this subsection. First we chose pairs of spectra for all stars observed more than once with the APO~2.5~m and where both spectra have S/N$>100$, STAR\_BAD==0, and calibrated values for all stellar parameters (if more than two such spectra were analyzed, we chose the two with highest S/N); this resulted in 15\,920 pairs of analyzed spectra of the same 15\,920 stars (7\,651 pairs of giants). The same was done for stars observed more than once with the LCO~2.5~m (529 pairs of spectra, 239 pairs of giants), as well as for stars where one spectrum was observed with the APO~2.5~m and the other with the LCO~2.5~m (341 pairs of spectra, 191 pairs of giants). The results of the comparison of the stellar parameters can be seen in Table \ref{tab:dup_params}.

\begin{deluxetable*}{lrcccccc}
\tablecaption{The differences in calibrated \teff, \logg~and \feh, as well as the ``abundance parameters'' \xm{C}, \xm{N}, and \am~for pairs of independently processed high-quality spectra of the same stars for different 2.5~m telescope combinations. The listed numbers denote the mean and the robust standard deviation (the median absolute deviation divided by 0.67449). \label{tab:dup_params}}
\tablehead{
\colhead{ } & \colhead{ } & \colhead{$\Delta$ \teff} & \colhead{$\Delta$ \logg} & \colhead{$\Delta$ \feh} & \colhead{$\Delta$ \xm{C}} & \colhead{$\Delta$ \xm{N}} & \colhead{$\Delta$ \am}
}
\startdata
Giants and dwarfs: \\
APO - APO & 15\,920 pairs & $  0 \pm  21$ & $  0.00 \pm   0.05$ & $  0.00 \pm   0.02$ & $  0.00 \pm   0.03$ & $  0.00 \pm   0.05$ & $  0.00 \pm   0.01$ \\
LCO - LCO &     529 pairs & $  4 \pm  21$ & $  0.01 \pm   0.06$ & $  0.00 \pm   0.02$ & $  0.00 \pm   0.03$ & $  0.01 \pm   0.06$ & $  0.00 \pm   0.01$ \\
APO - LCO &     341 pairs & $ -4 \pm  21$ & $ -0.01 \pm   0.06$ & $  0.02 \pm   0.02$ & $  0.01 \pm   0.03$ & $ -0.02 \pm   0.05$ & $  0.00 \pm   0.02$ \\
Only giants, \logg$<3.5$: \\
APO - APO &  7\,651 pairs & $ -2 \pm  14$ & $  0.00 \pm   0.04$ & $  0.00 \pm   0.02$ & $  0.00 \pm   0.02$ & $  0.00 \pm   0.03$ & $  0.00 \pm   0.02$ \\
LCO - LCO &     239 pairs & $ -2 \pm  14$ & $  0.00 \pm   0.06$ & $  0.00 \pm   0.02$ & $  0.00 \pm   0.03$ & $  0.00 \pm   0.03$ & $  0.00 \pm   0.02$ \\
APO - LCO &     191 pairs & $  2 \pm  13$ & $  0.00 \pm   0.05$ & $  0.01 \pm   0.02$ & $  0.01 \pm   0.02$ & $ -0.02 \pm   0.03$ & $  0.00 \pm   0.02$ \\
\enddata
\end{deluxetable*}

Comparing the ASPCAP results from these independently processed pairs of spectra leads to the following observations:
First the precision is fairly high, especially if only the giants are considered.
Second, the precision is about the same for all three telescope combinations, despite the different numbers of stars utilized in each set.
Finally, the precision in \xm{N} is somewhat lower when dwarfs are included in the comparison. This is expected, however, since the CN molecular lines --- which are the only N-indicators --- become very weak in FGK-dwarf stars.

Table \ref{tab:dup_abunds} lists the comparisons for the derived abundances for the same 7\,651, 239, and 191 pairs of giant spectra. The elements with the highest [X/Fe] precision in this table are C, Mg, Si, Ca, and Ni, which show scatters as low as 0.02 dex. The elements C I, N, O, Al, Ti, and Mn also yield a relatively high precision, in this table with a scatter less than 0.04 dex.  The elements S, K, Cr, and Co could be considered to be of medium precision, based on their scatter of 0.06-0.08 dex, along with Na, V, and Cu, which yield scatters of $\sim$0.10 dex. Not surprisingly, this exercise shows the elemental species with the least precise [X/Fe] to be P, Ti II, and Ce, which show scatters above 0.15 dex.

\begin{deluxetable*}{lrrr}
\tablecaption{The differences in calibrated abundances for pairs of high-quality giant (\logg$<3.5$) spectra of the same stars for different 2.5~m telescope combinations (7\,651, 239, and 191 pairs in total, respectively). The number of pairs for each individual element might however be lower, since only spectra with the relevant X\_FE\_FLAG==0 is used in the comparison. The listed numbers denote the mean and the robust standard deviation (the median absolute deviation divided by 0.67449), and in parenthesis the number of pairs used. \label{tab:dup_abunds}}
\tablehead{
\colhead{ } & \colhead{APO $-$ APO} & \colhead{LCO $-$ LCO} & \colhead{LCO $-$ APO}
}
\startdata
$\Delta$ [C/Fe]     & $  0.00 \pm   0.02$ (7643) & $  0.00 \pm   0.03$ (239) & $ -0.01 \pm   0.02$ (191) \\
$\Delta$ [C I/Fe]   & $  0.00 \pm   0.04$ (7643) & $  0.00 \pm   0.05$ (239) & $ -0.02 \pm   0.05$ (191) \\
$\Delta$ [N/Fe]     & $  0.00 \pm   0.03$ (7500) & $  0.00 \pm   0.03$ (235) & $  0.01 \pm   0.03$ (184) \\
$\Delta$ [O/Fe]     & $  0.00 \pm   0.03$ (7065) & $  0.00 \pm   0.02$ (234) & $  0.00 \pm   0.02$ (178) \\
$\Delta$ [Na/Fe]    & $ -0.01 \pm   0.12$ (7419) & $ -0.01 \pm   0.11$ (195) & $  0.02 \pm   0.11$ (146) \\
$\Delta$ [Mg/Fe]    & $  0.00 \pm   0.02$ (7649) & $  0.00 \pm   0.03$ (239) & $  0.00 \pm   0.03$ (191) \\
$\Delta$ [Al/Fe]    & $  0.00 \pm   0.04$ (7632) & $  0.01 \pm   0.03$ (234) & $  0.01 \pm   0.03$ (187) \\
$\Delta$ [Si/Fe]    & $  0.00 \pm   0.02$ (7651) & $  0.00 \pm   0.03$ (239) & $  0.02 \pm   0.02$ (191) \\
$\Delta$ [P/Fe]     & $ -0.01 \pm   0.17$ (7514) & $ -0.01 \pm   0.15$ (236) & $  0.02 \pm   0.14$ (188) \\
$\Delta$ [S/Fe]     & $  0.00 \pm   0.07$ (7596) & $  0.00 \pm   0.10$ (238) & $ -0.01 \pm   0.09$ (191) \\
$\Delta$ [K/Fe]     & $  0.00 \pm   0.06$ (7375) & $ -0.01 \pm   0.08$ (172) & $ -0.02 \pm   0.06$ (137) \\
$\Delta$ [Ca/Fe]    & $  0.00 \pm   0.02$ (7648) & $  0.00 \pm   0.03$ (239) & $  0.00 \pm   0.02$ (191) \\
$\Delta$ [Ti/Fe]    & $  0.00 \pm   0.04$ (6653) & $  0.01 \pm   0.05$ (140) & $  0.00 \pm   0.03$ (123) \\
$\Delta$ [Ti II/Fe] & $ -0.01 \pm   0.16$ (6764) & $  0.00 \pm   0.19$ (150) & $ -0.01 \pm   0.12$ (125) \\
$\Delta$ [V/Fe]     & $  0.00 \pm   0.10$ (5138) & $  0.00 \pm   0.09$ (208) & $ -0.01 \pm   0.08$ (142) \\
$\Delta$ [Cr/Fe]    & $  0.00 \pm   0.07$ (7615) & $  0.00 \pm   0.08$ (239) & $  0.00 \pm   0.08$ (191) \\
$\Delta$ [Mn/Fe]    & $  0.00 \pm   0.03$ (7182) & $ -0.01 \pm   0.04$ (159) & $  0.00 \pm   0.02$ (136) \\
$\Delta$ [Co/Fe]    & $ -0.01 \pm   0.08$ (7553) & $  0.00 \pm   0.07$ (235) & $  0.04 \pm   0.07$ (187) \\
$\Delta$ [Ni/Fe]    & $  0.00 \pm   0.02$ (7645) & $  0.00 \pm   0.03$ (239) & $ -0.01 \pm   0.03$ (191) \\
$\Delta$ [Cu/Fe]    & $  0.00 \pm   0.09$ (7187) & $ -0.01 \pm   0.12$ (160) & $ -0.02 \pm   0.11$ (136) \\
$\Delta$ [Ce/Fe]    & $  0.00 \pm   0.16$ (6511) & $  0.01 \pm   0.16$ (138) & $ -0.06 \pm   0.27$ (120) \\
\enddata
\end{deluxetable*}

\subsection{Parameter and abundance accuracy: Comparison to optical abundance results for individual stars}\label{sec:optical}

Precision is one thing, but it is also desirable that the stellar parameters and abundances are accurate. However, this is much harder to evaluate. As an example, the ``flat main sequence'' -- the lack of a correlation between \logg~ and \teff~ -- seen in the uncalibrated data in Figure \ref{fig:hrd} as well as in previous APOGEE DRs --- is a sign that we have a problem accurately determining \logg~for dwarf stars. The fact that the main sequence is {\it thin}, however, is a sign that the determination of \logg~is precise, something that is also seen in Table \ref{tab:dup_params}. The calibration of \teff~and \logg~is an attempt to remove systematic uncertainties from our supplied stellar parameters by making them compliant with other, objectively more accurate, values (in the case of \logg), or more widely trusted values (in the case of \teff).

When it comes to abundances, accuracy is even more difficult to assess, because there are few ``objectively accurate'' and/or ``widely trusted'' abundance measurements, especially for giant stars. In \citet{2018AJ....156..126J} we made an attempt to evaluate the accuracy of the abundances in DR13 and DR14 by a comparison to an optical analysis of a subsample of APOGEE giants, and to already published, independent abundance studies having significant overlap with the APOGEE stellar sample \citep{2016ApJS..225...32B,2015A&A...580A..24D,2017A&A...598A.100J}. Table \ref{tab:optical} presents the same evaluation for DR16 (compare to Tables 5-6 in \citealt{2018AJ....156..126J}). Note that the comparison sample of \citet{2017A&A...598A.100J} now is expanded with more elements from \citet{2019A&A...625A.141L} and  \citet{2019A&A...631A.113F}. The details of the different samples are given in detail in \citet{2018AJ....156..126J}, but in general the compared stars are giants with 4000~K$<$\teff $<$5500~K, $1.5<$\logg $<3.8$, and $-1<$\feh $<0.5$.  It is important to reiterate, however, that the optical studies are not necessarily guaranteed to present the ``true'' abundances.

\begin{deluxetable*}{lccccc}
\tablecaption{The differences in calibrated abundances for giants (\logg$<3.5$) in DR16 and independent, high-resolution, optical spectroscopic works. The listed numbers denote the mean and the standard deviation, and in parentheses the number of overlapping giants with that abundance determined. \label{tab:optical}}
\tablehead{
\colhead{ } & \colhead{Brewer+(2016)} & \colhead{da Silva+(2015)} & \colhead{J\"onsson+(2017)} & \colhead{J\"onsson+(2018)} & \colhead{All references}\\ 
\colhead{ } & \colhead{ } & \colhead{ } &\colhead{Lomaeva+(2019)} & \colhead{ } & \colhead{ }\\ 
\colhead{ } & \colhead{ } & \colhead{ } &\colhead{Forsberg+(2019)} & \colhead{ } & \colhead{ } 
}
\startdata
$\Delta$[C/H]      &  0.07 $\pm$ 0.07 (37) & -0.05 $\pm$ 0.05 (30) & ... & ... & 0.02 $\pm$ 0.08 (67)\\
$\Delta$[C I/H]    &  0.01 $\pm$ 0.07 (36) & -0.10 $\pm$ 0.08 (30) & ... & ... & -0.03 $\pm$ 0.09 (66)\\
$\Delta$[N/H]      &  0.00 $\pm$ 0.11 (36) & -0.31 $\pm$ 0.09 (28) &                   ... & ... & -0.15 $\pm$ 0.17 (64)\\
$\Delta$[O/H]      & -0.03 $\pm$ 0.10 (29) & -0.36 $\pm$ 0.06 (25) & -0.07 $\pm$ 0.12 (88) & -0.25 $\pm$ 0.17 (49) & -0.12 $\pm$ 0.17 (191)\\
$\Delta$[Na/H]     & -0.09 $\pm$ 0.12 (38) & -0.26 $\pm$ 0.20 (28) & ... & -0.10 $\pm$ 0.12 (76) & -0.11 $\pm$ 0.16 (142)\\
$\Delta$[Mg/H]     & -0.04 $\pm$ 0.04 (38) & -0.16 $\pm$ 0.06 (30) & -0.06 $\pm$ 0.06 (117) & -0.11 $\pm$ 0.07 (76) & -0.07 $\pm$ 0.07 (261)\\
$\Delta$[Al/H]     & -0.05 $\pm$ 0.06 (37) &                   ... & ... & -0.16 $\pm$ 0.06 (76) & -0.14 $\pm$ 0.08 (113)\\
$\Delta$[Si/H]     &  0.05 $\pm$ 0.12 (38) & -0.10 $\pm$ 0.04 (30) & ... & -0.09 $\pm$ 0.08 (76) & -0.07 $\pm$ 0.11 (144)\\
$\Delta$[S/H]      &                   ... &                   ... & ... & -0.07 $\pm$ 0.12 (63) & -0.07 $\pm$ 0.12 (63)\\
$\Delta$[K/H]      &                   ... &                   ... & ... & -0.27 $\pm$ 0.14 (56) & -0.27 $\pm$ 0.14 (56)\\
$\Delta$[Ca/H]     & -0.06 $\pm$ 0.10 (38) & -0.07 $\pm$ 0.06 (29) & -0.01 $\pm$ 0.06 (117) & -0.16 $\pm$ 0.07 (77) & -0.05 $\pm$ 0.09 (261)\\
$\Delta$[Ti I/H]   & -0.02 $\pm$ 0.16 (38) & -0.08 $\pm$ 0.10 (30) & 0.06 $\pm$ 0.11 (90) & -0.15 $\pm$ 0.11 (68) & -0.04 $\pm$ 0.14 (226)\\
$\Delta$[Ti II/H]  & -0.15 $\pm$ 0.26 (38) & -0.02 $\pm$ 0.21 (30) & 0.18 $\pm$ 0.14 (109) & -0.01 $\pm$ 0.16 (75) & 0.04 $\pm$ 0.22 (252)\\
$\Delta$[V/H]      &  0.11 $\pm$ 0.08 (12) & -0.03 $\pm$ 0.19 (17) & 0.14 $\pm$ 0.13 (102) & 0.01 $\pm$ 0.17 (63) & 0.09 $\pm$ 0.17 (194)\\
$\Delta$[Cr/H]     & -0.06 $\pm$ 0.15 (38) &                   ... & 0.06 $\pm$ 0.09 (116) & -0.06 $\pm$ 0.12 (77) & 0.00 $\pm$ 0.12 (231)\\
$\Delta$[Mn/H]     & -0.15 $\pm$ 0.08 (38) & -0.02 $\pm$ 0.06 (30) & 0.21 $\pm$ 0.09 (100) & 0.07 $\pm$ 0.11 (76) & 0.08 $\pm$ 0.16 (244)\\
$\Delta$[Fe/H]     & -0.13 $\pm$ 0.05 (38) & -0.05 $\pm$ 0.05 (30) & 0.04 $\pm$ 0.05 (117) & -0.10 $\pm$ 0.08 (77) & -0.03 $\pm$ 0.09 (262)\\
$\Delta$[Co/H]     &                   ... &                   ... & 0.02 $\pm$ 0.09 (116) & -0.05 $\pm$ 0.12 (77) & -0.00 $\pm$ 0.10 (193)\\
$\Delta$[Ni/H]     & -0.08 $\pm$ 0.06 (38) & -0.06 $\pm$ 0.05 (30) & 0.10 $\pm$ 0.06 (111) & -0.07 $\pm$ 0.10 (76) & -0.02 $\pm$ 0.12 (255)\\
$\Delta$[Cu/H]     &                   ... &  0.04 $\pm$ 0.11 (29) & ... & -0.03 $\pm$ 0.19 (76) & -0.02 $\pm$ 0.17 (105)\\
$\Delta$[Ce/H]     &                   ... &                   ... & -0.07 $\pm$ 0.16 (105) & ... & -0.07 $\pm$ 0.16 (105)\\
\enddata
\end{deluxetable*}

In general, the DR16 abundances are rather similar to those in DR14, but with a smaller scatter. Below we highlight the main differences between the DR16-optical comparison and the DR14-optical comparison in \citet{2018AJ....156..126J}:
\begin{itemize}
    \item In DR14 \xfe{N} had a systematic shift of 0.08 dex compared to the optical references, while Table \ref{tab:optical} shows that this value now is $-0.15$ dex. The trend of derived \xfe{N} with \feh~that was found based on the comparisons to optical measurements in \citet{2018AJ....156..126J} is still there in DR16. However, determining nitrogen abundances from optical spectra is challenging, and the trend might very well come from the comparison samples.
    \item The possible slight trend of derived \xfe{Mg} with metallicity compared to the optical reference abundances found for DR14 in \citet{2018AJ....156..126J} is essentially gone in DR16. 
    \item Regarding Al, there is a systematic difference as compared to the optical abundances that was not present in DR14 (see Table \ref{tab:optical}). The spread, in $\Delta$[Al/H], however, is lower in DR16. 
    \item The K abundances derived for both DR14 and DR16 are consistent with one another, but very different than those derived optically. The optical measurements follow a much more scattered \xfe{K}~vs. \feh-trend, indicating that the discrepancy might lay with the reference abundances and not APOGEE.
    \item The Ca abundances are very similar in DR14 and DR16, but the possible trend of derived Ca abundance with metallicity seen in DR14 when comparing to optical data seems to have disappeared in DR16.
    \item The possible trend of determined V abundance with metallicity seen in DR14 as compared to the optical studies has vanished in DR16. 
    \item Based on a smaller scatter and offset compared to the optical references, the DR16 abundances of Co are improved compared to those in DR14. 
\end{itemize}

There were no calibrated abundances supplied in DR14 for Cu, and regarding Ce, there were no such abundances in DR14 because these lines were not fully understood until later \citep{2017ApJ...844..145C}. Hence, there were no comparisons made for these elements in \citet{2018AJ....156..126J} and so a bit longer discussion is warranted here. 

\begin{figure*}
\epsscale{1.22}
\plotone{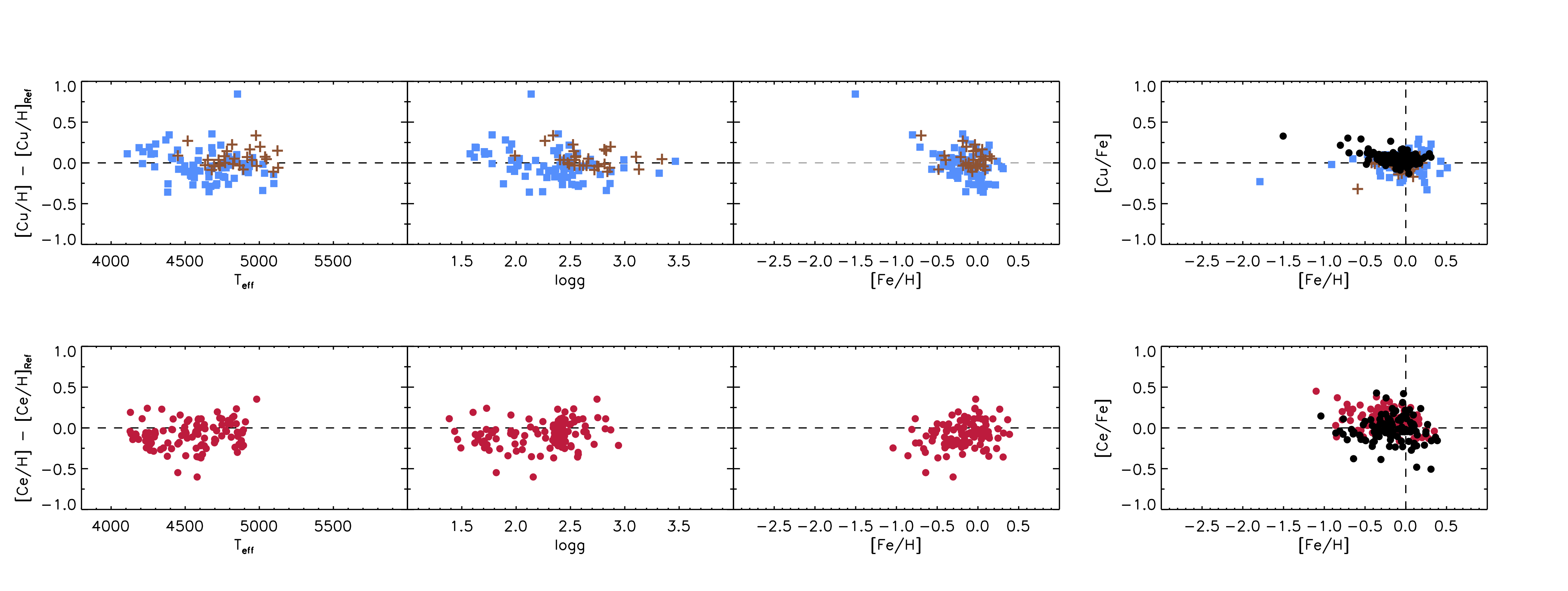}
\caption{Comparisons of DR16 abundances of copper and cerium compared to optical abundances for the very same stars. The left three panels show the difference in derived abundances in the sense DR16-reference as functions of DR16 stellar parameters, and the rightmost panels show the [X/Fe] vs. [Fe/H] trends for these very same stars. Abundances from \citet{2015A&A...580A..24D} are marked using brown crosses, values from \citet{2019A&A...631A.113F} are marked with red dots, values from \citet{2018AJ....156..126J} are marked using blue squares, and the DR16 values are marked with black dots.\label{fig:cuce}}
\end{figure*}

Table \ref{tab:optical} and the top rows in Figure \ref{fig:cuce} indicate a reasonable accuracy for \xfe{Cu} for stars with \feh$>-1$. However, the trend of \xfe{Cu} with \feh, presented in Figure \ref{fig:trends} below, shows a fair amount of scatter, and the rising trend at lower metallicities does not follow the expectations for a weak s-process element to be subsolar for lower metallicities 

The \xfe{Ce}~vs. \feh ~trend in Figure \ref{fig:trends} exhibits significant scatter, but does in principle show a reasonable (banana-like) shape for an s-process dominated element, something that is less obvious in the smaller sample plotted in Figure \ref{fig:cuce}.

In addition to the comparison works above with large overlap with APOGEE, five stars from the high quality sample of Gaia FGK benchmark stars \citep{2018RNAAS...2c.152J} are in APOGEE DR16, and the comparison for these is given in Table \ref{tab:gaiabench}. While this is a small sample of stars, the parameters and abundances in this table highlight some of the issues described in \citet{2018AJ....156..125H,2018AJ....156..126J}, and above:
\begin{itemize}
    \item The APOGEE uncalibrated \teff s are generally too high for metal-rich stars and too low for metal-poor stars.
    \item The APOGEE uncalibrated \logg s are generally too high for giants, while the calibration makes them lower.
    \item The APOGEE alpha-abundances are generally low compared to the optical measurements, except Ti I which has an unexpected rising trend with metallicity in APOGEE.
    \item The mismatch is particularly striking for HD122563, which is not unexpected, since this star has a benchmark-metallicity lower than the ASPCAP grid boundary of \mh=-2.5. However, it is worth noting that the uncalibrated parameters and some of the uncalibrated abundances are quite close to the benchmark values, suggesting that users interested in metal-poor giants might find values of use in some of the arrays with uncalibrated values (FPARAM and FELEM).
    \item The most metal-rich star in the sample, $\mu$Leo, shows several deviations. However, in the case of Mg and Si the values from APOGEE seem more like one would expect from other optical studies of dwarfs stars \citep[][among others]{2003MNRAS.340..304R,2012A&A...545A..32A,2014A&A...562A..71B}.
\end{itemize}

\begin{deluxetable*}{lccrrrrrrrrrr}
\tablecaption{A comparison to the giants among the Gaia FGK benchmark stars \citep{2018RNAAS...2c.152J} also within APOGEE DR16, sorted by benchmark \logg. For every star, there are three lines in the table: the benchmark values, the uncalibrated DR16-values, and the calibrated DR16 values, respectively. \label{tab:gaiabench}}
\tablehead{
\colhead{Star} & \colhead{\teff} & \colhead{\logg} & \colhead{\feh} & \colhead{[Mg/Fe]} & \colhead{[Si/Fe]} & \colhead{[Ca/Fe]} & \colhead{[Ti/Fe]} & \colhead{[V/Fe]} & \colhead{[Cr/Fe]} & \colhead{[Mn/Fe]} & \colhead{[Co/Fe]} & \colhead{[Ni/Fe]}
}
\startdata
$\alpha$Tau  &  3927  &  1.11  & -0.37  &  0.20  &  0.19  &  0.03  &  0.03  &  0.17  &  0.08  & -0.12  &  0.12  & -0.03\\
             &  3889  &  1.50  & -0.10  &  0.04  &  0.04  & -0.04  &  0.15  & -0.10  & -0.08  &  0.01  &  0.14  &  0.04\\
             &  3982  &  1.33  & -0.13  &  0.06  &  0.03  & -0.01  &  ...  &  0.12  & -0.03  &  ...  &  0.14  &  0.06\\
\hline
Arcturus     &  4286  &  1.60  & -0.52  &  0.36  &  0.27  &  0.11  &  0.21  &  0.08  & -0.06  & -0.37  &  0.11  &  0.03\\
             &  4180  &  1.91  & -0.53  &  0.24  &  0.22  &  0.09  &  0.09  & -0.28  & -0.07  & -0.22  &  0.16  &  0.10\\
             &  4291  &  1.75  & -0.55  &  0.25  &  0.20  &  0.10  &  ...  & -0.07  & -0.03  & -0.09  &  0.15  &  0.10\\
\hline
HD122563     &  4587  &  1.61  & -2.64  &  0.29  &  0.31  &  0.21  &  0.14  & -0.06  & -0.44  & -0.46  & -0.00  & -0.05\\
             &  4855  &  1.76  & -2.41  &  0.32  &  0.35  &  0.04  & -0.64  &  0.35  &  0.03  & -0.00  &  0.00  &  0.02\\
             &  5005  &  1.86  &  ...  &  ...  &  ...  &  ...  &  ...  &  ...  &  ...  &  ...  &  ...  &  ...\\
\hline
HD107328     &  4496  &  2.09  & -0.33  &  0.25  &  0.21  &  0.01  &  0.20  &  0.13  & -0.08  & -0.35  &  0.15  &  0.02\\
             &  4299  &  2.04  & -0.41  &  0.25  &  0.25  &  0.04  &  0.06  & -0.03  & -0.14  & -0.24  &  0.20  &  0.12\\
             &  4405  &  1.86  & -0.42  &  0.25  &  0.22  &  0.05  &  0.06  &  0.16  & -0.11  & -0.11  &  0.18  &  0.11\\
\hline
$\mu$Leo     &  4474  &  2.51  &  0.25  &  0.22  &  0.27  &  0.03  &  0.07  &  0.05  &  0.09  & -0.16  &  0.20  &  0.07\\
             &  4519  &  2.82  &  0.34  & -0.05  &  0.01  & -0.06  &  0.11  & -0.09  & -0.09  &  0.00  &  0.19  &  0.06\\
             &  4590  &  2.57  &  0.31  & -0.02  &  0.01  & -0.03  &  0.14  &  0.14  & -0.03  &  0.16  &  0.19  &  0.08\\
\enddata
\end{deluxetable*}

\subsection{Comparisons to cluster metallicities}
Star clusters offer another opportunity to check the DR16 results.  Two recently published studies have undertaken comparisons of APOGEE results to published cluster data.  

As part of the latest contribution from the APOGEE-based Open Cluster Chemical Analysis and Mapping Survey\footnote{\url{https://www.sdss.org/dr16/data_access/value-added-catalogs/?vac_id=open-cluster-chemical-abundances-and-mapping-catalog}} \citep[OCCAM,][]{2020arXiv200208980D}, the DR16-version APOGEE \feh ~were compared to literature metallicity values for six well-studied open clusters, and a mean difference of 0.004 dex was found, in the sense that the APOGEE metallicities are very slightly higher than the literature values. 

For more metal-poor stars, \citet{2019arXiv190103448N} made a comparison of DR16 APOGEE \feh ~and literature values for first-generation globular cluster stars, and found the APOGEE-values to be 0.06 dex higher in the mean, and with a spread of 0.09 dex (see their Figure 8). 

While absolute abundance scales are challenging to establish, these results suggest that APOGEE metallicities have little systematic errors for metal-rich stars, but may have systematic offsets for metal-poor stars at the 0.05-0.1 dex level.

\subsection{Individual elements}\label{sec:elements}
In this section we provide some summary notes about the DR16 abundances for each APOGEE-measured element, including the estimated quality of the abundances, any peculiarities about which the user should be aware, and the population of the named X\_FE tag for the element.

\subsubsection{Carbon, C}
APOGEE carbon abundances are determined largely from CO lines, in conjunction with the oxygen abundances (which are determined from OH lines).  Because the molecular data for both CO and OH lines were changed in DR16, the new carbon abundances are somewhat different from those in previous data releases, and these differences depend on stellar parameters.

Note that C (and N) abundances for dwarfs in the previous DR14 were incorrect because separate C and N dimensions were not included in the stellar parameters stage; this has been fixed for DR16. 

The carbon abundance as determined from molecular lines (C\_FE) is more precise than the carbon abundance determined from C I lines (CI\_FE), but both abundances seem accurate for giant stars according to the comparison to optically-derived abundances.  However, both C and C I abundances are affected by the ``finger'' feature affecting a very small number of giants with \teff$<4000$~K (see Figure \ref{fig:trends} and Section \ref{sec:trends}). For warmer and/or metal-poor stars the ``atomic'' carbon abundance is likely to be more accurate, since the molecular lines become very weak in those instances.

\subsubsection{Nitrogen, N}
The nitrogen abundances are determined from CN lines, in conjunction with the carbon abundance (which is determined from CO lines, as described above). While the CN line list was not modified for DR16, the CO and OH line lists were, and therefore the N abundances in DR16 differ from those in DR14. 

When compared to optical measurements, the precision of APOGEE nitrogen abundances appears good (i.e., there is little scatter), but there is an unexplained trend of \xfe{N} with \teff~when compared with the optical abundances. However, this might very well come from the optical measurements. Moreover, the CN lines become weak or non-existent for warmer stars, which makes the associated uncertainties very large.

\begin{figure}
\epsscale{1.15}
\plotone{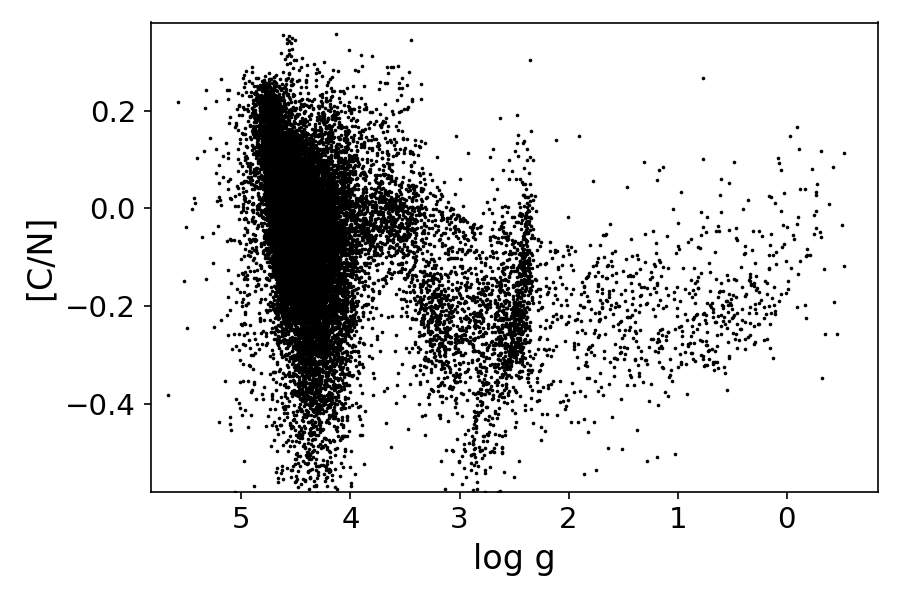}
\caption{[C/N] as a function of calibrated surface gravity for stars with near-solar metallicity in the vicinity of the Sun. The expected drop in [C/N] due to the first-dredge up at the base of the giant branch can be seen (at \logg$\sim3.5$), and the high-density region at \logg$\sim2.5$ which corresponds to the RC stars.\label{fig:cn}}
\end{figure}

Carbon and nitrogen in red giant stars is of particular interest because [C/N] has been found to be an indicator of stellar mass, which corresponds to stellar age for red giant stars \citep{2015MNRAS.453.1855M,2016ApJ...823..114N}. This is because material that has been processed through the CNO cycle is brought to the surface and leads to enhanced N abundances, with the level of enhancement depending on stellar mass and metallicity. Such canonical mixing appears near the base of the giant branch, but metal-poor stars may exhibit extra mixing as the stars ascend the red giant branch \citep{2019ApJ...872..137S}. At solar metallicity, [C/N] is expected to be relatively constant above the base of the giant branch. Figure \ref{fig:cn} shows [C/N] as a function of \logg ~for stars with $-0.1<$ \mh $<0.1$, and demonstrates the expected drop of [C/N] at the base of the giant branch (\logg$\sim3.5$).

The range of [C/N] abundances along the giant branch is plausible if there is a range of ages at solar metallicity. However, the rise of [C/N] at \logg$<1$ is unexpected, and occurs because \xfe{N} is dropping (see Figure \ref{fig:logg_abun}) at low surface gravity, which perhaps makes the measurements of N (and corresponding [C/N]) suspicious in this regime. Also worth noting is the locus at \logg$\sim2.5$ which corresponds to the RC stars, and the fact that some of them show unexpectedly high [C/N]-values above 0.1 \citep{2017MNRAS.464.3021M,2019ApJ...872..137S}.

\subsubsection{Oxygen, O}
The oxygen abundance have been determined using OH lines, which, however, become weak at higher \teff; hence the O\_FE tag is not populated for stars with \teff$>5000$ K.

Compared with optical studies, the derived oxygen abundances have little scatter (i.e., high precision), and appear also to be accurate (i.e., showing small offsets) when referenced to two of the studies, but less accurate compared to the other two (see Table \ref{tab:optical}). Given the difficulty of determining oxygen abundances in general \citep{2004A&A...417..751A,2016MNRAS.455.3735A}, it is not clear which studies (including APOGEE) are the most correct.

A small number of giant stars with \teff$<4000$~K likely have systematic errors that contribute to the formation of the finger of high \xfe{O} stars at high metallicity (see Figure \ref{fig:trends} and Section \ref{sec:trends}).

\subsubsection{Sodium, Na}
Sodium in APOGEE is measured using only two relatively weak lines. Because of this, sodium is one of the least precisely determined element abundances in APOGEE DR16.

Furthermore, one or both of these lines sometimes fall in a region of the spectrum that is potentially affected by poor telluric emission subtraction; 
in ASPCAP such portions of the spectra are treated by dramatically inflating the flux uncertainties, leading to the underlying stellar lines being essentially ignored in the fit for these stars, both in stellar parameters and abundance fits. For Na, this means that for some stars both lines are used in the abundance derivation, while for other stars only a single line is used, leading to even larger sodium abundance uncertainties. 

The NA\_FE tag is only populated for \teff$>3750$ K in giants, and not at all for dwarfs.

For the next data release, NLTE abundance determination of Na is planned to be included \citep{2020arXiv200313353O}. We also hope to be able to identify better those stars for which the abundances are affected by the masking of spectral regions affected by sky lines.

\subsubsection{Magnesium, Mg}
Based on the comparison to optical references, magnesium, together with silicon, is the most precisely determined element in DR16, and also the most accurate element. Mg is also the only $\alpha$-element seemingly free from the ``finger'' feature affecting giants with \teff$<4000$~K (e.g., Figure \ref{fig:trends}). In dwarfs, however, there seems to be a slight systematic effect giving rise to the ``belly'' at slightly subsolar metallicities in the \xfe{Mg}~vs. \feh~plot (Figure \ref{fig:trends}) around \xfe{Mg}$=-0.2$ and \feh$=-0.2$. 

Such systematic shifts between different types of stars \emph{might} be due to NLTE effects. An NLTE abundance determination of Mg is planned to be included in the next data release \citep{2020arXiv200313353O}. These NLTE calculations will differ from the ones of \citet{2017ApJ...835...90Z} in the sense that updated atomic data, in particular for the collisions involving the higher energy levels that form the lines in the $H$-band, will be included.

\subsubsection{Aluminium, Al}
APOGEE Al abundances are determined from three relatively strong lines in the longer wavelength portion of the APOGEE spectra.

Users should be aware that there has been a large zero-point calibration applied to the aluminium abundances for the giants, but not for the dwarfs; this difference possibly indicates the influence of NLTE effects. The derived aluminum abundances seem to be precise, however, as the scatter compared with optical abundances is low. Curiously, the DR16 systematic difference compared to the optical abundances is $-0.14$ dex, which is very close to the zero-point calibration applied for giants ($-0.15$ dex, see Table \ref{tab:abundcal}). If the different zero-point calibration of giants and dwarfs in DR16 \emph{are} indeed due to NLTE-effects, these effects seem to impact similarly the $H$-band lines we use and the optical lines used in the references. For some of the optical lines and the one $H$-band line that have been investigated in \citet[][ see their Figure 13]{2017A&A...607A..75N}, this indeed is the case. 

Aluminum is affected by the unexplained bimodality present in some giants with \teff$<4000$~K (compare Section \ref{sec:trends}). In the case of aluminium, this bimodality is not as obvious in Figure \ref{fig:trends} due to the rather scattered trend, but can be seen as a collection of points at \xfe{Al}=-0.2 for large metallicities around \feh=0.5.
The AL\_FE tag is only populated for \teff$>3400$ K in giants.

\subsubsection{Silicon, Si}
Silicon is one of the most precisely determined elements in DR16. However, a small fraction of stars with \teff$<4000$~K create the ``finger'' feature in Figure \ref{fig:trends} discussed above.

\citet{2016ApJ...833..137Z} calculated NLTE corrections for $H$-band Si lines and found that the corrections should be of the order of $-0.2$ dex for some of the Si I lines in the APOGEE windows, although we do not see this effect when comparing to optical abundances.

\subsubsection{Phosphorous, P}
Phosphorous is measured from a few very weak lines, and is the least precisely determined element abundance in DR16.  It also has had a very large zero-point shift calibration of +0.183 dex applied for giants.  The phosphorous abundances also show some multi-modalities at low temperatures that are most likely non-physical.  Presently there is no known optical comparison sample against which to compare the DR16 values. Hence, for all of these reasons, the DR16 P abundances probably should be avoided or, at minimum, used with extreme caution.

\subsubsection{Sulfur, S}
The DR16 sulfur abundances are determined from just two lines, and exhibit a moderate amount of scatter. Unfortunately, not many optical measurements are available, and we have found no samples with overlap enough to evaluate the accuracy of the APOGEE measurements.
For dwarfs, the S\_FE tag is populated only for \teff$> 4260$ K.

\subsubsection{Potassium, K}
Potassium abundances are determined from two moderately strong lines in the APOGEE wavelength range.  Thus, the abundances are determined with medium precision in DR16. The zero-point offset calibration is about $-0.1$ dex for both giants and dwarfs, perhaps indicating some kind of systematic effect affecting giants and dwarfs similarly. 
Not many optical measurements of K exist to assess the accuracy of the APOGEE abundances, and the one optical comparison sample used is systematically different to APOGEE by 0.27 dex (see \ref{tab:optical}). However, the optical K lines have been shown to suffer large NLTE-effects \citep{2019A&A...627A.177R}.

The K\_FE tag is populated only for \teff$>3900$ in giants, and only for 4000~K$<$\teff$<$6000~K in dwarfs. NLTE abundance determination of K is planned to be included in the next data release \citep{2020arXiv200313353O}.

\subsubsection{Calcium, Ca}
The APOGEE calcium abundances are of high precision. However, their assessed accuracy varies with the optical comparison sample used. A small number of giants with \teff$<4000$~K have the unexplained ``finger'' feature in Figure 
\ref{fig:trends}. For the dwarfs, the \xfe{Ca}~vs. \feh~trend is the tightest among the $\alpha$-elements.

As is the case for Mg, NLTE determination of Ca abundances is planned to be included in the next APOGEE data release \citep{2020arXiv200313353O}, however, these calculations will differ from the ones of \citet{2019ApJ...881...77Z} because updated atomic data for the $H$-band line transitions will be included.

\subsubsection{Titanium, Ti}
The Ti I measurements show low scatter when compared with the optical measurements but, as in previous data releases, the APOGEE \xfe{Ti I} vs. \feh~ trend differs from that seen in optical data: instead of the canonical $\alpha$-element ``knee''-trend, the trend we find is rising with metallicity. This may be due to issues with some of the lines used for measuring the Ti abundance \citep{2016A&A...594A..43H}, or by the combination of a strong \teff ~dependence of the Ti I lines and the use of uncalibrated \teff ~in the abundance measurement \citep{2018AJ....156..126J}.

On the other hand, the Ti II abundances show large scatter when compared to the optical abundances, which is expected since we only have one Ti II line available in the APOGEE wavelength range. The \xfe{Ti II}~vs. \feh~trend does, however, show the expected --- albeit scattered --- ``knee'' trend for giant stars.  Meanwhile, for dwarfs, the Ti II abundances are very scattered.

In conclusion, the DR16 Ti I abundances probably should be avoided, while the Ti II abundances from giant stars might be used with caution in some instances. 

The TI\_FE tag is populated in giants only for \teff$>4200$ K because of apparent trends at cooler \teff, and in dwarfs only for 4000~K$<$\teff$<$6000~K.

\subsubsection{Vanadium, V}
Vanadium is one of the less precise and least accurate abundances in DR16 and, just like phosphorus, shows some odd structure in the \xfe{V}~vs. \feh~trend in Figure \ref{fig:trends}. There is a large zero-point calibration applied for the giants, but not for the dwarfs, possibly on account of differential NLTE effects.  DR16 vanadium abundances should be used with caution.

The V\_FE tag is populated in giants only for \teff$<$4800 and in dwarfs for 4800~K$<$\teff$<$5500~K.

\subsubsection{Chromium, Cr}
Chromium is the element that probably is most affected by the unexpected and unexplained bimodality present in some of the elements for giants with \teff$<$4000~K (see Figure \ref{fig:trends} and Section \ref{sec:trends}). Other than that, Cr is measured with medium-precision and medium-accuracy in DR16, based on the comparison with optical studies.

\subsubsection{Manganese, Mn}
The APOGEE Mn abundances in DR16 have been calibrated using a rather large zero-point abundance shift for both giants and dwarfs (see Table \ref{tab:abundcal}), but the resulting \xfe{Mn}~vs. \feh~trends are very tight. The precision is high, but the accuracy varies a lot depending on which optical comparison sample is used, which makes it difficult to reach a conclusion regarding the accuracy for this element in APOGEE DR16. A lot of the spread in the APOGEE-optical abundance comparison for Mn seems to come from the optical measurements, possibly indicating that the APOGEE abundances are even more precise. The differences between different analyses might arise from different NLTE-effects for the lines used \citep{2019A&A...627A..40E}.

The MN\_FE tag is populated only for \teff$>$4000~K in giants.

\subsection{Iron, Fe}
Iron abundances are determined using windows as for the other elements. In general, the resulting [Fe/H] are very close to the [M/H] determined at the atmospheric parameter-determination stage. However, there are a small number of cooler stars that seem to show [Fe/H] offsets from \mh, with a bimodality similar to that seen in some other elements (see Section \ref{sec:trends}); as with those elements, this is likely some artifact of the analysis.

Stars with \xh{Fe} that differ from \mh ~by more than 0.1 dex have the PARAM\_MISMATCH\_WARN bit in the iron ELEMFLAG set, and those that differ by more than 0.25 dex have the PARAM\_MISMATCH\_BAD bit set. The FE\_H tag is not populated for stars that have either of these bits set. In turn, this will imply that such a star has \emph{none} of the ``named'' tags (C\_FE, N\_FE, O\_FE, etc.) populated.

\subsubsection{Cobalt, Co}
The cobalt abundances are derived from a single line, and therefore it is not unexpected that they show significant scatter. This is especially the case for dwarfs, which contribute most of the scatter in the \xfe{Co}~vs. \feh~trend in Figure \ref{fig:trends}. Compared to the optical abundances, the DR16 Co abundances are of surprisingly high accuracy for giants, but Co is somewhat affected by the bimodality present in some of the elements for giants with \teff$<4000$~K (see Figure \ref{fig:trends} and Section \ref{sec:trends}).

The CO\_FE tag is populated for 3300~K$<$\teff$<$6500~K in giants.

\subsubsection{Nickel, Ni}
Nickel is one of the most precise DR16 abundances, but is unfortunately affected by the bimodality for cool giants with \teff$<4000$~K (see above). There are some systematic differences compared with the optical comparison studies, but, all-in-all, we deem Ni to be the most accurate and precisely measured of the APOGEE iron-peak elements, aside from Fe itself.

\subsubsection{Copper, Cu}
The copper abundances have a larger zero-point calibration offset for the dwarfs as compared to the giants.  Moreover, the precision as well as the accuracy of the copper abundances is rather low. The \xfe{Cu}~vs. \feh~trend shows unexpectedly high \xfe{Cu} for \feh$<-1$. The DR16 copper abundances should be used with caution.

The CU\_FE tag is populated for \teff$>$4000~K in giants.

\subsubsection{Cerium, Ce}
DR16 is the first APOGEE data release to provide cerium abundances, thereby providing APOGEE access to the neutron-capture nucleosynthesis pathway. The cerium abundances were derived from a single line at 15784.8 \AA~(air, 15789.1 \AA~in vacuum), and, because of this, it is not unexpected that the Ce uncertainties are high. In coming data releases we hope to be able to use more of the Ce II lines available in the APOGEE spectral region (presented in \citealt{2017ApJ...844..145C}), but doing so would require a change of our current methodology that uses the \mh~ dimensions when determining Ce abundances, because all of the other Ce II lines unfortunately are blended with another element varying with the same \mh ~grid dimension. 

Compared to the only optical sample with Ce abundances, the DR16 values are systematically offset by 0.07 dex (see Table \ref{tab:optical}). This could be because of a problem in the optical analysis, or perhaps because of systematic uncertainties in the (uncalibrated) surface gravities of 
DR16:  The Ce II lines are sensitive to the adopted \logg, and  --- as previously described --- we use the spectroscopic, uncalibrated \logg~values when determining abundances.

The CE\_FE tag is populated only for 4000~K$<$\teff$<$ 5000~K in giants. Also, as mentioned previously, we do not populate the CE\_FE tag for stars with \vrad$>$120 km/s because, for these stars, the window for the single Ce line that is used shifts into wavelengths that fall in one of the gaps between the APOGEE detectors.

\section{Conclusions} \label{sec:conclusions}
We have presented the data and analysis of the APOGEE-2/SDSS-IV Data Release 16. This includes all data that have been collected by APOGEE and APOGEE-2 up to August 2018, all of which have been processed using the newest versions of the reduction and analysis pipelines. This has resulted in a database with 473\,307 analyzed APOGEE spectra, which includes observations from both the northern and southern hemispheres. Some significant changes/additions for APOGEE DR16 include:
\begin{itemize}
    \item The first data from the APOGEE-S instrument on the du Pont 2.5m telescope at LCO are included. This makes available the first APOGEE observations of the Magellanic Clouds as well as the southern Galactic disk, globular clusters, dwarf satellite galaxies, and more extensive coverage of the Galactic bulge.
    \item Abundances for the neutron-capture element Ce are included for the first time. Note, however, that they are uncertain (see Section \ref{sec:discussion}). 
    \item An all-MARCS grid of model atmospheres has been used, which improves the stellar parameter measurements for cooler stars with \teff$\sim 3500$~K, essentially removing the discontinuity at this \teff ~that was present in APOGEE DR14. 
    \item A new scheme for filling ``holes'' in the spectral grid due to non-converged model atmospheres has led to a significant improvement of the analysis of cool giants with \teff$<3500$~K.
    \item Calibrated surface gravities (\logg ~values) are supplied for dwarf stars.
    \item Various updates to the ASPCAP pipeline (including normalization and $\chi^2$-minimizing algorithm) have led to more precise results.
\end{itemize}

APOGEE DR16 provides abundances for 20 different elements. The most reliable abundances are provided in ``named" tags, X\_FE, in the summary data files (see Section \ref{sec:namedtags}. Users should be aware that the population of abundances in these named tags is subject to several criteria, and that these criteria may bias samples constructed from objects using these abundances, in exchange for providing more reliable abundances.  The abundances that are provided in the X\_H and X\_M arrays in the summary data files are not restricted by these additional criteria.

Based on the analysis done in this paper, we recommend caution in the use of the APOGEE abundances of Na, P, Ti, V, and Cu, even if the abundances in the ``named'' tags (NA\_FE, P\_FE, etc.) are used.

The APOGEE reduction and analysis pipelines will continue to develop for the final data release of APOGEE, DR17, currently planned for public release in mid-2021. 
Possible areas of improvement include abundances derived using NLTE analysis for Na, Mg, K, and Ca, modifications in the derivation of Ce abundances to exploit more of the available Ce II lines, and improvements to the radial velocity code, especially for faint stars.

\acknowledgments
H. J. acknowledges support from the Crafoord Foundation, Stiftelsen Olle Engkvist Byggm\"astare, and Ruth och Nils-Erik Stenb\"acks stiftelse.

J.H. and D.B. acknowledge partial support from NSF grant AST-1715898, and S.R.M. from NSF grants AST-1616636 and AST-1909497.

S.H. is supported by an NSF Astronomy and Astrophysics Postdoctoral Fellowship under award AST-1801940.

C.A.P., D.A.G.H., T.M., Y.O., and O.Z. acknowledge support from the State Research Agency (AEI) of the Spanish Ministry of Science, Innovation and Universities (MCIU) and the European Regional Development Fund (FEDER) under grants AYA2017-86389-P and AYA2017-88254-P.

SzM has been supported by the J{\'a}nos Bolyai Research Scholarship of the Hungarian Academy of
Sciences, by the Hungarian NKFI Grants K-119517 and GINOP-2.3.2-15-2016-00003 of the Hungarian National
Research, Development and Innovation Office, and by the {\'U}NKP-19-4 New National Excellence Program of the 
Ministry for Innovation and Technology.

Funding for the Sloan Digital Sky Survey IV has been provided by the Alfred P. Sloan Foundation, the U.S. Department of Energy Office of Science, and the Participating Institutions. SDSS acknowledges support and resources from the Center for High-Performance Computing at the University of Utah. The SDSS web site is www.sdss.org.

SDSS is managed by the Astrophysical Research Consortium for the Participating Institutions of the SDSS Collaboration including the Brazilian Participation Group, the Carnegie Institution for Science, Carnegie Mellon University, the Chilean Participation Group, the French Participation Group, Harvard-Smithsonian Center for Astrophysics, Instituto de Astrof\'isica de Canarias, The Johns Hopkins University, Kavli Institute for the Physics and Mathematics of the Universe (IPMU) / University of Tokyo, Lawrence Berkeley National Laboratory, Leibniz Institut f\"ur Astrophysik Potsdam (AIP), Max-Planck-Institut f\"ur Astronomie (MPIA Heidelberg), Max-Planck-Institut f\"ur Astrophysik (MPA Garching), Max-Planck-Institut f\"ur Extraterrestrische Physik (MPE), National Astronomical Observatories of China, New Mexico State University, New York University, University of Notre Dame, Observat\'orio Nacional / MCTI, The Ohio State University, Pennsylvania State University, Shanghai Astronomical Observatory, United Kingdom Participation Group, Universidad Nacional Aut\'onoma de M\'exico, University of Arizona, University of Colorado Boulder, University of Oxford, University of Portsmouth, University of Utah, University of Virginia, University of Washington, University of Wisconsin, Vanderbilt University, and Yale University.

This publication made use of NASA's Astrophysics Data System.

\facilities{Sloan (APOGEE)}

\software{
ASPCAP \citep{2016AJ....151..144G},
AstroLines \citep{2015ascl.soft02022B}, 
FERRE  \citep{2006ApJ...636..804A}, 
MARCS \citep{2008A&A...486..951G}, 
Turbospectrum \citep{1998A&A...330.1109A,2012ascl.soft05004P}
}

\bibliographystyle{yahapj}
\bibliography{papers.bib,others.bib}

\end{document}